\documentclass[12pt]{article}
\usepackage{graphicx} 
\usepackage{natbib}
\usepackage{url}
\usepackage{subfig}
\usepackage{dirtytalk}
\usepackage[table]{xcolor}
\usepackage{amsmath,amssymb,amsfonts}%
\usepackage{amsthm}%
\usepackage{mathrsfs}%
\usepackage{booktabs}%
\usepackage{multicol}
\usepackage{multirow}
\usepackage{hyperref}

\renewcommand{\vec}[1]{\mathbf{#1}}
\newcommand{\mean}{\boldsymbol{\mu}}
\newcommand{\meang}{\boldsymbol{\mu}_{g}}
\newcommand{\meangi}{\boldsymbol{\mu}_{g[\vec{w}_{i}]}}

\newcommand{\sigu}{\boldsymbol{\vec{\Sigma}}}
\newcommand{\sigug}{\boldsymbol{\vec{\Sigma}}_{g}}
\newcommand{\sigugii}{\boldsymbol{\vec{\Sigma}}_{g[\vec{w}_{i}, \vec{w}_{i}]}}
\newcommand{\sigugest}{\widehat{\boldsymbol{\vec{\Sigma}}}_{g}}

\newcommand{\tr}{\text{tr}}

\newcommand{\blind}{0}

\begin{document}

\def\spacingset#1{\renewcommand{\baselinestretch}%
{#1}\small\normalsize} \spacingset{1}

\if0\blind
{
  \title{\bf Cellwise outlier detection in heterogeneous populations}
    \author{Giorgia Zaccaria\footnote{Corresponding author: \href{giorgia.zaccaria@unimib.it}{giorgia.zaccaria@unimib.it}}\\
    \normalsize{Department of Statistics and Quantitative Methods, University of Milano-Bicocca} \\
    and \\
    Luis A. García-Escudero \\ 
    \normalsize{Department of Statistics and Operational Research, University of Valladolid} \\
    and \\
    Francesca Greselin \\  
    \normalsize{Department of Statistics and Quantitative Methods, University of Milano-Bicocca}  \\
    and \\
    Agustín Mayo-Íscar \\ 
    \normalsize{Department of Statistics and Operational Research, University of Valladolid}}
    \date{}
  \maketitle
} \fi

\if1\blind
{
  \bigskip
  \bigskip
  \bigskip
  \begin{center}
    {\LARGE\bf Title}
\end{center}
  \medskip
} \fi

\vspace{-0.5cm}
\begin{abstract}
Real-world applications may be affected by outlying values. In the model-based clustering literature, several methodologies have been proposed to detect units that deviate from the majority of the data (rowwise outliers) and trim them from the parameter estimates. However, the discarded observations can encompass valuable information in some observed features. Following the more recent cellwise contamination paradigm, we introduce a Gaussian mixture model for cellwise outlier detection. The proposal is estimated via an Expectation-Maximization (EM) algorithm with an additional step for flagging the contaminated \textit{cells} of a data matrix and then imputing -- instead of discarding -- them before the parameter estimation. This procedure adheres to the spirit of the EM algorithm by treating the contaminated cells as missing values. We analyze the performance of the proposed model in comparison with other existing methodologies through a simulation study with different scenarios and illustrate its potential use for clustering, outlier detection, and imputation on three real data sets. Additional applications include socio-economic studies, environmental analysis, healthcare, and any domain where the aim is to cluster data affected by missing information and outlying values within features. 
\end{abstract}

\bigskip
\noindent%
\vfill

\newpage
\spacingset{2} 
\section{Introduction} \label{sec: intro}
Real data often contain outlying and missing values. For instance, health records tracking the presence of a disease could be affected by missing or noisy measurements, while socio-economic surveys about individuals or firms might contain intentionally incorrect responses. In robust statistical literature, several methodologies have been proposed to prevent biased parameter estimates by detecting and downweighting or discarding contaminated cases within a data matrix. These cases typically correspond to entire rows, therefore referred to as rowwise/casewise outliers \citep{H:1964}, which are assumed not to follow the distribution of the majority of the data. In recent years, the component-wise contamination model proposed by \citet{AVAYZ:2009} has garnered increasing attention from researchers. This introduced a new contamination paradigm that assumes some \textit{cells} of a data matrix have been replaced by arbitrary values. According to it, a low percentage of cellwise contamination potentially corrupts many rows, or even all of them, as the number of variables increases. Consequently, casewise trimming would discard valuable information encompassed in the uncontaminated cells of the rows or be unfeasible. In the single-population framework, the cellwise Minimum Covariance Determinant estimator \citep[cellMCD,][]{RR:2023} represents the most recent proposal for robustly estimating the location and scale parameters in the presence of cellwise contamination, forerunning by the Detection-Imputation (DI) method \citep{RR:2021a} and other methodologies for several purposes, such as principal component analysis (\citealp{HRvdB:2019}; see \citealp{RR:2024}, for a comprehensive review). Both cellMCD and DI, being based on the Expectation-Maximization (EM) algorithm \citep{DLR:1977}, can further handle missing values, faithfully adhering to its spirit by treating contaminated cells as missing information to be imputed. However, they assume one single homogeneous population and cannot deal with heterogeneity in the data, which is common in practice. 

For uncovering contamination in heterogeneous populations, a first class of models has been proposed which accommodate to the presence of outliers by partially relaxing the normality assumption and considering heavy-tailed distributions for the components. This class encompasses the mixture of $t$ distributions \citep{PMcL:2000} and the mixtures of contaminated normal distributions \citep{PMN:2016} as examples, both extended to handle Missing At Random \citep[MAR,][]{R:1976} values in the data by \citet{WZLW:2004} and \citet{TT:2022}, respectively, as is common in other statistical methodologies \citep{LR:2019}. To deal with contamination in general position in model-based clustering, \citet{GEGMMI:2008} introduced TCLUST, which extended the Minimum Covariance Determinant \citep[MCD,][]{R:1984, R:1985} estimator for the location and covariance matrix through the use of classification trimmed likelihoods. Specifically, TCLUST includes a \textit{Concentration} step (C-step) in the EM algorithm, similar to that in the faster MCD algorithm (FAST-MCD, \citealp{RvD:1999}), where a fixed fraction of observations whose contribution to the objective function is the smallest are considered as outliers and discarded from the parameter estimation. An analogous trimming approach was considered in a mixture likelihood model-based framework in \citet{NFDN:2007} and \citet{GEGMY:2014}.

The goal of this paper is to provide a model-based extension of cellMCD in the clustering framework for coping with cellwise contamination and MAR information. The proposal, called cellwise Gaussian Mixture Model (cellGMM), is based upon the maximization of the log-likelihood via a mixture EM algorithm with a fixed number of components and including constraints to avoid spurious solutions \citep{GEGMY:2014}. The difficult challenge of simultaneously detecting clustering structures via a model-based approach and identifying outlying cells position, influenced by the variable dependence within sub-populations, has been previously addressed by \citet{F:2014}. His snipping proposal for Gaussian mixture models (sclust) involves the removal of contaminated cells from the E- and M-step of the algorithm. Nonetheless, sclust has some limitations that our proposal seeks to address. Specifically, cellGMM supplies imputation, say \say{correction}, for outlying values, as well as for missing values whose positions are known. This improves the efficiency of the parameter estimators and eliminates the need for additional correction procedures to avoid undesirable bias (shrinkage) of the component covariance matrices that can occur in discarding approaches. Additionally, cellGMM, unlike sclust, can automatically determine the contamination level in the data matrix while retaining a certain fraction of observations considered non-contaminated per variable. These features enable cellGMM to obtain better results in terms of clustering performance, recovery of outlying cells, and efficiency in parameter estimation, as shown in the simulation study presented in Section \ref{sec: simulation}. It is worth noting that the initialization of cellGMM is a crucial issue, which we will discuss in Section \ref{subsec: computationalasp} and in detail in the Supplementary Material available online. Other approaches for robust clustering based on cellwise trimming can be found in \citet{F:2013} and \citet{GERGMIO:2021}. These approaches are no longer be considered in this work because they do not provide estimators of the component covariance matrices and are essentially based on searching for approximating sets made of $G$ points or $G$ affine sub-spaces in the observed data.

The rest of the paper is organized as follows. We describe our proposal in Section \ref{sec: cellGMM}, along with its computational aspects. Sections \ref{sec: simulation} and \ref{sec: application} provide evidence of the cellGMM performance compared to other existing methodologies for casewise and cellwise outlier detection through a simulation study and three real-world applications involving spectral data, image reconstruction and data on cars' features. Finally, a discussion reviews the obtained results and presents open research lines for future investigation in Section \ref{sec: discussion}.  

\section{Gaussian mixture models with cellwise outliers} \label{sec: cellGMM}
The proposed methodology follows in the footsteps of cellMCD by leveraging the EM algorithm in the model-based clustering framework. Consider a $p$-dimensional random vector drawn from a mixture of $G$ multivariate Gaussian distributions, whose probability density function (pdf) is given by
\begin{equation}\label{eqn: GMM}
    f(\vec{x}; \vec{\Psi}) = \sum_{g = 1}^{G} \pi_{g} \phi_{p} \big(\vec{x}; \meang, \sigu_{g} \big),
\end{equation}
where $\phi_{p}(\cdot; \mean,\sigu)$ is the pdf of the $p$-variate normal with mean $\mean$ and covariance matrix $\sigu$. $\vec{\Psi} = \{ \boldsymbol{\pi}, \boldsymbol{\theta}\}$ is the overall parameter set composed of $\boldsymbol{\pi} = \{\pi_{g}\}_{g = 1}^{G}$ with weights $\pi_{g} \in (0, 1]$ and such that $\sum_{g = 1}^{G} \pi_{g} = 1$, and $\boldsymbol{\theta} = \{\meang, \sigug \}_{g = 1}^{G}$ with component mean vectors $\meang \in \mathbb{R}^{p}$ and symmetric positive definite component covariance matrices $\sigug \in \mathbb{R}^{p \times p}$. 

Let $\vec{X} = [x_{ij}: i = 1, \dots, n, j = 1, \ldots, p]$ be a data matrix, whose rows $\vec{x}_{1}, \ldots, \vec{x}_{n}$, with $\vec{x}_{i} \in \mathbb{R}^{p},$ are supposed to be a random sample drawn from GMM in (\ref{eqn: GMM}). It can happen that some individual measurements or cells, i.e., some $x_{ij}$, have been replaced by outlying values. We refer to the latter as \textit{contaminated} or \textit{unreliable} cells, that can be distributed everywhere throughout the sample. To track their pattern, we define the matrix $\vec{W} = [w_{ij}: i = 1, \ldots, n, j = 1, \ldots, p]$, where $w_{ij} = 1$ if $x_{ij}$ is reliable, and $w_{ij} = 0$ if not because it has been contaminated. We use herein the notation $\vec{w}_{i}, i = 1, \ldots, n$, when referring to the rows of $\vec{W}$, but we denote with $\vec{W}_{\cdot j}$ the $j$th column of that matrix (equivalent notation will be used for other matrices considered in this work). According to the values of $\vec{w}_{i}$, we partition $\vec{x}_{i}$ into $\vec{x}_{i[\vec{w}_{i}]}$ and $\vec{x}_{i[\vec{1}_{p} - \vec{w}_{i}]}$ (respectively, reliable and unreliable cells in the observation $\vec{x}_{i}$), where $\vec{1}_{p}$ is a $p$-dimensional unitary vector. Henceforth, we refer to $\vec{1}_{p} - \vec{w}_{i}$ as $\vec{w}_{i}^{c}$ for simplicity. In a similar manner, given a vector $\mean \in \mathbb{R}^{p}$, $\mean_{[\vec{w}_{i}]}$ stands for its sub-vector in $\mathbb{R}^{p[\vec{w}_{i}]}$, where $p[\vec{w}_{i}] = \sum_{j = 1}^{p} w_{ij}$, i.e., the sub-vector corresponding to the cells of $\vec{x}_{i}$ for which $w_{ij} = 1$. Analogously, $\sigu_{[\vec{w}_{i},\vec{w}_{i}]}$ symbolizes the $p[\vec{w}_{i}] \times p[\vec{w}_{i}]$ sub-matrix of $\sigu \in \mathbb{R}^{p \times p}$ obtained by keeping only the rows and columns of $\sigu$ whose indexes $j$ satisfy $w_{ij} = 1$ in $\vec{w}_{i}$. Finally, with $\mean_{[j]}$ ($\sigu_{[j, j]}$) we extract the $j$th ($jj$th) element of $\mean$ ($\sigu$).  

Considering the notation introduced, the objective function of the Gaussian Mixture Model with cellwise outliers (cellGMM) to maximize is 
\begin{equation}\label{eqn: obj}
    \ell(\vec{\Psi}, \vec{W}; \vec{X}) = \sum_{i = 1}^{n} \ln \Big[ \sum_{g = 1}^{G} \pi_{g} \phi_{p[\vec{w}_{i}]} (\vec{x}_{i[\vec{w}_{i}]}; \meangi, \sigugii) \Big], 
\end{equation}
subject to constraints
\begin{align}
   &\lVert \vec{W}_{\cdot j} \rVert_{0} \geq h, \forall j = 1, \ldots, p, \label{eqn: constrW} \\
   &\dfrac{\max\limits_{g = 1, \ldots, G}\max\limits_{j = 1, \ldots, p} \lambda_{j}(\sigug)}{\min\limits_{g = 1, \ldots, G}\min\limits_{j = 1, \ldots, p} \lambda_{j}(\sigug)} \leq c, \label{eqn: constrS}
\end{align}
where $\lambda_{j}(\sigug)$ is the $j$th eigenvalue of the covariance matrix $\sigug$ and $c \geq 1$ is a fixed constant \citep{GEGMMI:2008}. The constraint in (\ref{eqn: constrW}) requires at least $h$ reliable cells in each column, so we set $h \geq 0.75n$ \citep[see][]{RR:2023}. The eigenvalue-ratio constraint in (\ref{eqn: constrS}) addresses the problem of the unboundedness of (\ref{eqn: obj}) and the spurious solutions that may result from its maximization, while maintaining the positive definiteness of $\sigug$.

To solve the constrained maximization in (\ref{eqn: obj}), we consider an adaptation of the EM algorithm -- typically used for mixture modeling and handling of missingness in the data -- which allows to simultaneously detect the outlying cells and find the maximum likelihood estimates of the model parameters. In this EM framework, there are multiple sources of unknown information beyond the model parameters: firstly, the outlying cells of $\{ \vec{x}_{i} \}_{i = 1}^{n}$ corresponding to the zeros in $\{ \vec{w}_{i} \}_{i = 1}^{n}$, and, secondly, the indicator to the population belonging, also called unit-component membership, reported in $\vec{Z} = [z_{ig}: i = 1, \ldots, n, g = 1, \ldots, G]$, where $z_{ig} = 1$ if the $i$th observation belongs to the $g$th component of the mixture, and $z_{ig} = 0$ otherwise. Let $\vec{z}_{i}$ and $\vec{x}_{i[\vec{w}_{i}^{c}]}$ be the \say{missing} data, and $\vec{x}_{i[\vec{w}_{i}]}$ the observed data. The cellGMM complete data log-likelihood is 
\begin{equation}\label{eqn: lc}
    \ell_{c}(\vec{\Psi}, \vec{W}, \vec{Z}; \vec{X}) = \sum_{i = 1}^{n} \sum_{g = 1}^{G} z_{ig} \Big[  \ln(\pi_{g}) +  \ln \big( \phi_{p} (\vec{x}_{i[\vec{w}_{i}]}, \vec{x}_{i [\vec{w}_{i}^{c}]}; \meang, \sigug) \big) \Big] 
\end{equation}
subject to constraints (\ref{eqn: constrW}) and (\ref{eqn: constrS}). In the following, we detail the steps of the EM algorithm for the estimation of the cellGMM parameters. The algorithm starts with initial solutions for the parameters, which will be discussed in Section \ref{subsec: computationalasp}. The C-step involves estimating $\vec{W}$, whereas the E-step consists of computing expectations of the remaining missing data. In the M-step, we derive the estimates of the overall parameter set $\vec{\Psi}$.

\textbf{Update of $\vec{W}$.}
At iteration $(t+1)$, given $\widehat{\vec{\Psi}}^{(t)}$ and $\vec{Z}^{(t)}$, we update the configuration of $\vec{W}$ column-by-column. To simplify the notation, we use an intermediate $\widetilde{\vec{W}}$ matrix before providing the updated $\vec{W}^{(t+1)}$. Consequently, we start from $\widetilde{\vec{W}} = \vec{W}^{(t)}$ and we update $\widetilde{\vec{W}}_{\cdot j}$ sequentially for $j = 1, \ldots, p$, by considering the previously estimated $(j-1)$ columns of $\widetilde{\vec{W}}$ as fixed. To this aim, we define the contribution of the $i$th observation to the cellGMM objective function in (\ref{eqn: obj}) as
\begin{equation}\label{eqn: li}
    \ell_{(i)}(\vec{\Psi}, \vec{w}_{i}; \vec{x}_{i}) = \ln \Big[ \sum_{g = 1}^{G} \pi_{g} \phi_{p[\vec{w}_{i}]} (\vec{x}_{i[\vec{w}_{i}]}; \meangi, \sigugii) \Big].
\end{equation}
For the $j$th column of $\widetilde{\vec{W}}$, we compare the contribution of the $i$th unit to the observed log-likelihood in (\ref{eqn: li}) when we modify the $j$th element of $\widetilde{\vec{w}}_{i}$ such that its value for the $i$th observation is considered reliable ($w_{ij} = 1$) or contaminated ($w_{ij} = 0$), while the other terms in $\widetilde{\vec{w}}_{i}$ remain unchanged. Therefore, we compute
\begin{equation}\label{eqn: deltaZ}
   \Delta_{ij} = \ell_{(i)}(\widehat{\boldsymbol{\Psi}}^{(t)}, \widetilde{\vec{w}}_{i}; \vec{x}_{i}, \widetilde{w}_{ij} = 1) - \ell_{(i)}(\widehat{\boldsymbol{\Psi}}^{(t)}, \widetilde{\vec{w}}_{i}; \vec{x}_{i}, \widetilde{w}_{ij} = 0).
\end{equation}
With the aim of maximizing the observed log-likelihood $\sum_{i = 1}^{n} \ell_{(i)}(\widehat{\vec{\Psi}}^{(t)}, \vec{w}_{i}; \vec{x}_{i})$ subject to constraint (\ref{eqn: constrW}), we attain the optimum by setting $\widetilde{w}_{ij} = 1$ for all indexes $i$ corresponding to the $h$ largest values of $\{ \Delta_{ij} \}_{i = 1}^{n}$, $0$ otherwise. Repeating this procedure for all columns of $\widetilde{\vec{W}}$ we obtain $\vec{W}^{(t+1)} = \widetilde{\vec{W}}$. As shown in \citet{RR:2023}, the chosen order for the update of $\vec{W}$ does not seem to significantly affect the performance of the procedure. It is worth noting that the rationale for updating $\vec{W}$ is inspired by the approach proposed in Raymaekers and Rousseeuw (2023). Nevertheless, it is tailored to a Gaussian mixture model framework.

\textbf{Update of $\vec{Z}$ and $\vec{X}_{[\vec{W}^{c}]}$.}
At iteration $(t+1)$, we update the membership matrix $\vec{Z}$ and the unreliable data in $\vec{X}$ corresponding to the zero elements into $\vec{W}^{(t+1)}$. For this purpose, we consider the outlying cells uncovered in $\vec{W}^{(t+1)}$ as missing values. We compute the expected value of the complete data log-likelihood in (\ref{eqn: lc}) conditional on the observed and reliable data, given $\vec{W}^{(t+1)}$ and the current estimate of the parameter set in $\widehat{\boldsymbol{\Psi}}^{(t)}$, i.e. $\mathbb{E} \big[ \ell_{c}(\vec{\Psi}, \vec{W}^{(t+1)}, \vec{Z}; \vec{X}) \vert \vec{X}_{[\vec{W}^{(t+1)}]}; 
    \widehat{\boldsymbol{\Psi}}^{(t)} \big]$, as follows      
\begin{align}\label{eqn: Qfun}
    &Q(\vec{\Psi} ; \widehat{\vec{\Psi}}^{(t)}) = \sum_{i = 1}^{n} \sum_{g = 1}^{G} \mathbb{E}[Z_{ig} \vert  \vec{x}_{i[\vec{w}_{i}^{(t+1)}]}; \widehat{\vec{\Psi}}^{(t)}] \times \Bigg\{ \ln (\pi_{g}) - \dfrac{1}{2} \Big[ \ln (\lvert \sigug \rvert) + \tr \Big( \sigug^{-1} \nonumber \\
    &\mathbb{E} \big[ \big( (\vec{x}_{i[\vec{w}_{i}^{(t+1)}]}^{\prime}, \vec{X}_{i[\vec{w}_{i}^{(t+1)c}]}^{\prime})^{\prime} - \meang \big) \big( (\vec{x}_{i[\vec{w}_{i}^{(t+1)}]}^{\prime}, \vec{X}_{i[\vec{w}_{i}^{(t+1)c}]}^{\prime})^{\prime} - \meang \big)^{\prime} \big\vert \vec{x}_{i[\vec{w}_{i}^{(t+1)}]}, z_{ig} = 1; \widehat{\vec{\Psi}}^{(t)} \big] \Big) \Big] \Bigg\},
\end{align}
where we omit the constant term of the normal distribution not depending on the model parameters. The computation of all the conditional expectations required in (\ref{eqn: Qfun}), which correspond to those presented by \citet{GJ:1994} for GMM with missing data, is reported in Section 1.1 of the Supplementary Material. These expectations constitute the E-step of the proposed algorithm, while the M-step maximizes $Q(\vec{\Psi} ; \widehat{\vec{\Psi}}^{(t)})$ with respect to $\vec{\Psi}$ by updating the estimates of the cellGMM parameters. The latter correspond to the usual estimates of the GMM parameters computed on the \textit{completed} data $\Big\{ \big\{ \Tilde{\vec{x}}_{i(g)}^{(t+1)} = (\vec{x}_{i[\vec{w}_{i}^{(t+1)}]},  \widehat{\vec{x}}_{i[\vec{w}_{i}^{(t+1)c}](g)}^{(t+1)})^{\prime} \big\}_{g = 1}^{G} \Big\}_{i = 1}^{n}$ (see Section 1.1 of the Supplementary Material for details). 

\subsection{Penalized log-likelihood approach for cellGMM}\label{subsec: pencellGMM}
The EM algorithm described for estimating the cellGMM parameters uncovers a fixed number $h$ of hopefully reliable cells per variable through $\vec{W}$. However, the cellwise outlier contamination can affect the variables differently. To prevent excessive cell flagging, we add penalty terms to the log-likelihood in (\ref{eqn: obj}) as follows
\begin{equation}\label{eqn: objpen}
    \ell_{\text{pen}}(\vec{\Psi}, \vec{W}; \vec{X}) = \sum_{i = 1}^{n} \ln \Big[ \sum_{g = 1}^{G} \pi_{g} \phi_{p[\vec{w}_{i}]} (\vec{x}_{i[\vec{w}_{i}]}; \meangi, \sigugii) \Big] - \sum_{j = 1}^{p} \sum_{i = 1}^{n} q_{ij} (1 - w_{ij}),
\end{equation}
where $\vec{Q} = [q_{ij}: i = 1, \ldots, n, j = 1, \ldots, p]$ represents a tuning matrix (see Section \ref{subsec: computationalasp} for the details on its setting). Maximizing (\ref{eqn: objpen}) under constraints (\ref{eqn: constrW}) and (\ref{eqn: constrS}) can avoid detecting an unnecessary high number of contaminated cells for an appropriate $\vec{Q}$ matrix. It is worth noting that the penalty term plays a crucial role in the update of $\vec{W}$, while it can be ignored in the other steps of the cellGMM algorithm. Specifically, the penalized version of the C-step is described below.

\textbf{Penalized update of $\vec{W}$.}
Inspired by the rationale in \citet{RR:2023}, we propose a modified $\Delta_{ij}$, denoted as $\widetilde{\Delta}_{ij}$, which is defined from
\begin{equation}\label{eqn: lc_cell}
    \widetilde{\ell}_{(i)}(\vec{\Psi}, \vec{w}_{i}, \vec{z}_{i}; \vec{x}_{i}) = \sum_{g = 1}^{G} z_{ig} \Big[  \ln(\pi_{g}) +  \ln \big( \phi_{p[\vec{w}_{i}]} (\vec{x}_{i[\vec{w}_{i}]}; \meangi, \sigugii) \big) \Big]-\sum_{j=1}^p q_{ij}(1-w_{ij}),
\end{equation}
as
\begin{align}\label{eqn: deltaZ2}
   \widetilde{\Delta}_{ij} &= \widetilde{\ell}_{(i)}(\widehat{\boldsymbol{\Psi}}^{(t)}, \widetilde{\vec{w}}_{i}, \vec{z}_{i}^{(t)}; \vec{x}_{i}, \widetilde{w}_{ij} = 1) - \widetilde{\ell}_{(i)}(\widehat{\boldsymbol{\Psi}}^{(t)}, \widetilde{\vec{w}}_{i}, \vec{z}_{i}^{(t)}; \vec{x}_{i}, \widetilde{w}_{ij} = 0) \nonumber \\
   &= - \dfrac{1}{2} \Bigg\{ \sum_{g = 1}^{G} z_{ig}^{(t)} \Big[ \ln(\widehat{C}_{ij(g)}) + \dfrac{(x_{ij} - \widehat{x}_{ij(g)})^{2}}{\widehat{C}_{ij(g)}} \Big] + \ln(2\pi) \Bigg\} + q_{ij},
\end{align}
where $\widehat{x}_{ij(g)} = \widehat{\mean}_{g[j]}^{(t)} + \widehat{\sigu}_{g[j, \widetilde{\vec{w}}_{i}]}^{(t)} (\widehat{\sigu}_{g [\widetilde{\vec{w}}_{i}, \widetilde{\vec{w}}_{i}]}^{(t)})^{-1} (\vec{x}_{i[\widetilde{\vec{w}}_{i}]} - \widehat{\mean}_{g[\widetilde{\vec{w}}_{i}]}^{(t)})$ and $\widehat{C}_{ij(g)} = \widehat{\sigu}_{g [j, j]}^{(t)} - \widehat{\sigu}_{g [j, \widetilde{\vec{w}}_{i}]}^{(t)}$ $(\widehat{\sigu}_{g[\widetilde{\vec{w}}_{i}, \widetilde{\vec{w}}_{i}]}^{(t)})^{-1}$ $\widehat{\sigu}_{g [\widetilde{\vec{w}}_{i}, j]}^{(t)}$ are the expectation and the variance, respectively, of the cell $X_{ij}$ for the $g$th mixture component conditional on the reliable cells in the same row $i$, excluding the $j$th variable. We obtain this result thanks to the additive property of the Mahalanobis distance and log-likelihood reported in \citet[][Proposition 5]{RR:2023}, which stem from the Gaussian distribution properties.

In this case, the update of $\widetilde{\vec{W}}_{\cdot j}$, and precisely the number of cells flagged as outliers, depends on the values of $\widetilde{\Delta}_{ij}$: a) if the number of nonnegative $\widetilde{\Delta}_{ij}$ is \textit{greater than h}, i.e., $\#\{\widetilde{\Delta}_{ij} \geq 0\} > h$, we set to one the corresponding cells $\widetilde{w}_{ij}$ for which $\widetilde{\Delta}_{ij} \geq 0$, and to zero the others; b) if the number of nonnegative $\widetilde{\Delta}_{ij}$ is \textit{lower than or equal to h}, i.e., $\#\{\widetilde{\Delta}_{ij} \geq 0\} \leq h$, we set to one the cells $\widetilde{w}_{ij}$ corresponding to the $h$ highest $\widetilde{\Delta}_{ij}$, and to zero the other cells. Consequently, a higher number of cells can be considered as reliable in scenario a), whereas this number is $h$ in scenario b). The penalized log-likelihood approach aims to enable the cellGMM algorithm to recover cells \textit{wrongly} flagged by the unpenalized cellGMM, potentially improving estimation accuracy, as we will see in Section \ref{sec: simulation}.

\subsection{Computational issues} \label{subsec: computationalasp}
\textbf{Initialization.} In the implementation of the EM-type algorithm for estimating the cellGMM parameters, their initialization is pivotal. In Section 1.2 of the Supplementary Material, we detail the procedure proposed to obtain initial solutions for $\vec{W}$ and $\vec{\Psi}$, along with the tuning parameters' setting used in our experiments. The initialization involves applying TCLUST individually to each variable and pairs of variables with a fixed trimming level for the initial solution of $\vec{W}$ and computing TCLUST on random subsets of variables, followed by a trimmed $k$-means \citep{CAGM:1997} type algorithm to obtain the initial solution for $\vec{\Psi}$. The provided procedure has proved to be compelling aligning closely with the rationale underlying cellGMM. Nonetheless, alternative approaches could be implemented to tackle this issue, that remains open for future work.

\noindent \textbf{Missing information.} The cellGMM algorithm can be applied to data with missing information. Specifically, the cells of $\vec{W}$ corresponding to the missing values in $\vec{X}$ are set to $0$, and the update of $\vec{W}$ is computed only on the observed cells. 

\noindent \textbf{Stopping criterion and monotonicity.} The stopping criterion adopted in the cellGMM algorithm is the Aitken acceleration procedure (\citealt{Betal:1994}; \citealt[][Section 2.11]{MLP:2000}; \citealt[][Section 4.9]{MK:2008}). Moreover, the cellGMM algorithm is monotone, as the objective function is non-decreasing in every iteration. Further details on these aspects are provided in Section 1.3 of the Supplementary Material.

\textbf{Setting of the penalization tuning matrix $\vec{Q}$.} Following \citet{RR:2023}, we set the generic element of $\vec{Q}$ as
\begin{equation}\label{eqn: cut}
     q_{ij} = \frac{1}{2} \Bigg[ \sum_{g = 1}^{G} \hat{z}_{ig} \ln \Bigg( \dfrac{1}{(\sigugest^{-1})_{[j,j]}} \Bigg)  + \chi^{2}_{1, 1-\alpha} + \ln(2\pi) \Bigg],
\end{equation}
where $\hat{z}_{ig}$ and $\sigugest$ are the estimates obtained at convergence by cellGMM with no penalty (i.e., $\vec{Q} = \vec{0}$), and $\chi^{2}_{1, 1-\alpha}$ is the quantile of the chi-squared distribution with one degree of freedom and probability $1-\alpha$ ($\alpha = 0.01$ in our experiments). As we will show in Section \ref{sec: simulation}, the use of the penalized log-likelihood approach for cellGMM usually increases its effectiveness. 

\section{Simulation study} \label{sec: simulation}
We carry out herein a simulation study to assess the performance of cellGMM in clustering and parameter estimation recovery. We generate random samples from (\ref{eqn: GMM}) by considering three scenarios: \textit{Scenario 1} and \textit{Scenario 2} with $n = 200, p = 5, G = 2$, non-spherical with $\sigu_{1} = \sigu_{2} = [\sigma_{ij} = 0.9^{\lvert i-j \rvert}: i, j = 1, \ldots, p]$, unbalanced ($\boldsymbol{\pi} = [0.3, 0.7]$), and well-separated and close components, respectively; \textit{Scenario 3} with $n = 400, p = 15, G = 4$, non-spherical (like Scenarios 1 and 2 with $4$ components), unbalanced  ($\boldsymbol{\pi} = [0.2, 0.2, 0.3, 0.3]$), and well-separated components. The component mean vectors are generated from uniform distributions. Specifically, $\mean_{1} = \vec{0}$, and the elements of $\mean_{g}, g = 2, \ldots, G,$ are drawn from a uniform distribution in $[0, 10]$ in the well-separated case, and from $[1, 3]$ in the close case. In the former, we assess whether the distance between the component mean vectors is less than 5, and re-generate them if this occurs. The components' configuration is controlled through the overlapping measure $\omega$ introduced by \citet{MM:2010}, where well-separated and close components correspond to $\omega_{\text{max}} < 0.01$ and $0.05 <\omega_{\text{max}} < 0.06$, respectively. For each scenario, we obtain $100$ data matrices that we contaminate with $0\%, 5\%$, and $10\%$ of outlying cells randomly drawn from a uniform distribution in the interval $[-10, 10]$, ensuring that the contaminated observations do not lie within the $99$th percentile ellipsoid of any component. Therefore, we consider $900$ random samples altogether. Additional scenarios with missing data, more extreme contamination, and structural outliers are reported in the Supplementary Material.

\begin{figure}
    \centering
    \includegraphics[width=\linewidth]{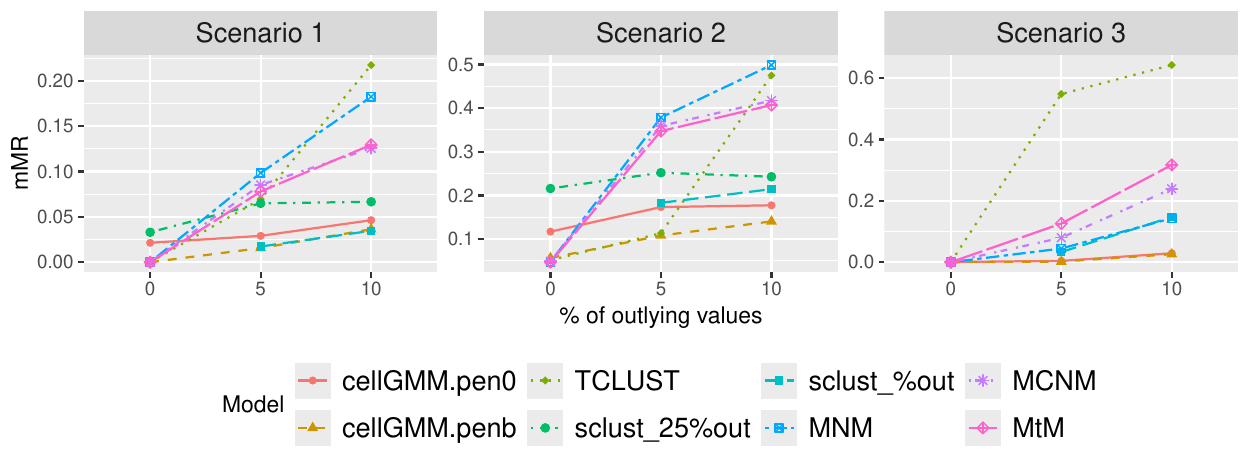}
    \caption{Results of the simulation study: mean of the Misclassification Rate (mMR) per scenario, percentage of contamination, and model}
    \label{fig: simstudy_results_MR}
\end{figure}

\begin{figure}
    \centering
    \includegraphics[width=\linewidth, height=0.6\textheight]{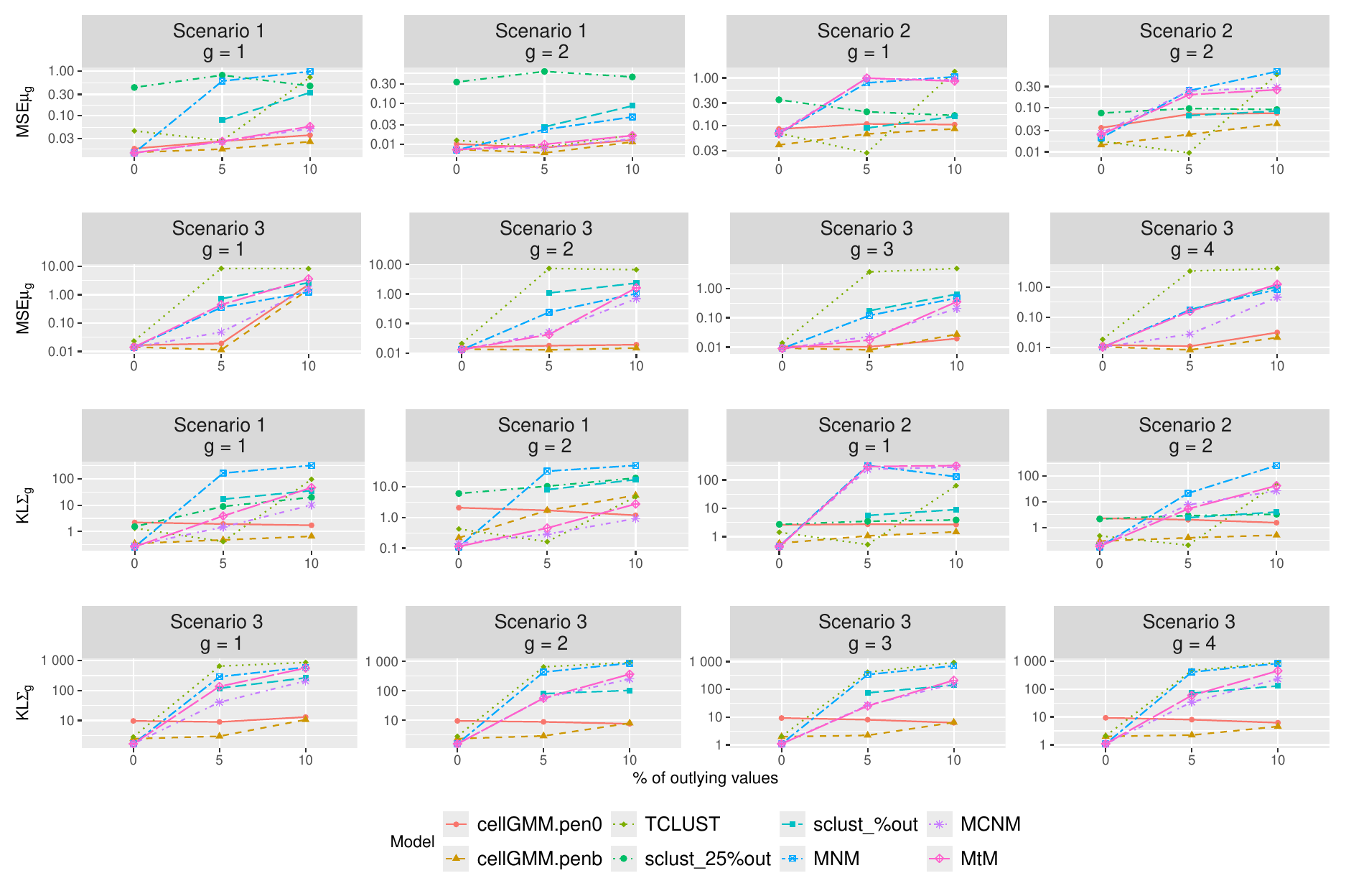}
    \caption{Results of the simulation study: Mean Squared Error (MSE) of the component mean vectors and Kullback-Leibler (KL) discrepancy for the component covariance matrices per scenario, percentage of contamination, and model. The values are represented via log-transformation, while the y-axis ticks are labeled using the original scale}
    \label{fig: simstudy_results_Parameters}
\end{figure}

First, we apply cellGMM without any penalty (cellGMM.pen0). Afterward, we run the penalized version of the algorithm (cellGMM.penb) with the same initialization, using the penalty defined in (\ref{eqn: cut}), which is derived from the parameters estimated at the convergence of the unpenalized cellGMM. The proposed methodology is also compared with those illustrated in Section \ref{sec: intro}: i) TCLUST (R package \texttt{tclust}) with $25\%$ of contamination, where we determine the membership of the trimmed units, which are not assigned to any cluster, by computing their posterior probabilities using the parameters estimated by TCLUST at convergence; ii) sclust (R package \texttt{snipEM}) with the suggested initialization for $\vec{W}$ by considering $25\%$ and the true percentage of outliers per variable; iii-v) Multivariate Normal Mixture (MNM), Multivariate Contaminated Normal Mixture (MCNM) and Multivariate $t$ Mixture with $25\%$ of contamination (M$t$M), which are implemented in the R package \texttt{MixtureMissing}. The classification performance of the models is evaluated through the Misclassification Rate, while the recovery of the parameters is assessed via the Mean Squared Error for the component mean vectors and the Kullback-Leibler discrepancy for the component covariance matrices (see the Supplementary Material for additional evaluation indices). To correctly compute the latter, we need to solve the label switching problem. With this aim, we order the estimated components via the complete likelihood-based labelling method introduced by \citet{Y:2015}. The results reported in Figures \ref{fig: simstudy_results_MR} and \ref{fig: simstudy_results_Parameters} show that the penalized version of cellGMM usually outperforms the unpenalized one, especially when the number of variables is higher and the components are well-separated (Scenario 3). In this case, cellGMM benefits from the relationships among variables in the imputation of the outlying cells, which in turn affects the parameter estimates, unlike sclust which trims them out. Additionally, the higher the contamination level, the greater the difference between cellGMM.penb and the methodologies for rowwise outlier detection, i.e., TCLUST, MCNM and M$t$M, or the non-robust GMM, i.e. MNM. This is due to the fact that cellwise outliers spread out through the observations as the contamination level increases. Therefore, they can affect more than half of the rows, leading to severely biased parameter estimates, even for methods like TCLUST (see the results for $10\%$ of contamination in Figure \ref{fig: simstudy_results_Parameters}).

\begin{table}[!htbp]
\centering
\caption{Outlier detection and imputation: percentage of True and False Positive ($\%$TP and $\%$FP), Mean Absolute and Root Mean Squared Error (MAE and RMSE) for comparing the original and imputed data matrices per scenario, percentage of contamination, and model}\label{tab: sim_W_Imp}
\resizebox{0.93\linewidth}{!}{
\begin{tabular}{cr|cccc|cccc|cccc}
\hline
& & \multicolumn{4}{c|}{Scenario 1} & \multicolumn{4}{c|}{Scenario 2} & \multicolumn{4}{c}{Scenario 3} \\
\hline
$\%$ out. & Method & \%TP & \%FP & MAE & RMSE & \%TP & \%FP & MAE & RMSE & \%TP & \%FP & MAE & RMSE \\
\hline
\multirow{8}{*}{0} & cellGMM.pen0 & - & 25.00 & 0.22 & 0.55 & - & 25.00 & 0.21 & 0.46 & - & 25.00 & 0.20 & 0.44 \\ 
& cellGMM.penb & - & 2.26 & 0.02 & 0.15 & - & 3.31 & 0.04 & 0.20 & - & 2.78 & 0.03 & 0.16 \\ 
& TCLUST & - & 25.00 & - & - & - & 25.00 & - & - & - & 25.00 & - & -\\ 
& sclust\_25 & - & 25.00 & - & - & - & 25.00 & - & - & - & - & - & - \\
& MCNM & - & 6.38 & - & - & - & 4.57 & - & - & - & 40.69 & - & - \\ 
& M$t$M & - & 26.05 & - & - & - & 25.14 & - & - & - & 25.12 & - & - \\ 
& cellMCD & - & 13.76 & 0.51 & 1.47 & - & 2.17 & 0.02 & 0.18 & - & 5.52 & 0.30 & 1.39 \\ 
& DI & - & 12.66 & 0.46 & 1.36 & - & 1.93 & 0.02 & 0.19 & - & 15.24 & 0.68 & 1.91 \\  
\cmidrule{2-14}
\multirow{9}{*}{5} & cellGMM.pen0 & 96.20 & 21.25 & 0.24 & 0.68 & 97.30 & 21.19 & 0.22 & 0.51 & 96.64 & 21.23 & 0.22 & 0.56 \\ 
& cellGMM.penb & 96.56 & 3.56 & 0.10 & 0.56 & 96.80 & 5.37 & 0.10 & 0.40 & 95.47 & 3.78 & 0.06 & 0.29 \\ 
& TCLUST & 99.56 & 21.08 & - & - & 99.66 & 21.07 & - & - & 54.65 & 23.44 & - & - \\ 
& sclust\_25 & 71.61 & 22.55 & - & - & 92.36 & 21.45 & - & - & - & - & - & - \\ 
& sclust\_5 & 54.67 & 2.39 & - & - & 77.71 & 1.17 & - & - & 45.64 & 2.86 & - & - \\
& MCNM & 93.58 & 17.29 & - & - & 57.02 & 10.28 & - & - & 75.33 & 35.55 & - & - \\ 
& M$t$M & 99.54 & 36.48 & - & - & 59.86 & 32.31 & - & - & 92.51 & 48.36 & - & - \\ 
& cellMCD & 94.20 & 13.71 & 0.53 & 1.49 & 96.76 & 2.20 & 0.05 & 0.23 & 92.18 & 3.87 & 0.22 & 1.10 \\ 
& DI & 92.12 & 5.29 & 0.19 & 0.64 & 96.78 & 1.97 & 0.05 & 0.24 & 90.03 & 12.37 & 0.57 & 1.68 \\ 
\cmidrule{2-14}
\multirow{9}{*}{10} & cellGMM.pen0 & 94.98 & 17.22 & 0.29 & 0.89 & 95.47 & 17.17 & 0.22 & 0.55 & 93.47 & 17.39 & 0.29 & 0.84 \\ 
& cellGMM.penb & 94.31 & 6.31 & 0.20 & 0.84 & 94.87 & 6.33 & 0.15 & 0.52 & 92.01 & 6.13 & 0.17 & 0.68 \\   
& TCLUST & 64.94 & 20.56 & - & - & 65.35 & 20.52 & - & - & 44.14 & 22.87 & - & - \\ 
& sclust\_25 & 70.30 & 19.97 & - & - & 89.45 & 17.84 & - & - & - & - & - & - \\ 
& sclust\_10 & 59.12 & 4.54 & - & - & 78.97 & 2.34 & - & - & 48.88 & 5.68 & - & - \\ 
& MCNM & 92.77 & 31.86 & - & - & 50.94 & 16.14 & - & - & 70.27 & 45.49 & - & - \\ 
& M$t$M & 96.21 & 40.92 & - & - & 60.56 & 34.19 & - & - & 77.39 & 52.65 & - & - \\ 
& cellMCD & 89.46 & 13.32 & 0.68 & 1.82 & 94.62 & 2.07 & 0.07 & 0.28 & 88.55 & 3.28 & 0.21 & 1.00 \\ 
& DI & 89.45 & 2.53 & 0.12 & 0.49 & 94.67 & 1.92 & 0.07 & 0.28 & 88.56 & 5.68 & 0.28 & 1.06 \\ 
\hline
\end{tabular}%
}
\end{table}

To address the correct detection of outliers and data matrix imputation, we compute the percentage of True and False Positives on one hand, and the Mean Absolute Error and Root Mean Squared Error between the original and imputed data matrices on the other hand. To this aim, we also consider cellMCD and DI from the R package \texttt{cellWise}. For the model-based clustering methodologies with rowwise outlier detection, we build a $\vec{W}$ matrix by setting to zero all the cells of the rows that have been flagged as contaminated. Table \ref{tab: sim_W_Imp} illustrates the results. In this case as well, the penalized version of cellGMM outperforms the unpenalized one in detecting the unreliable cells, with a drastic decrease in the percentage of false positives and a slight decrease in the true positive due to the enlarged number of cells flagged in cellGMM.pen0 by constraint. The accurate identification of outliers is reflected in enhanced values for the indices assessing the imputation of contaminated cells, with improvements between cellGMM.pen0 and cellGMM.penb. Due to this better performance, we report the results of cellGMM.penb directly in Section \ref{sec: application}.

In Table \ref{tab: sim_W_Imp}, the difference between the proposal and the model-based clustering methodologies sharpens as the contamination level increases and components overlap, up to the higher-dimensional case (Scenario 3), where cellGMM generally outperforms the competitors significantly. The single-population methods have better results in Scenario 2 with close components, as expected, since it is more likely that the components are so overlapping that they appear as a single population. In contrast, their performance unsurprisingly worsens in Scenario 1 and Scenario 3, both in terms of \%TP and \%FP. However, it is worth noting that in Scenario 1 with $10\%$ of contamination, DI has lower \%FP and imputation indices than cellGMM, despite having a lower \%TP. This is riskier for a robust model since the failure to detect outliers can heavily affect the parameter estimates. The peculiar behavior of cellGMM compared to DI in Scenario 1 with $10\%$ of contamination, especially in the imputation results, is due to a higher misclassification rate of the former, which can favor DI over cellGMM. Indeed, if the components are well-separated and a unit is misclassified, the imputation within the wrong component is worse than the DI's imputation between the two components. The additional scenarios illustrated in the Supplementary Material provide further insights into the performance of cellGMM and its competitors.

\vspace{-0.4cm}
\section{Real data examples} \label{sec: application}
\subsection{Homogenized Meat Data Set}\label{subsec: data1}
\begin{figure}[t]
\centering
\subfloat[][\label{fig: meat_pairs}] 
{\includegraphics[width=0.46\linewidth]{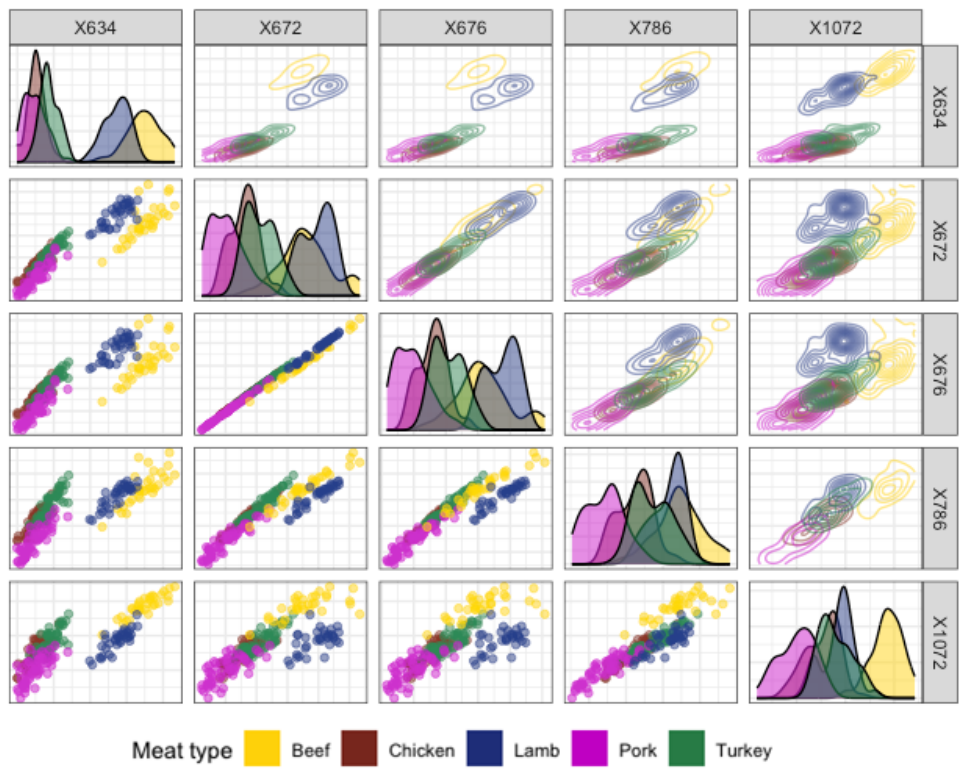}} \,
\subfloat[][\label{fig: meat_3out}] 
{\includegraphics[width=0.50\linewidth]{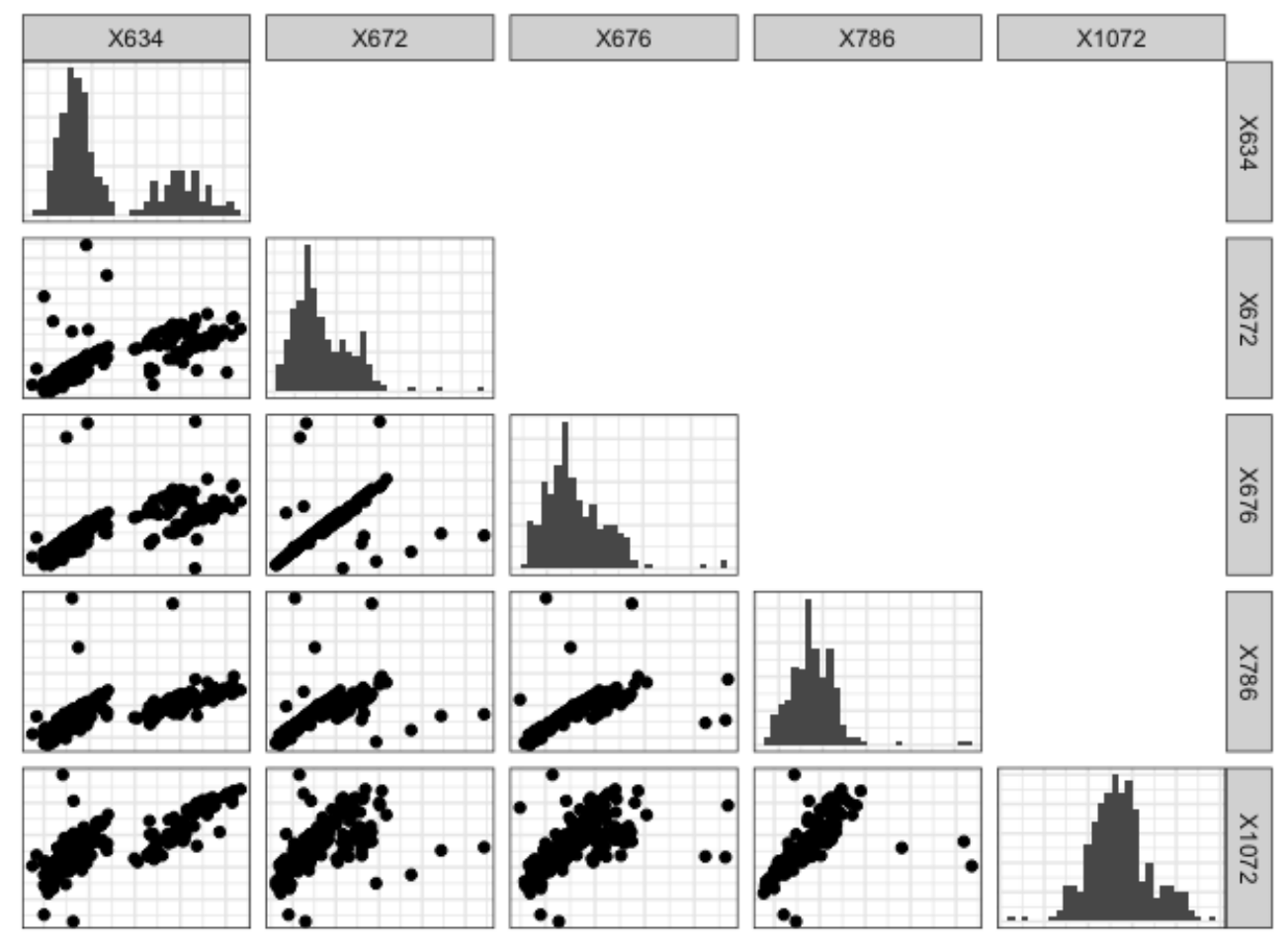}} \,
\caption{Homogenized Meat data set. (a) Pair plot of $5$ variables from the original data: true classification in five homogenized meat types; (b) Pair plot of the data with $3\%$ of cellwise contamination}
\label{fig: meat}
\end{figure}

In this section, we analyze the homogenized meat data set presented by \citet{MDF:1999} to evaluate the cellGMM performance in correctly classifying the samples from their visible and infrared spectra. For each of the $231$ homogenized meat samples, we consider the five relevant wavelengths which span the protein spectral region \citep{CGM:2021}: the first four belonging to the visible part of the spectrum (634 nm, 672 nm, 676 nm and 786 nm) and the remaining one within the near-infrared part (1072 nm). It is worth noting that, even if the original classes are five (beef, chicken, lamb, pork, turkey), some of the selected variables, such as the wavelengths $634$ nm and $676$ nm, turn out to be important to distinguish red meats (beef and lamb) from white meats (chicken, pork and turkey). This evidence was also highlighted by \citet{MDR:2010} and can be seen in Figure \ref{fig: meat_pairs} as well. Therefore, we consider the classification of samples in two classes and randomly adulterate them via $3\%$ (without and with $2\%$ of missing) and $10\%$ of cellwise contamination in the range $[0.73, 1.25]$ (see Figure \ref{fig: meat_3out} as an example). 

\begin{table}[t]
\caption{Misclassification rate comparing the theoretical and the estimated classification of meat samples in two classes per model and percentage of contamination and missing values, i.e. $(a\%, b\%)$}
\label{tab: meat_MR_G2}
\centering
\resizebox{0.5\textwidth}{!}{
\begin{tabular}{lccccccc}
\hline
& cellGMM & TCLUST & sclust & MNM & MCNM & M$t$M \\
\hline
$(0\%, 0\%)$ & 0.01 & 0.00 & 0.00 & 0.00 & 0.00 & 0.00 \\
$(3\%, 0\%)$ & 0.01 & 0.03 & 0.04 & 0.05 & 0.05 & 0.06 \\
$(3\%, 2\%)$ & 0.01 & - & - & 0.05 & 0.06 & 0.06 \\
$(10\%, 0\%)$ & 0.02 & 0.16 & 0.07 & 0.15 & 0.25 & 0.20 \\
\hline
\end{tabular}
}
\end{table}

We run cellGMM, TCLUST, MNM, MCNM, and M$t$M with the same settings used in the simulation study. For sclust, we use the theoretical level of contamination, except when no missing data or outliers are introduced; in such case, the trimming (snipping) level is set to $0.01$. As reported in Table \ref{tab: meat_MR_G2}, as the level of contamination increases, the performance gap between cellGMM and its competitors becomes evident. Indeed, with $10\%$ of contamination, cellGMM maintains good classification results, while the performance of the competitors deteriorates. Additional results can be found in Section 3.1 of the Supplementary Material.

\subsection{Carina Nebula Data Set}\label{subsec: data2}
\begin{figure}[t]
\centering
\subfloat[][\textit{Original picture} \label{fig: cnr_or}] 
{\includegraphics[width=0.3\linewidth]{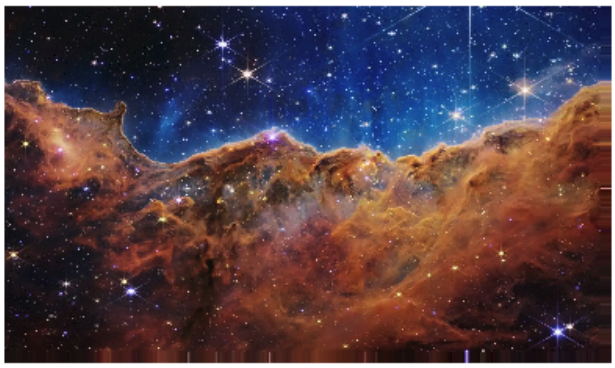}} \,
\subfloat[][\textit{Reduced picture} \label{fig: cnr_reduced}] 
{\includegraphics[width=0.3\linewidth]{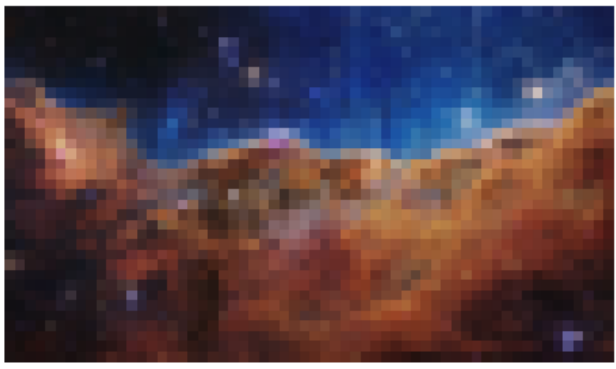}} \,
\subfloat[][\textit{$3\%$ of contamination} \label{fig: cnr_contaminated}] 
{\includegraphics[width=0.3\linewidth]{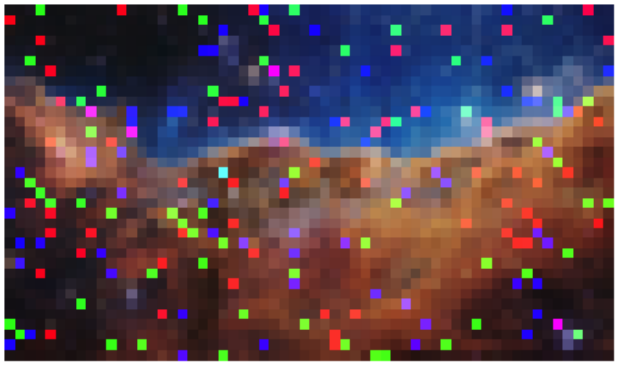}} \\ 
\subfloat[][\textit{cellGMM with $G = 5$} \label{fig: CN_cellGMMG5_Imp}] 
{\includegraphics[width=0.3\linewidth]{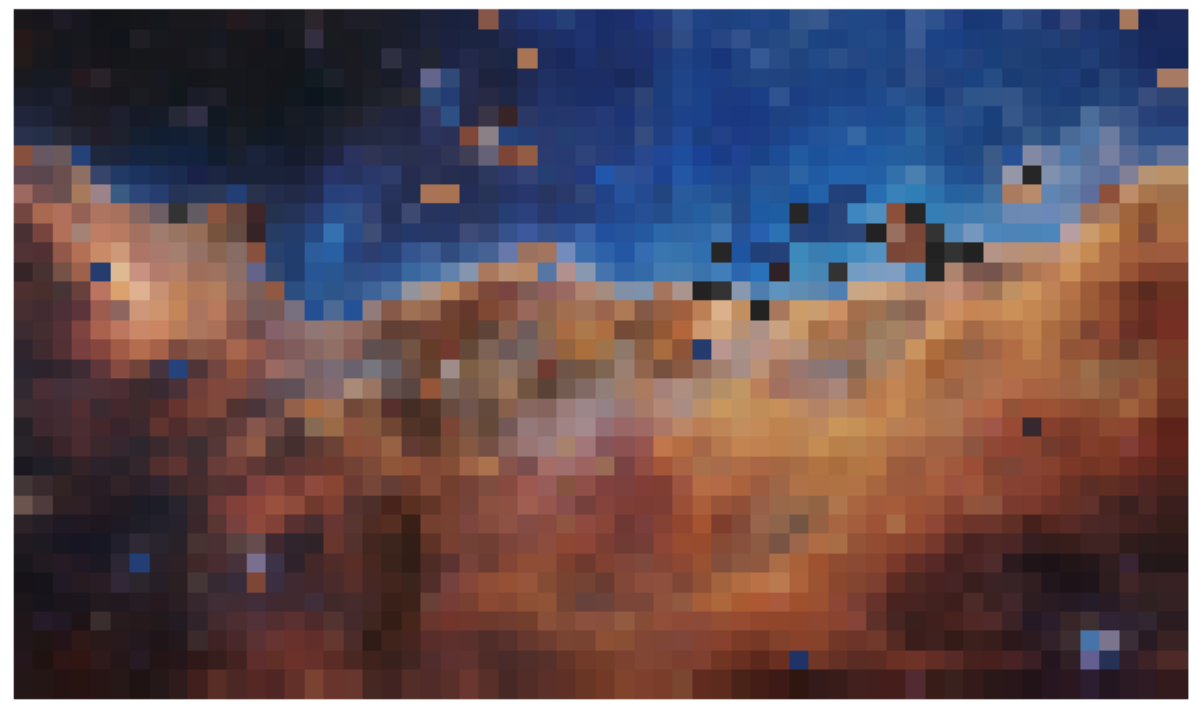}} \,
\subfloat[][\textit{cellMCD} \label{fig: cn_MCD_Imp}] 
{\includegraphics[width=0.3\linewidth]{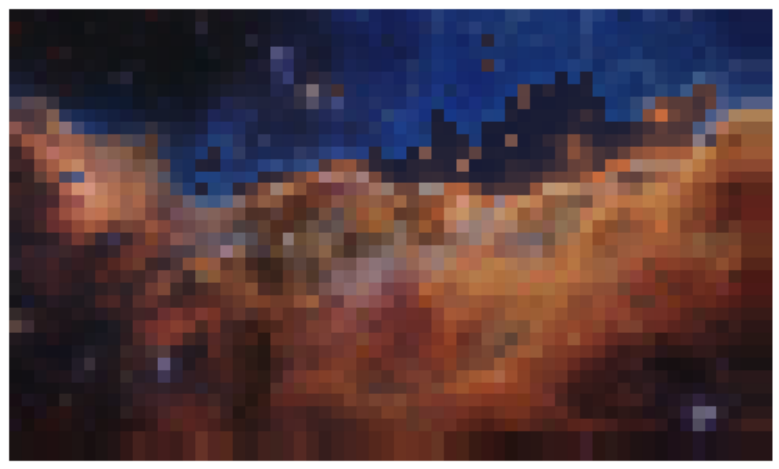}} \,
\subfloat[][\textit{DI} \label{fig: cn_DI_Imp}] 
{\includegraphics[width=0.3\linewidth]{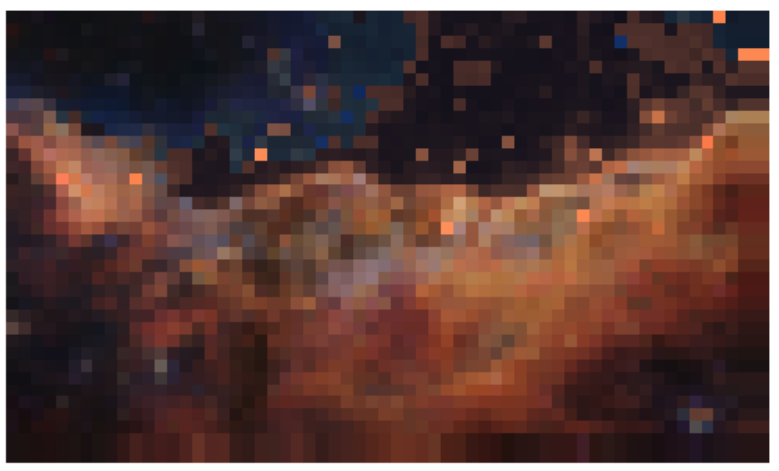}} 
\caption{Carina Nebula birth. The methods' imputation are shown in (d)-(f).}
\label{fig: Carina Nebula}
\end{figure}

\begin{figure}[t]
\centering
\subfloat[][\textit{Contaminated data set} \label{fig: pp_cont}] 
{\includegraphics[width=0.4\linewidth]{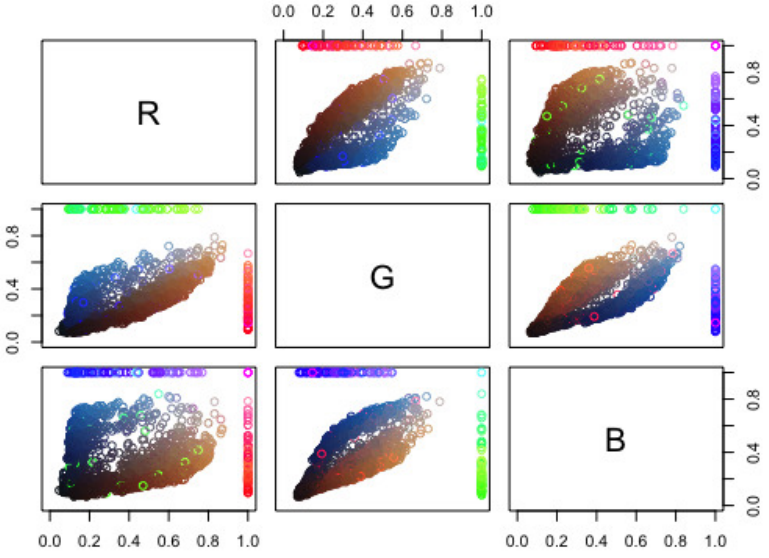}}
\subfloat[][\textit{cellGMM with $G = 5$} \label{fig: pp_cellGMMG5}] 
{\includegraphics[width=0.4\linewidth]{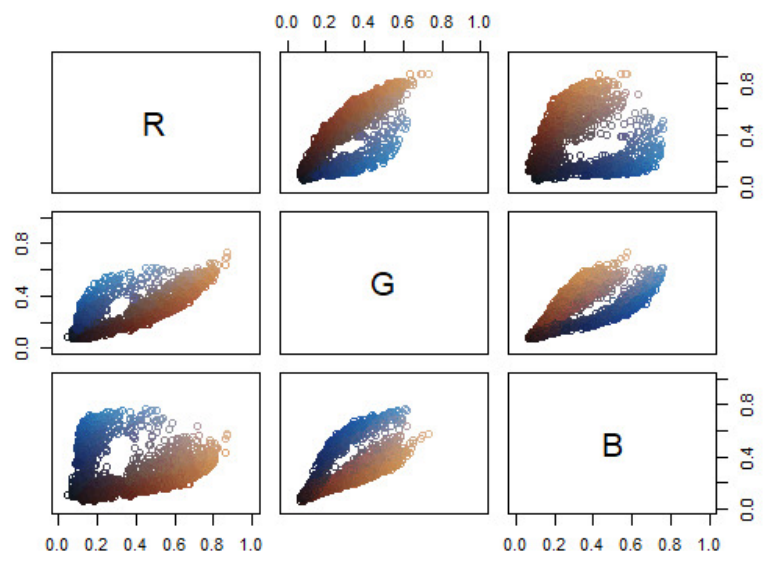}}  \\
\subfloat[][\textit{cellMCD} \label{fig: pp_cellMCD}] 
{\includegraphics[width=0.4\linewidth]{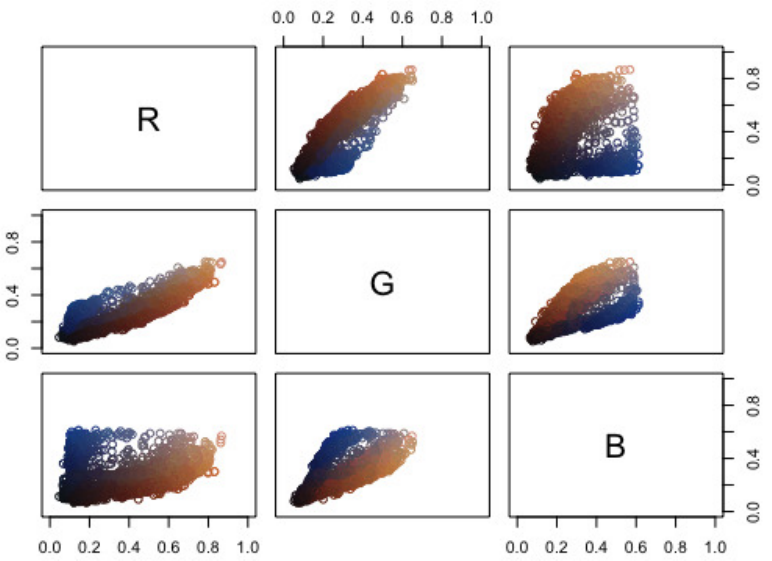}} 
\subfloat[][\textit{DI} \label{fig: pp_DI}] 
{\includegraphics[width=0.4\linewidth]{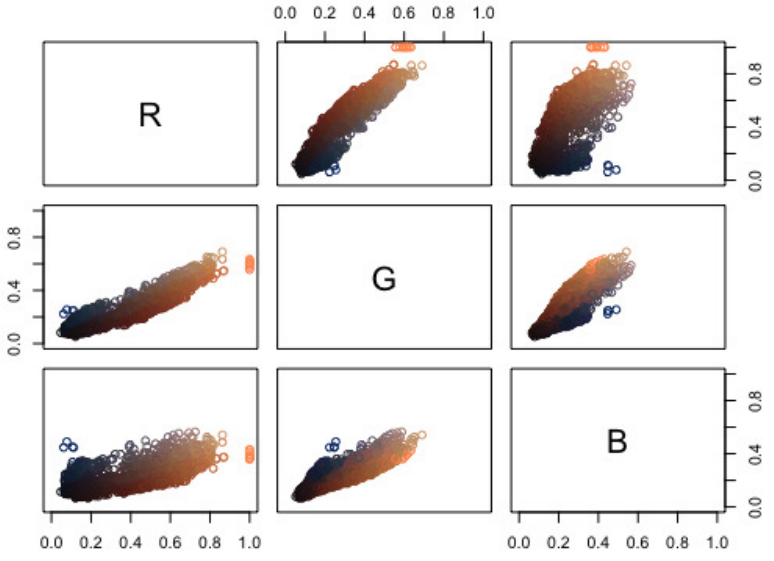}} 
\caption{Pair plots of the Carina Nebula data set. The methods' imputation are shown in (b)-(d).}
\label{fig: Carina Nebula_PP}
\end{figure}

The proposed methodology is also suitable for image reconstruction. In this framework, we evaluate the detection and \say{correction} of contaminated cells through their imputation. Specifically, we analyze an image captured by the James Webb Space Telescope, which depicts the edge of NGC 3324 located in the northwest corner of the Carina Nebula. The image of the stars' birth is available at \href{https://science.nasa.gov/universe/stars/}{https://science.nasa.gov/universe/stars/}. The original picture has $1600 \times 927$ pixels (Figure \ref{fig: cnr_or}), corresponding to more than one million units measured in three variables (RGB). Due to its high size, we reduce it into $60 \times 35$ pixels (Figure \ref{fig: cnr_reduced}), which we then corrupt with $3\%$ of contamination, as depicted in Figures \ref{fig: cnr_contaminated} and \ref{fig: pp_cont}. 

Given our goal, the most appropriate methods to compare to cellGMM are cellMCD and DI, as they provide imputation for contaminated cells. As shown in Figures \ref{fig: CN_cellGMMG5_Imp}-\ref{fig: cn_DI_Imp} and \ref{fig: pp_cellGMMG5}-\ref{fig: pp_DI}, cellGMM demonstrates better potential in image reconstruction than the single-population methods. Moreover, as expected, the higher the number of components, the better the classification performance of cellGMM. This is evident by considering the RMSE between the imputed and the original data sets, which is $0.073$ for cellGMM with $G = 2$ and $0.059$ for cellGMM with $G = 5$, both lower than $0.079$ for cellMCD and $0.127$ for DI.

\subsection{Top Gear Data Set}\label{subsec: data3}
\begin{table}[t]
\centering
\caption{Ten representative cars per cluster}\label{tab: TopGear_10best}
\resizebox{0.8\linewidth}{!}{
\begin{tabular}{l|l|l|l}
\hline
Cluster 1 & Cluster 2 & Cluster 3 & Cluster 4 \\
\hline 
Audi A4 & Chevrolet Spark & Aston Martin DB9 & BMW X5 \\
Jaguar XF Sportbrake & Hyundai i10 & Aston Martin DB9 Volante & BMW X6 \\
Kia Optima & Kia Picanto & Aston Martin V12 Zagato & Hyundai i800 \\
Lexus GS & Peugeot 107 & Aston Martin Vanquish & Jeep Grand Cherokee \\
Mercedes-Benz E-Class Coupé & Proton Savvy & Aston Martin Vantage & Land Rover Discovery 4 \\
Skoda Octavia & SEAT Mii & Aston Martin Vantage Roadster & Land Rover Range Rover \\
Vauxhall Cascada & Suzuki Alto & Audi R8 & Mercedes-Benz G-Class \\
Vauxhall Insignia Sports Tourer & Toyota AYGO & Audi R8 V10 & Mercedes-Benz GL-Class \\
Volkswagen CC & Toyota iQ & Bentley Continental & Porsche Cayenne \\
Volkswagen Passat & Volkswagen Up & Bentley Continental GTC & Toyota Land Cruiser V8 \\
\hline
\end{tabular}}
\end{table}

In this section, we analyze the car data set available in the R package \texttt{robustHD} \citep{A:2016}, called \textit{TopGear}, which contains authentic missing values ($2.74\%$ of the cells) and potential outliers. We focus on the eleven numerical variables by removing the two cars with more missing than observed values, resulting in a final sample size of $n = 295$, and transforming the highly skewed variables using their logarithms, as reported in \citet{RR:2023}. Unlike these authors, we provide a cluster-oriented approach to this data set by setting $G = 4$, which identifies meaningful groups with distinct features. The cars are divided into \say{compact and mid-size sedans and crossovers} (Cluster 1); \say{economy and city cars} (Cluster 2); \say{luxury and high-performance cars} (Cluster 3); \say{large SUVs and off-roaders} (Cluster 4). A list of ten representative cars per cluster, selected as those with the highest posterior probabilities, is provided in Table \ref{tab: TopGear_10best}, while a comprehensive overview of the clusters is reported in Section 3.2 of the Supplementary Material. 

\begin{figure}[!htbp]
\centering
\subfloat[][\textit{Weight} \label{fig: TG_StandRes_Weight}] 
{\includegraphics[width=0.75\linewidth]{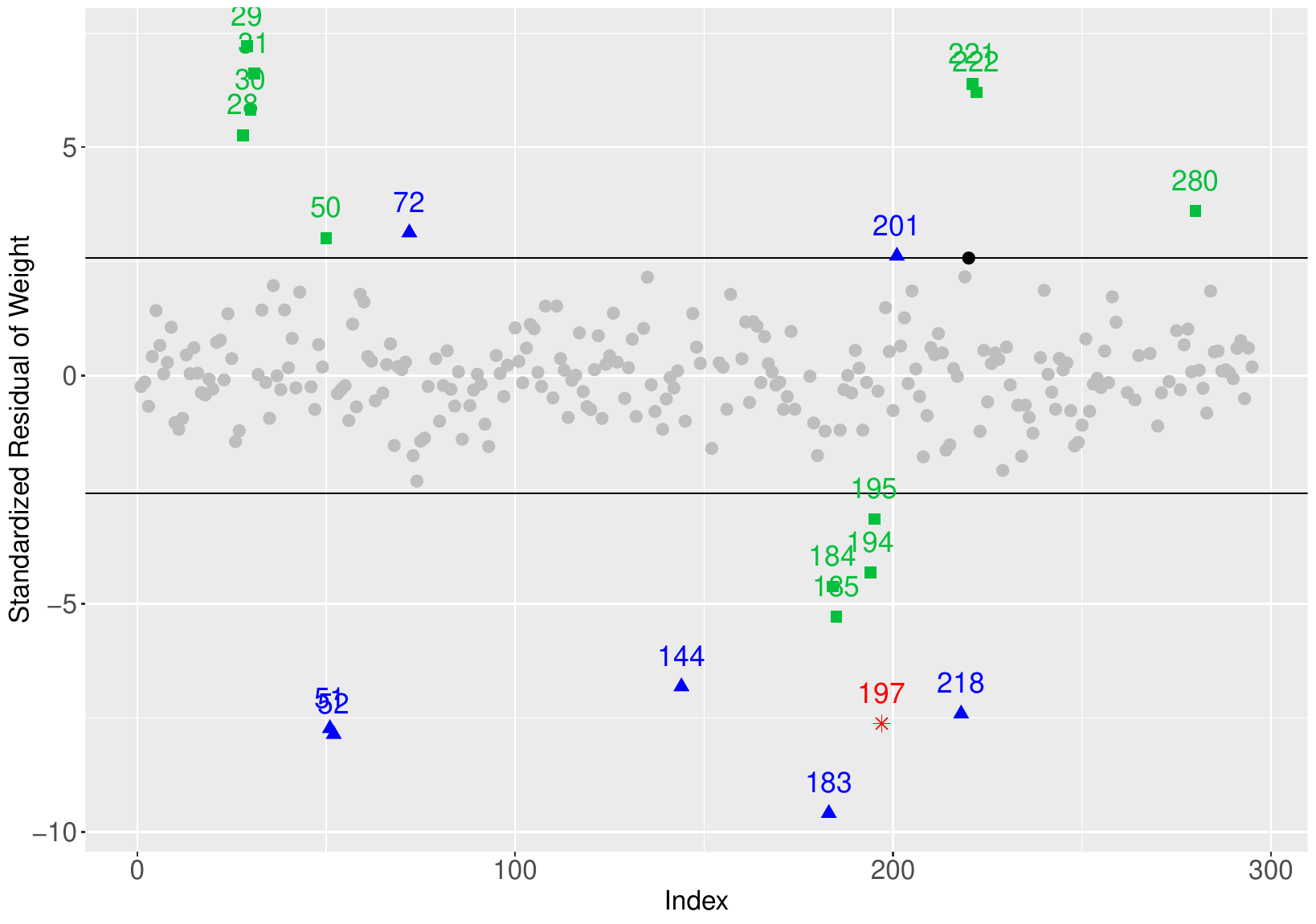}} \\
\subfloat[][\textit{Length} \label{fig: TG_StandRes_Length}] 
{\includegraphics[width=0.75\linewidth]{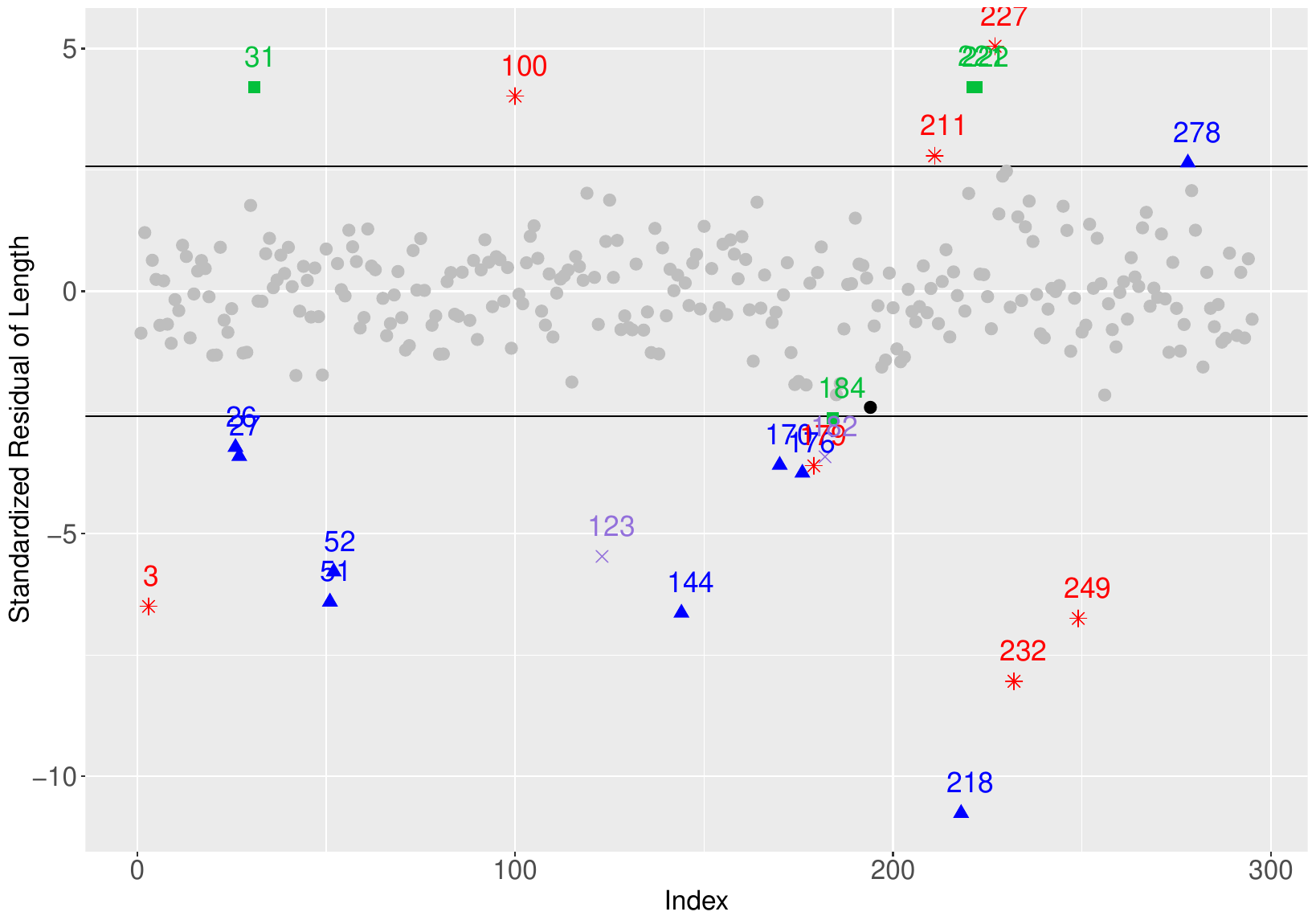}} 
\caption{Top Gear data set: index plot of the standardized residuals per variable (Cluster 1: blue - triangle; Cluster 2: red - asterisk; Cluster 3: green - square; Cluster 4: purple - cross)}
\end{figure}

\begin{figure}[!htbp]\ContinuedFloat
\centering
\subfloat[][\textit{Width} \label{fig: TG_StandRes_Width}] 
{\includegraphics[width=0.75\linewidth]{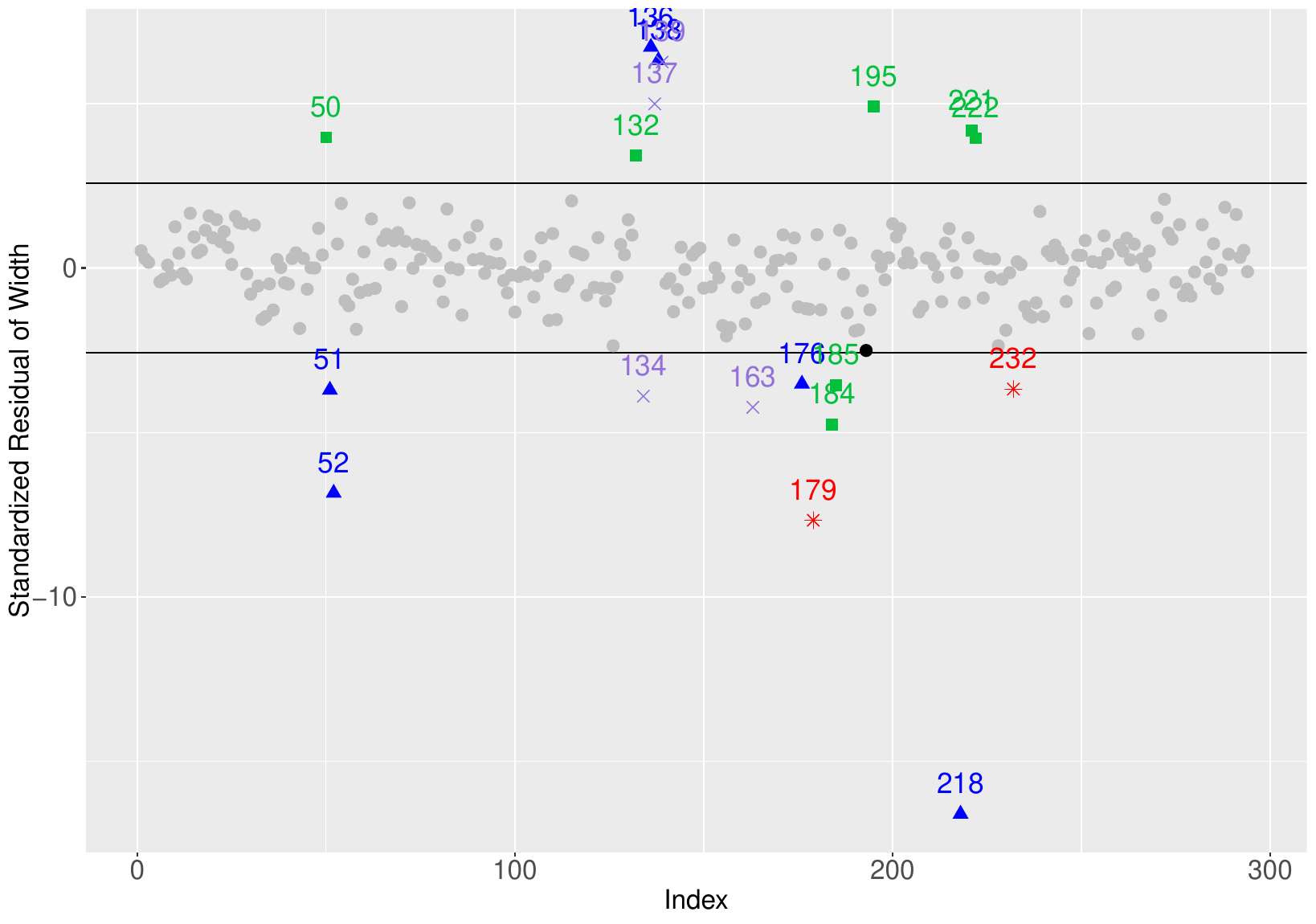}} \\
\subfloat[][\textit{Height} \label{fig: TG_StandRes_Height}] 
{\includegraphics[width=0.75\linewidth]{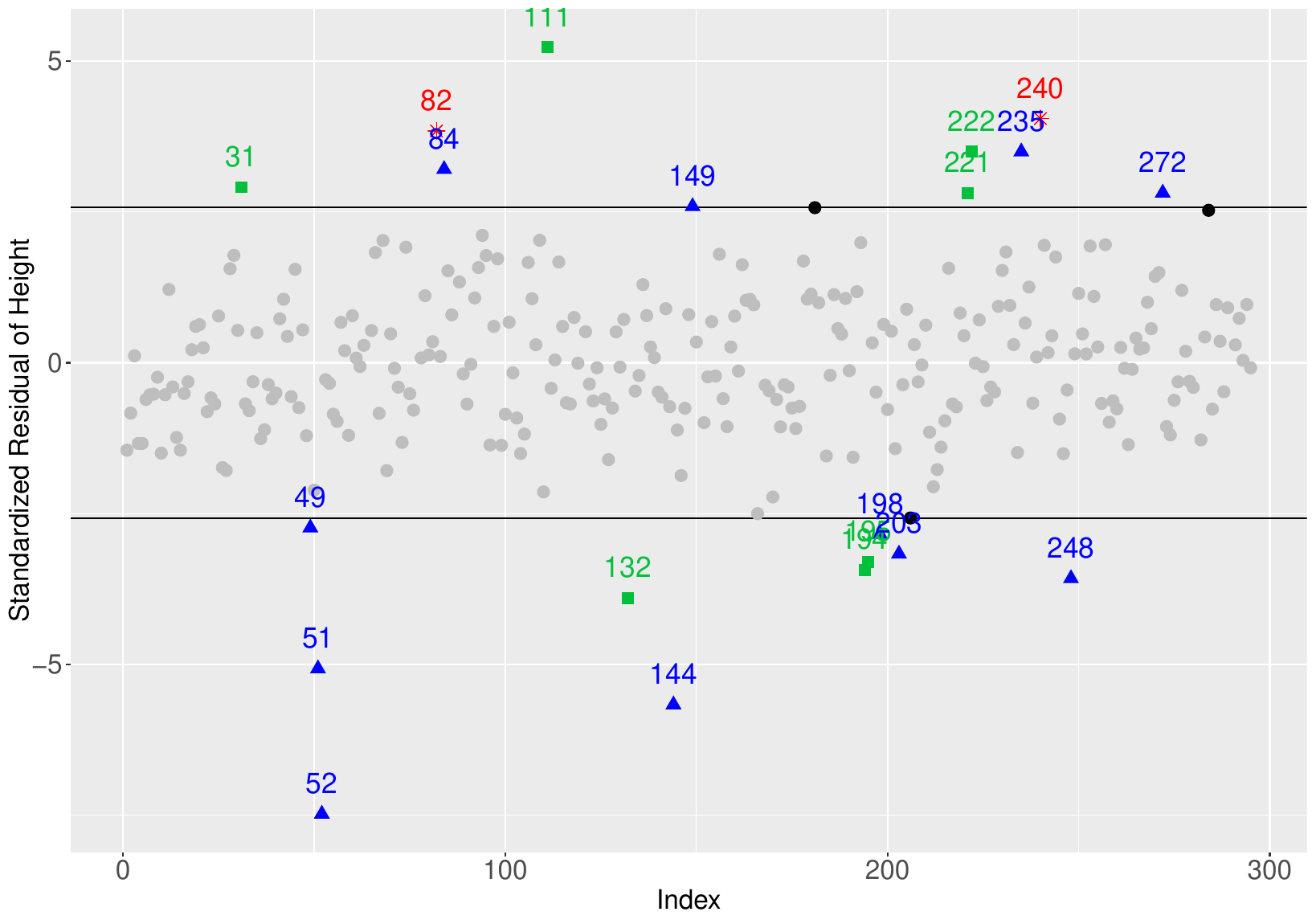}} 
\caption{Top Gear data set: index plot of the standardized residuals per variable (Cluster 1: blue - triangle; Cluster 2: red - asterisk; Cluster 3: green - square; Cluster 4: purple - cross)}
\label{fig: TG_StandRes}
\end{figure}

\begin{table}[t]
\centering
\caption{Top Gear data set: ID and name of the cars resulting with high standardized residuals in absolute terms}
\resizebox{\linewidth}{!}{
\begin{tabular}{llllllll}
\hline
ID & Car name & ID & Car name & ID & Car name & ID & Car name \\
\hline
3 & Aston Martin Cygnet & 84 & Fiat Doblo & 170 & Mercedes-Benz SLK & 211 & Proton GEN-2  \\
26 & Audi TT Coupe & 100 & Honda Insight &  176 & Mini John Cooper Works & 218 & Renault Twizy \\
27 & Audi TT Roadster & 111 & Infiniti EX & 179 & Mitsubishi i-MiEV & 221 & Rolls-Royce Phantom \\
28 & Bentley Continental & 123 & Jeep Wrangler & 182 & Mitsubishi Shogun & 222 & Rolls-Royce Phantom Coupe \\
29 & Bentley Continental GTC & 132 & Lamborghini Aventador & 183 & Morgan 3 Wheeler & 227 & SEAT Toledo \\
30 & Bentley Flying Spur & 134 & Land Rover Defender & 184 & Morgan Aero & 232 & Smart fortwo \\
31 & Bentley Mulsanne & 136 & Land Rover Freelander 2 & 185 & Morgan Roadster & 235 & Subaru Forester \\
49 & BMW Z4  & 137 & Land Rover Range Rover & 194 & Noble M600 & 240 & Suzuki Jimny \\
50 & Bugatti Veyron & 138 & Land Rover Range Rover Evoque & 195 & Pagani Huayra & 248 & Toyota GT 86 \\
51 & Caterham CSR & 139 & Land Rover Range Rover Sport  & 197 & Peugeot 107 & 249 & Toyota iQ  \\
52 & Caterham Super 7 & 144 & Lotus Elise & 198 & Peugeot 207 CC & 272 & Vauxhall Zafira Tourer \\
72 & Citroen DS5 & 149 & Mazda CX-5 & 201 & Peugeot 3008 & 278 & Volkswagen Jetta \\
82 & Fiat 500L & 163 & Mercedes-Benz G-Class & 203 & Peugeot 308 CC & 280 & Volkswagen Phaeton \\
\hline
\end{tabular}}
\end{table}

\begin{figure}[!t]
\centering
\subfloat[][\textit{Cluster 1} \label{fig: TG_W_G1}] 
{\includegraphics[width=0.52\linewidth]{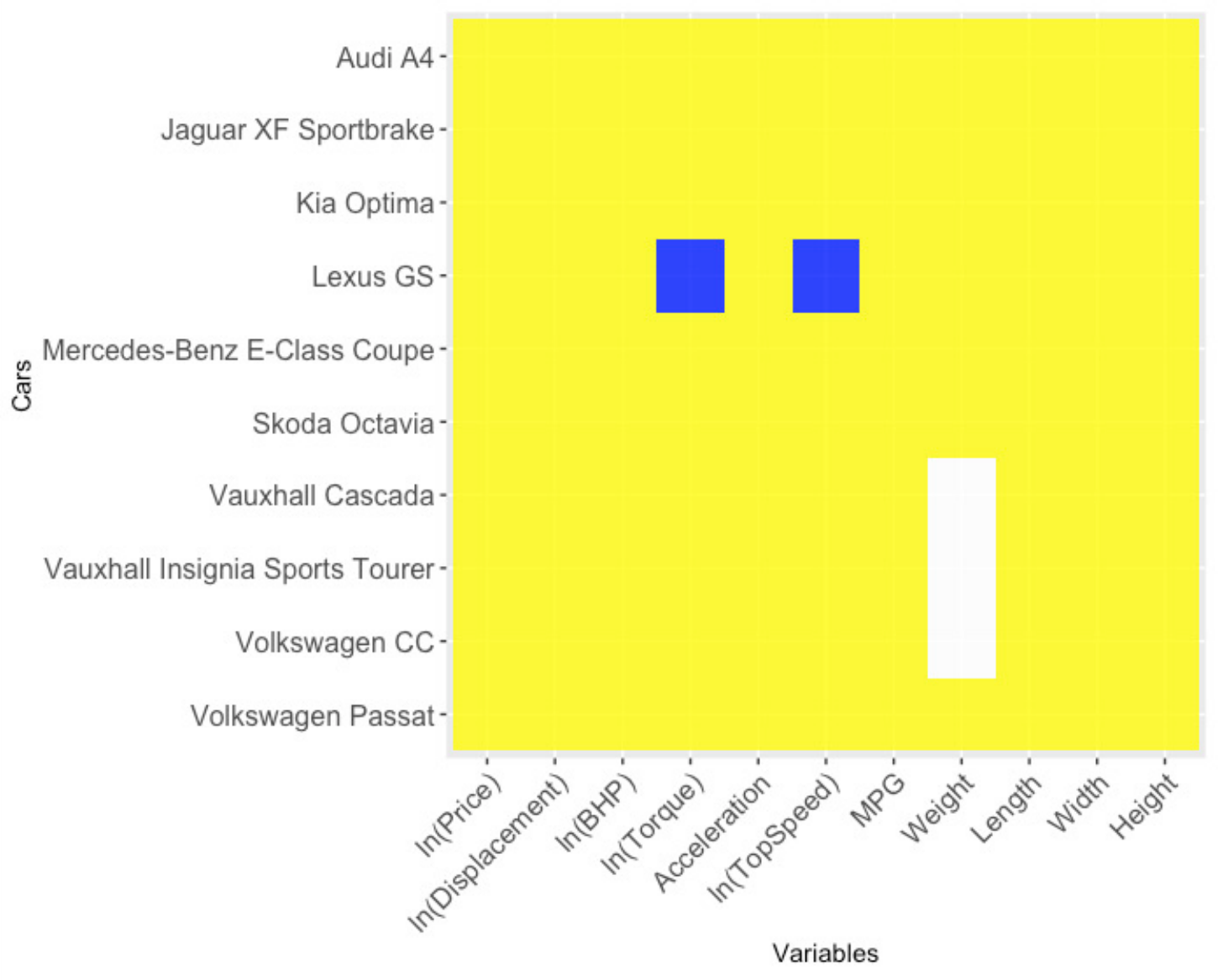}} \hfill \quad
\subfloat[][\textit{Cluster 2} \label{fig: TG_W_G2}] 
{\includegraphics[height=0.42\linewidth, width=0.42\linewidth]{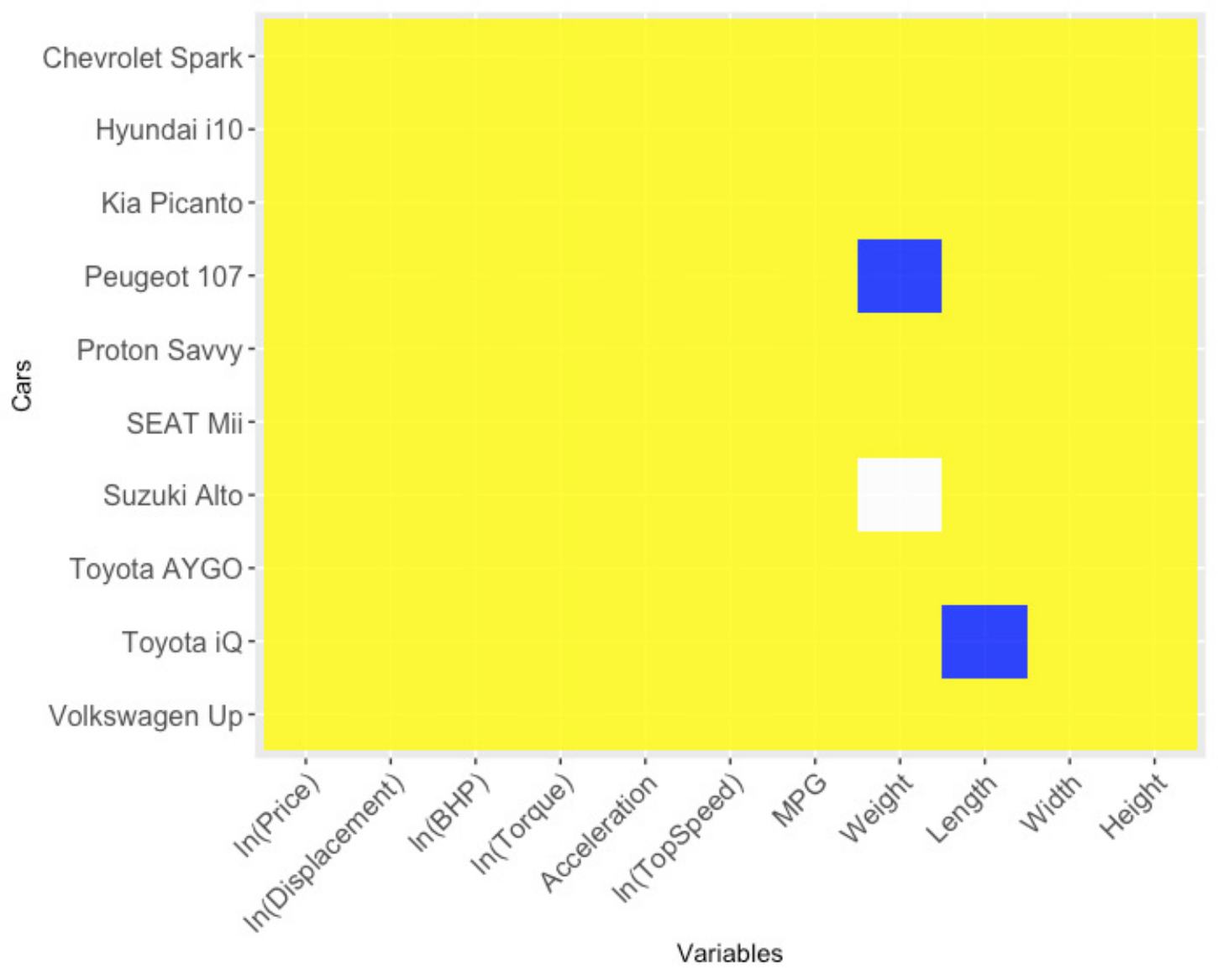}}  \\
\subfloat[][\textit{Cluster 3} \label{fig: TG_W_G3}] 
{\includegraphics[width=0.52\linewidth]{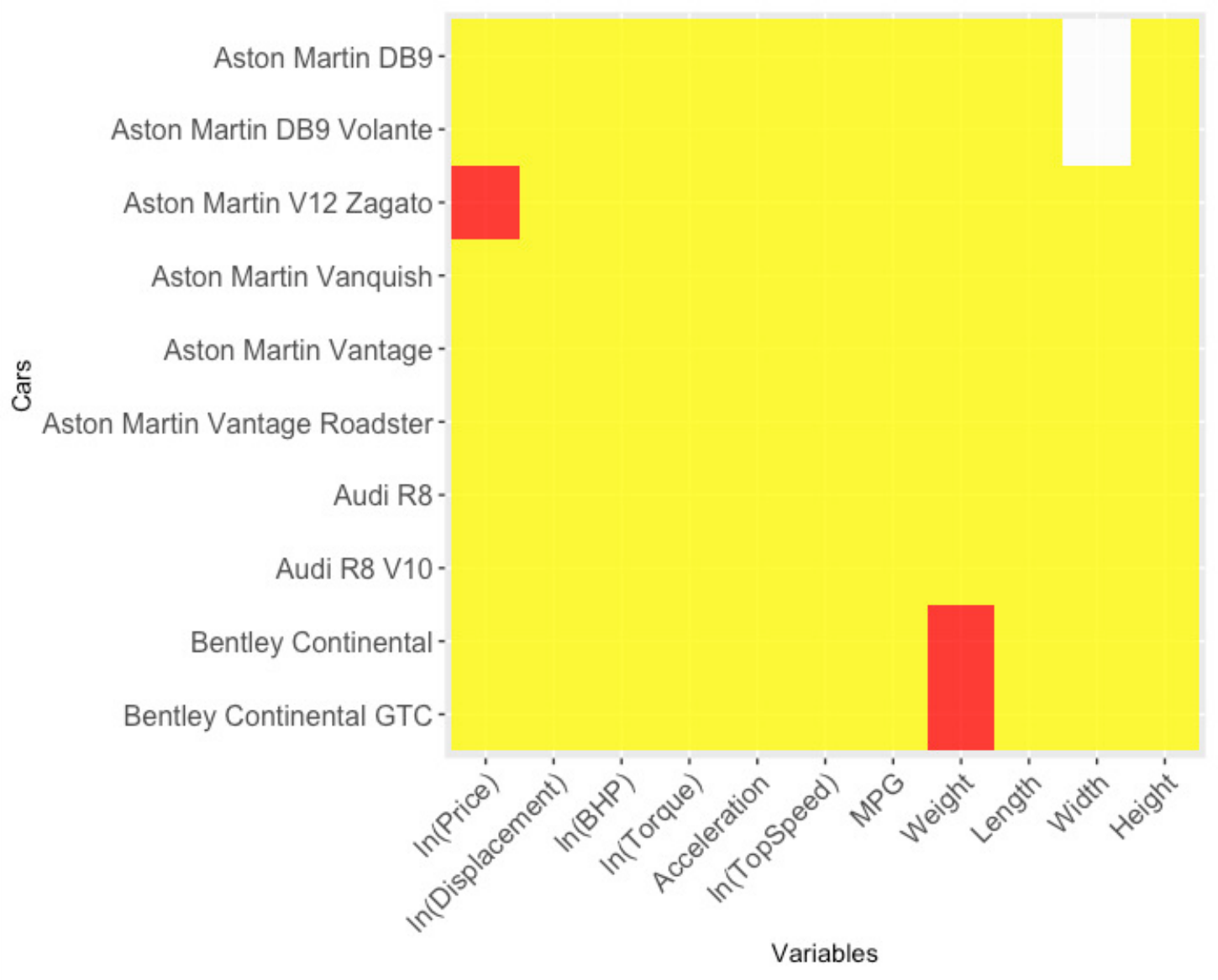}} \hfill
\subfloat[][\textit{Cluster 4} \label{fig: TG_W_G4}] 
{\includegraphics[height=0.42\linewidth, width=0.48\linewidth]{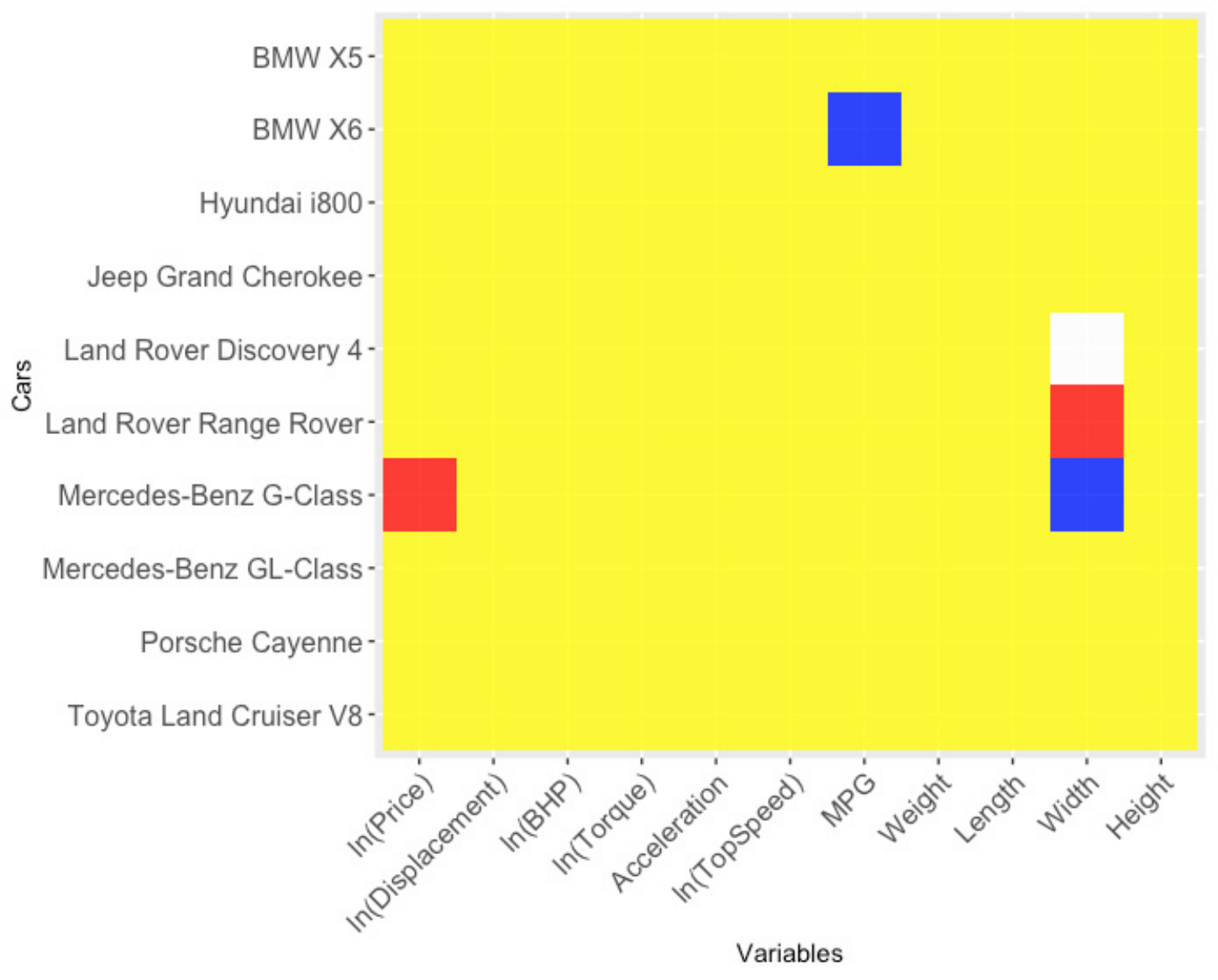}} 
\caption{Top Gear data set: representation of cellwise outliers and their imputation for the ten representative cars per cluster. Yellow: reliable cells; blue: contaminated cells imputed with a higher value than the original one; red: contaminated cells imputed with a lower value than the original one; white: missing values}
\label{fig: TG_out}
\end{figure}

All the variables are affected by unreliable cells, particularly \textit{Weight} ($17.63\%$), \textit{Width} ($11.86\%$), \textit{Length} ($11.19\%$),  \textit{Height} ($11.19\%$), \textit{MPG} ($8.47\%$), and \textit{ln(Price)} ($8.14\%$). For the other five variables, i.e., \textit{ln(Displacement)}, \textit{ln(BHP)}, \textit{ln(Torque)}, \textit{Acceleration} and \textit{ln(TopSpeed)}, the percentage of reliable cells ranges from $94\%$ to $99\%$ and refers to both contaminated and missing values, except for \textit{ln(Price)} and \textit{Acceleration}, which are fully observed. Considering separately the four variables mostly affected by unreliable cells, Figure \ref{fig: TG_StandRes} shows the standardized cellwise residuals, calculated as $(x_{ij} - \widehat{x}_{ij(g)})/\sqrt{\widehat{C}_{ij(g)}}$, where $g = \underset{g^{\prime} = 1, \ldots, G}{\text{arg}\max} \; \hat{z}_{ig^{\prime}}$. Missing data are not displayed in the plots, and the horizontal lines represent $\pm \sqrt{\chi^{2}_{1, 0.99}}$. For instance, the Bentley Continental, Bentley Continental GCT, Bentley Mulsanne, Flying Spur, Rolls-Royce Phantom, Rolls-Royce Phantom Coupé, and Volkswagen Phaeton feature soundproofing materials and structural reinforcements that increase their weight, among other features (Figure \ref{fig: TG_StandRes_Weight}); the Aston Martin Cygnet, Smart fortwo, and Toyota iQ are shorter than typical city cars, whereas the SEAT Toledo, Honda Insight, and Proton GEN-2 are longer (Figure \ref{fig: TG_StandRes_Length}). Other examples are the Land Rover Range Rover and Land Rover Range Rover Sport for Cluster 4, which have a width greater than expected for large SUVs or off-roaders (Figure \ref{fig: TG_StandRes_Width}), while the Fiat Doblò and Subaru Forester have a height higher than that of crossovers (Figure \ref{fig: TG_StandRes_Height}). It should be noted that for each variable, some cars may not be marginal outliers but could still be flagged as contaminated by cellGMM (black points in Figure \ref{fig: TG_StandRes}) based on the cluster configuration, meaning they might actually be points lying between clusters. Finally, an illustration of the unreliable cells for the most representative cars per cluster, as listed in Table \ref{tab: TopGear_10best}, is shown in Figure \ref{fig: TG_out}. Here, it is notable that some cars, such as the Peugeot 107 in Cluster 2, display only one contaminated cell, whereas others have multiple. For instance, conditional on their other $9$ variables, the Lexus GS in Cluster 1 has \textit{ln(Torque)} and \textit{ln(Top Speed)} significantly lower than expected in the cluster including mostly mid-side sedans (Figure \ref{fig: TG_W_G1}), and the Mercedes-Benz G-Class, although more expensive than expected for off-roaders, has a lower width (Figure \ref{fig: TG_W_G4}).

\section{Discussion} \label{sec: discussion}
The cellGMM methodology introduced in this paper is designed to detect and handle contaminated cells in heterogeneous populations, as well as manage missing values. Following the new paradigm of cellwise contamination and the rationale of the EM, we have proposed an algorithm that includes a concentration step to identify unreliable cells before the E- and M-steps, where the corresponding values are imputed and the Gaussian mixture model parameters are estimated, respectively. The estimation is constrained to ensure robustness and avoid degeneracies and spurious solutions. Compared to the existing methodology for model-based clustering in the presence of cellwise adulteration, cellGMM leverages the relationships among variables -- especially when they are strong -- by imputing the unreliable cells. In contrast, sclust sets these cells to NA in the parameter estimation.

While the proposal was initially illustrated by constraining the same proportion of contaminated cells per variable, a penalized version of cellGMM allows for automatic adjustments to avoid discarding valuable information and improve efficiency. The performance of these two approaches, in terms of clustering, parameter recovery, and data imputation, has been tested in a simulation study where cellGMM was compared to other robust and non-robust methodologies across both single and heterogeneous population frameworks. The examined scenarios cover simple and complex situations, considering different cluster configurations (well-separated and close components), data dimensionality (smaller and higher), outlier generation and magnitude (less and more extreme), and information removal (missing data). Additionally, three real data applications illustrate the results of cellGMM in different fields and for various purposes, such as the classification of spectral data, image reconstruction, and outlier detection in a data set where contaminated and missing values are not artificially generated. These are a few examples of possible applications of cellGMM, which can be implemented in any domain where the goal is to cluster data affected by missing information and potential outlying values within the variables.

Notwithstanding the results shown throughout the paper, several issues remain open for future research as the proposal is one of the first attempts to handle cellwise outlier detection in mixture models. First and foremost, we will delve into a criterion for selecting the number of components, which is rather fixed in our experiments. This criterion should account for the number of cells detected as contaminated, which can vary in the penalized approach. It is worth noting that cellGMM was developed under a frequentist framework, but it could potentially be developed within a Bayesian framework, where the number of mixture components might be estimated directly within the model, as in an infinite Gaussian mixture model \citep{F:1973}. Moreover, although the proposal can sometimes handle heavy tails through the cellwise approach, it is not explicitly designed to address skewed and/or multi-modal cluster distributions. This limitation could be addressed by extending the model to assume a $t$ distribution for each mixture component in a frequentist approach, or by considering infinite mixture of infinite Gaussian mixtures, as proposed by \citet{YRD:2014} in a Bayesian context. Additionally, the procedure for the initialization of the cellGMM algorithm may be improved by performing a detailed study on the tuning parameter settings to balance performance and avoid trapping initial solutions in specific regions of the parameter space. Finally, the cellwise properties of the parameter estimators will be theoretically inspected.

\subsubsection*{Supplementary Materials}
{\bf Supplementary Material:} This .pdf file contains insights into the cellGMM algorithm, including computations for the E- and M-step, initialization, and monotonicity (Section 1); additional results from the simulation study, along with computational time and complexity (Section 2); and from real data applications (Section 3).\\
{\bf Code:} \texttt{R} code for the implementation of cellGMM and replication of the analyses in Section \ref{sec: simulation} (Figures \ref{fig: simstudy_results_MR}-\ref{fig: simstudy_results_Parameters} and Table \ref{tab: sim_W_Imp}) and the results in Section \ref{subsec: data2} (Figure \ref{fig: Carina Nebula_PP}) is available in the .zip file and at \href{https://github.com/giorgiazaccaria/cellGMM}{https://github.com/giorgiazaccaria/cellGMM}.\\
{\bf Data:} The sources for the Carina Nebula and Top Gear data sets are reported in Sections \ref{subsec: data2} and \ref{subsec: data3}, respectively, while the Homogenized Meat data set is available in the Supplemental Content of \citet{MDR:2010}.

\subsubsection*{Acknowledgments}
The Authors sincerely thank the Editor, Associate Editor and two anonymous Reviewers for their constructive comments and suggestions.

\subsubsection*{Disclosure Statement}
The authors report there are no competing interests to declare.

\subsubsection*{Funding}
The research of Giorgia Zaccaria and Francesca Greselin was supported by Milano-Bicocca University Fund for Scientific Research, 2023-ATE-0448. Francesca Greselin's research was also supported by PRIN2022 - 2022LANNKC. The research of Luis A. García-Escudero and Agustín Mayo-Íscar was partially supported by grant PID2021-128314NB-I00 funded by MCIN/AEI/10.13039/501100011033/FEDER, UE and Junta Castilla y Le\'{o}n grant VA064G24. 

\bibliographystyle{jabes}
\bibliography{technometrics-bib}
\end{document}


\def\spacingset#1{\renewcommand{\baselinestretch}%
{#1}\small\normalsize} \spacingset{1}

\if0\blind
{
  \title{\bf Supplementary Material to \say{Cellwise outlier detection in heterogeneous populations}}
    \author{Giorgia Zaccaria$^1$, Luis A. García-Escudero$^2$, Francesca Greselin$^1$, \\ and Agustín Mayo-Íscar$^2$ \\
    $^1$Department of Statistics and Quantitative Methods \\
    University of Milano-Bicocca \\
    $^2$Department of Statistics and Operational Research \\ University of Valladolid}
    \date{}
  \maketitle
} \fi

\if1\blind
{
  \bigskip
  \bigskip
  \bigskip
  \begin{center}
    {\LARGE\bf Title}
\end{center}
  \medskip
} \fi

\bigskip
\spacingset{2} 

This document includes the supplementary material to the main article \say{Cellwise outlier detection in heterogeneous populations}. Specifically, Section \ref{sec: algorithm_sup} provides details on the cellGMM algorithm. Additional scenarios for the simulation study are presented in Section \ref{sec: simulation_sup}, along with the computational time and complexity of the proposed methodology. Finally, Section \ref{sec: app_sup} contains an in-depth analysis of the Homogenized Meat data set and the Top Gear data set examined in the main article.

\section{Details on the cellGMM algorithm}\label{sec: algorithm_sup}
\subsection{Computations for the E- and M-step} \label{subsec: math_sup}
In this section, we illustrate the computations required in the E- and M-step of the cellGMM algorithm. In the E-step, we compute the expected values $\mathbb{E}[Z_{ig} \vert \vec{x}_{i[\vec{w}_{i}^{(t+1)}]}; \widehat{\vec{\Psi}}^{(t)}]$, $\mathbb{E}[\vec{X}_{i [\vec{w}_{i}^{(t+1)c}]} \vert \vec{x}_{i[\vec{w}_{i}^{(t+1)}]}, z_{ig} = 1; \widehat{\vec{\Psi}}^{(t)}]$ and $\mathbb{E}[(\vec{X}_{i [\vec{w}_{i}^{(t+1)c}]} - \mean_{g [\vec{w}_{i}^{(t+1)c}]}) (\vec{X}_{i [\vec{w}_{i}^{(t+1)c}]} - \mean_{g [\vec{w}_{i}^{(t+1)c}]})^{\prime}$ $\vert \vec{x}_{i[\vec{w}_{i}^{(t+1)}]}, z_{ig} = 1; \widehat{\vec{\Psi}}^{(t)}]$, which can be easily derived following \citet{GJ:1994}:

\begin{align}
    &\mathbb{E}[Z_{ig} \vert \vec{x}_{i[\vec{w}_{i}^{(t+1)}]}; \widehat{\vec{\Psi}}^{(t)}] = \dfrac{\widehat{\pi}_{g}^{(t)} \phi_{p[\vec{w}_{i}^{(t+1)}]} \Big(\vec{x}_{i[\vec{w}_{i}^{(t+1)}]}; \meangestw^{(t)}, \sigugestw^{(t)} \Big)}{\sum_{h = 1}^{G} \widehat{\pi}_{h}^{(t)} \phi_{p[\vec{w}_{i}^{(t+1)}]} \Big(\vec{x}_{i[\vec{w}_{i}^{(t+1)}]}; \meanhestw^{(t)}, \siguhestw^{(t)} \Big)} := z_{ig}^{(t+1)} \\
    &\mathbb{E}[\vec{X}_{i [\vec{w}_{i}^{(t+1)c}]} \vert \vec{x}_{i[\vec{w}_{i}^{(t+1)}]}, z_{ig} = 1; \widehat{\vec{\Psi}}^{(t)}] = \meangestwc^{(t)} + \sigugestwcba^{(t)} \big( \sigugestw^{(t)} \big)^{-1}  \nonumber \\
    &\hspace{6.5cm} (\vec{x}_{i[\vec{w}_{i}^{(t+1)}]} - \meangestw^{(t)}) := \widehat{\vec{x}}_{i[\vec{w}_{i}^{(t+1)c}](g)}^{(t+1)} 
\end{align}
\begin{align}
    &\mathbb{E}[(\vec{X}_{i [\vec{w}_{i}^{(t+1)c}]} - \mean_{g [\vec{w}_{i}^{(t+1)c}]}) (\vec{X}_{i [\vec{w}_{i}^{(t+1)c}]} - \mean_{g [\vec{w}_{i}^{(t+1)c}]})^{\prime} \vert \vec{x}_{i[\vec{w}_{i}^{(t+1)}]}, z_{ig} = 1; \widehat{\vec{\Psi}}^{(t)}] = \widetilde{\vec{C}}_{i(g)}^{(t+1)} + \overset{\approx \quad \quad}{\vec{C}^{(t+1)}_{i(g)}},
\end{align}
where $\widetilde{\vec{C}}_{i(g)}^{(t+1)} := \sigugestwcbb^{(t)} - \sigugestwcba^{(t)} \Big( \sigugestw^{(t)} \Big)^{-1} \sigugestwcab^{(t)}$ and $\overset{\approx \quad \quad}{\vec{C}^{(t+1)}_{i(g)}} = \Big(\widehat{\vec{x}}_{i[\vec{w}_{i}^{(t+1)c}](g)}^{(t+1)} - \meangestwc^{(t)}\Big)$ $\Big(\widehat{\vec{x}}_{i[\vec{w}_{i}^{(t+1)c}](g)}^{(t+1)} - \meangestwc^{(t)}\Big)^{\prime}$. 

The M-step provides the parameter estimates by considering the \textit{completed} data $\Big\{ \big \{ \Tilde{\vec{x}}_{i(g)}^{(t+1)} = (\vec{x}_{i[\vec{w}_{i}^{(t+1)}]},  \widehat{\vec{x}}_{i[\vec{w}_{i}^{(t+1)c}](g)}^{(t+1)})^{\prime} \big \}_{g = 1}^{G} \Big\}_{i = 1}^{n}$. 

\medskip
\textbf{Update of $\boldsymbol{\pi}$.}
The weights $\boldsymbol{\pi} = \{ \pi_{g} \}_{g = 1}^{G}$ are updated as 
\begin{equation}\label{eqn: piest}
    \widehat{\pi}_{g}^{(t+1)} = \dfrac{\sum_{i = 1}^{n} z_{ig}^{(t+1)}}{n}.
\end{equation}

\medskip
\textbf{Update of $\boldsymbol{\theta}$.}
The component mean vectors and covariance matrices in $\boldsymbol{\theta} = \{\meang, \sigug \}_{g = 1}^{G}$ are updated as
\begin{align}\label{eqn: muest}
    &\meangest^{(t+1)} = \dfrac{\sum_{i = 1}^{n} z_{ig}^{(t+1)} \Tilde{\vec{x}}_{i(g)}^{(t+1)}}{\sum_{i = 1}^{n} z_{ig}^{(t+1)}}, \\
    &\sigugest^{(t+1)} = \dfrac{\sum_{i = 1}^{n} z_{ig}^{(t+1)} \Big[ \big(\Tilde{\vec{x}}_{i(g)}^{(t+1)} - \meangest^{(t+1)}\big) \big(\Tilde{\vec{x}}_{i(g)}^{(t+1)} - \meangest^{(t+1)}\big)^{\prime} + \widetilde{\vec{C}}_{i(g)}^{(t+1)} \Big]}{\sum_{i = 1}^{n} z_{ig}^{(t+1)}}.
\end{align}
If the eigenvalue-ratio constraint is not satisfied, truncated eigenvalues that comply with the constraint are obtained using the efficient procedure proposed by \citet{FGEMI:2013}. The corresponding component covariance matrices are then computed by substituting the truncated eigenvalues into their singular value decomposition.

\subsection{Initialization} \label{subsec: initialization_sup}
The procedure proposed for the initialization of the cellGMM algorithm is detailed below. This is heavily based on several applications of the TCLUST method, which is implemented via the R package \texttt{tclust} \citep{FGEMI:2012}.
\begin{enumerate}
    \item[Step 1] \textbf{Initialization of $\vec{W}$:} the initial solution for $\vec{W}$ is obtained by applying TCLUST individually to each variable and to pairs of variables with a fixed $\alpha_{\text{tclust}}$ trimming level. The estimated parameters per component from these TCLUST applications are used to compute the Mahalanobis distances. Roughly speaking, the smallest Mahalanobis distance across components for each unit is considered in both univariate and bivariate TCLUST applications. Proportions $\alpha_{1}$ and $\alpha_{2}$ of cells per variable are flagged as outliers based on the distribution of the Mahalanobis distances computed via univariate and bivariate TCLUST, respectively. Unreliable cells per variable previously identified through univariate TCLUST are excluded from the bivariate TCLUST outlier detection. A more detailed pseudo-code description of this procedure is provided in Algorithm \ref{algo: W_init}. Notice that in the pseudo-code we refer to the Mahalanobis distances as MD.
    \newpage
    \item[Step 2] \textbf{Initialization of $\vec{\Psi}$:} 
    \begin{enumerate}
    \item[2.1] $n_\text{rep}$ subsets of $q$ out of $p$ variables are randomly selected. For each reduced set of $q$ variables, TCLUST is implemented with a fixed trimming level $\alpha_{A_1}$ on the resulting completely reliable observations to obtain $G$ component mean vectors and covariance matrices. It is worth noting that reducing the set of variables considered increases the number of completely reliable observations as input for TCLUST, where the reliability of an observation is based upon the initial configuration of $\vec{W}$ (see Step 1). The two sets composed of $n_\text{rep}$ parameters each resulting from TCLUST are stored into $\textsf{tclust}_{\mean}$ ($n_\text{rep}$ location vectors)  and $\textsf{tclust}_{\sigu}$  ($n_\text{rep}$ scatter matrices), respectively. 
    
    \item[2.2] In this part, a version of trimmed $k$-means \citep{CAGM:1997} accommodated for the presence of missing values is implemented to obtain a partition of the $n_\text{rep}$ elements in $\textsf{tclust}_{\mean}$ into $G$ groups. The corresponding \say{centers of centers} are then used to initialize $\{\meang \}_{g = 1}^{G}$. Specifically, according to a multi-start procedure, a matrix $\vec{N}$ representing $G$ centers of center groups is initialized by one of the $n_\text{rep}$ elements in $\textsf{tclust}_{\mean}$. Then, $\vec{N}$ is iteratively updated across $n_{\text{iter}}$ iterations by assigning the $n_{\text{rep}}$ elements in $\textsf{tclust}_{\mean}$ to the nearest center of centers in $\vec{N}$ and recomputing $\vec{N}$ accordingly. The latter is obtained as the mean of centers in $\textsf{tclust}_{\mean}$ per group, after having discarded a proportion $\alpha_{\text{A2}}$ of them according to their squared Euclidean distances. Among several random starts ($n_\text{start}$), the selected solution for $\{\meang \}_{g = 1}^{G}$ is $\vec{N}$ that minimizes the total squared Euclidean distances between the centers in $\textsf{tclust}_{\mean}$ and the centers of center groups in $\vec{N}$. The resulting partition in groups is also used for computing the $G$ component covariance matrices $\{\sigug \}_{g = 1}^{G}$ as the means of the elements in $\textsf{tclust}_{\sigu}$ per group. It is worth noting that this procedure does not necessarily guarantee the positive definiteness of $\sigug$, as the mean of covariance matrices in $\textsf{tclust}_{\sigu}$ refers to different subsets of variables. Therefore, we impose on $\{ \sigug \}_{g = 1}^{G}$ the eigenvalue-ratio constraint reported in (4) of the main article, which entails obtaining positive definite matrices. This constraint is implemented throughout the cellGMM algorithm via the efficient procedure reported in \citet{FGEMI:2013}. The weights $\boldsymbol{\pi} = \{ \pi_{g} \}_{g = 1}^{G}$ are figured up after assigning each observation to a component by considering the minimum squared Euclidean distance from $\{ \meang \}_{g = 1}^{G}$. A more detailed pseudo-code description of the procedure proposed in this Step 2.2 is illustrated in Algorithm \ref{algo: Psi_init}.
    \end{enumerate}
\end{enumerate}

The initialization described herein depends on several tuning parameters. In our experiments, we set those regarding the trimming levels according to the true level of contamination, denoted as $\alpha_{\text{true}}$. Specifically, we set $\alpha_{\text{tclust}} = 2 \cdot \alpha_{\text{true}}$, $\alpha_{1} = \alpha_{2} = \alpha_{\text{true}}$, $\alpha_{\text{A1}} = \alpha_{\text{true}}$ and $\alpha_{\text{A2}} = 2 \cdot \alpha_{\text{true}}$. If $\alpha_{\text{true}}$ is unknown, as it is usually the case in real-world applications, we suggest considering a conservative level of contamination as the true value, e.g., $0.03$ or $0.05$ (i.e., assuming $3\%$ or $5\%$ of contamination). The tuning parameter $q$, which controls the number of variables selected in the first step of the initialization of $\vec{\Psi}$, is fixed to $\lfloor \frac{p}{2} \rfloor + 1$. This choice guarantees that the subsets of sampled variables overlap for at least one variable. The remaining parameters for the initialization of $\vec{\Psi}$ are set as follows: $n_{\text{rep}} = 40$, $n_{\text{start}} = 10$ and $n_{\text{iter}} = 10$. This configuration of the tuning parameters proves to be effective in our experiments. However, their setting could be improved by performing a specific simulation study on the initialization, which is out of the scope of this paper primarily focused on the main part of the cellGMM algorithm. 

\begin{algorithm}
\caption{Initialization of $\vec{W}$ (Step 1)}\label{algo: W_init}
\begin{algorithmic}[1]
\STATE \textbf{Input:} $\vec{X}$, $\alpha_{\text{tclust}}$, $\alpha_{1}$, $\alpha_{2}$, $G$
\STATE $p \gets$ number of columns (variables) in $\vec{X}$ 

\noindent \underline{\textbf{Univariate TCLUST}}
\FOR{$j_{1} \gets 1:p$}
    \STATE $\texttt{tc}_{1}[[j_{1}]] \gets$ \texttt{tclust}($\vec{X}[, j_{1}]$, $\texttt{k} = G$, $\texttt{alpha} = \alpha_{\text{tclust}}$), where only observed units in variable $j_{1}$ are considered
\ENDFOR 

\FOR{$j_{1} \gets 1:p$}
    \STATE MD$_{1} \gets$ compute MD for each unit per component according to the parameters obtained from univariate TCLUST
    \STATE $\texttt{f}_{1}[, j_{1}] \gets$ select the minimum MD$_{1}$ across components 
\ENDFOR

\noindent \textbf{\underline{Detection of cellwise outliers through univariate TCLUST}}

\STATE $\texttt{qq}_{1} \gets$ \texttt{quantile}($\texttt{f}_{1}, 1-\alpha_{1}$, \texttt{na.rm = TRUE})
\STATE $\vec{W}_{(1)}[\texttt{f}_{1} > \texttt{qq}_{1}] \gets 0$, where $\vec{W}_{(1)}$ has been initialized as $\vec{1}_{n}\vec{1}_{p}^{\prime}$ (i.e., an  $n\times p$ matrix of ones)

\noindent \underline{\textbf{Bivariate TCLUST}}
\FOR{$j_{1} \gets 1:(p-1)$}
    \FOR{$j_{2} \gets (j_{1}+1):p$}
    \STATE $\texttt{tc}_{2}[[j_{1}]][[j_{2}]] \gets$ \texttt{tclust}($\texttt{cbind}(\vec{X}[, j_{1}], \vec{X}[, j_{2}])$, $\texttt{k} = G$, $\texttt{alpha} = \alpha_{\text{tclust}}$), where only observed units in both variables $j_{1}$ and $j_{2}$ are considered
\algstore{W_init}
\end{algorithmic}
\end{algorithm}

\begin{algorithm}                     
\begin{algorithmic} [1]                  
\algrestore{W_init} 
 \ENDFOR
\ENDFOR
\FOR{$j_{1} \gets 1:(p-1)$}
    \FOR{$j_{2} \gets (j_{1}+1):p$}
        \STATE MD$_{2} \gets$ compute MD for each unit per component according to the parameters from bivariate TCLUST
        \STATE $\texttt{f}_{2}[, j_{1}, j_{2}] \gets$ select the minimum MD$_{2}$ across components 
    \ENDFOR
\ENDFOR

\noindent \textbf{\underline{Detection of cellwise outliers through bivariate TCLUST by excluding the} \underline{already flagged cells}}
\FOR{$j \gets 1:p$}
    \STATE $\texttt{f}_{2}[\vec{W}_{(1)}[, j] == 0, j, ] \gets$ NA
    \STATE $\texttt{f}_{2}[\vec{W}_{(1)}[, j] == 0, , j] \gets$ NA
\ENDFOR
  
\FOR{$j \gets 1:p$}
    \STATE $\texttt{ff}_{2}[, j] \gets$ \texttt{apply}($\texttt{f}_{2}[, j, ], 1$, \texttt{sum, na.rm = TRUE}) + \texttt{apply}($\texttt{f}_{2}[, , j], 1$, \texttt{sum, na.rm = TRUE})
\ENDFOR

\STATE $\texttt{ff}_{2}[\vec{W}_{(1)} == 0] \gets$ NA

\STATE $\texttt{qq}_{2} \gets$ \texttt{quantile}($\texttt{ff}_{2}, 1 - \frac{\alpha_{1}}{1 - \alpha_{2}}$, \texttt{na.rm = TRUE})
\algstore{W_init}
\end{algorithmic}
\end{algorithm}

\begin{algorithm}                     
\begin{algorithmic} [1]                  
\algrestore{W_init} 
\STATE $\vec{W}_{(2)}[\texttt{ff}_{2} > \texttt{qq}_{2}] \gets 0$, where $\vec{W}_{(2)}$ has been initialized as $\vec{1}_{n}\vec{1}_{p}^{\prime}$ and NA values in $\vec{W}_{(2)}$ are replaced by $1$
\STATE $\vec{W} \gets \vec{W}_{(1)} \odot  \vec{W}_{(2)}$ (Hadamard or element-wise product), where cells corresponding to NA values in $\vec{X}$ are set to $0$
\STATE \textbf{Output:} $\vec{W}$
\end{algorithmic}
\end{algorithm}

\begin{algorithm}
\caption{Step 2.2 of the initialization of $\vec{\Psi}$}\label{algo: Psi_init}
\begin{algorithmic}[1]
\STATE \textbf{Input:} $\vec{X}, G, \alpha_{\text{A2}}, n_{\text{start}}, n_{\text{iter}}, \textsf{tclust}_{\mean}, \textsf{tclust}_{\sigu}$ ($n_{\text{iter}}$, $\textsf{tclust}_{\mean}$ and $\textsf{tclust}_{\sigu}$ come from Step 2.1)
\STATE $\texttt{obj.best} \gets + \infty$
\FOR{$\texttt{start} \gets 1:n_{\text{start}}$}
    \STATE Sample one of the $n_\text{rep}$ subsets of variables used for obtaining $\textsf{tclust}_{\mean}$ and $\textsf{tclust}_{\sigu}$ and initialize the centers of center groups in $\vec{N}$ as the $\textsf{tclust}_{\mean}$ corresponding to the selected repetition 
    \FOR{$\texttt{iter} \gets 1:n_{\text{iter}}$}
        \STATE Compute distances between the centers $\textsf{tclust}_{\mean}$ and the centers of center groups in $\vec{N}$
        \STATE Assign each center in $\textsf{tclust}_{\mean}$ to the corresponding group $g \in \{1, \ldots, G\}$ according to the minimum distance
        \STATE Update the rows of $\vec{N}$ as the mean of the corresponding centers in $\textsf{tclust}_{\mean}$ assigned to the $g$th group, $g = 1, \ldots, G$, by excluding a proportion $\alpha_{\text{A2}}$ of centers in $\textsf{tclust}_{\mean}$ with the most \say{extreme} distances
    \ENDFOR
    \STATE $\texttt{obj} \gets$ sum of distances across $n_{\text{rep}}$ repetitions and $G$ groups
    \IF{$\texttt{obj} < \texttt{obj.best}$}
        \STATE Update $\{ \meang \}_{g = 1}^{G}$ and group assignment of the centers in $\textsf{tclust}_{\mean}$ 
        \STATE $\texttt{obj.best} \gets \texttt{obj}$
    \ENDIF
\ENDFOR
\algstore{Param_init}
\end{algorithmic}
\end{algorithm}

\begin{algorithm}              
\begin{algorithmic}[1]                  
\algrestore{Param_init}
    \STATE Compute $\{ \sigug \}_{g = 1}^{G}$ as the mean of the covariance matrices in $\textsf{tclust}_{\sigu}$ according to the group assignment obtained for the centers in $\textsf{tclust}_{\mean}$ and check for the eigenvalue-ratio constraint
    \STATE Compute the weights $\{ \pi_{g} \}_{g = 1}^{G}$ according to the minimum squared Euclidean distance of each observation from $\{ \meang \}_{g = 1}^{G}$ 
\STATE \textbf{Output:} initial parameters $\vec{\Psi} = \{ \boldsymbol{\pi}, \boldsymbol{\theta}\}$, where $\boldsymbol{\pi} = \{\pi_{g}\}_{g = 1}^{G}$ and $\boldsymbol{\theta} = \{\meang, \sigug \}_{g = 1}^{G}$
\end{algorithmic}
\end{algorithm}

\subsection{Algorithm monotonicity}\label{subsec: algo_sup}
In the following theorem, we state the monotonicity of the cellGMM algorithm. 

\begin{theorem}
    The cellGMM objective function, subject to the constraint on the number of reliable cells per variable and the eigenvalue-ratio constraint, is non-decreasing at each iteration of the EM-type algorithm. 
\end{theorem}
\begin{proof}
    We first show the monotonicity of the log-likelihood, i.e., $\ell(\widehat{\vec{\Psi}}^{(t+1)}, \vec{W}^{(t+1)}; \vec{X}) \geq \ell(\widehat{\vec{\Psi}}^{(t)}, \vec{W}^{(t)}; \vec{X}), \forall t = 0, 1, 2, \ldots$. To establish this, we need to verify that the log-likelihood does not decrease at each step of the algorithm. Specifically, in the C-step, we sequentially update each column of $\vec{W}^{(t)}$, keeping all the other columns fixed, by maximizing the objective function under the constraint on the number of reliable cells per variable. Therefore, for each column we do not decrease the objective function and this guarantees that, eventually, $\ell(\widehat{\vec{\Psi}}^{(t)}, \vec{W}^{(t+1)}; \vec{X}) \geq \ell(\widehat{\vec{\Psi}}^{(t)}, \vec{W}^{(t)}; \vec{X})$. The other two steps correspond to the traditional E- and M-step of the EM algorithm with missing data \citep{GJ:1994} for which, given $\vec{W}^{(t+1)}$, the objective function does not decrease \citep[see][Section 11.3]{LR:2019}. The same holds for the penalized approach, where $\ell_{\text{pen}}(\widehat{\vec{\Psi}}^{(t+1)}, \vec{W}^{(t+1)}; \vec{X}) \geq \ell_{\text{pen}}(\widehat{\vec{\Psi}}^{(t)}, \vec{W}^{(t)}; \vec{X}), \forall t = 0, 1, 2, \ldots$.
\end{proof}

\newpage
The stopping criterion used for the cellGMM algorithm considers Aitken's acceleration procedure (\citealp{Betal:1994}; \citealp[Section 2.11]{MLP:2000}; \citealp[Section 4.9]{MK:2008}). This is based on the Aitken-accelerated estimate of the log-likelihood -- for simplicity, we denote hereinafter $\ell(\vec{\Psi}, \vec{W}; \vec{X})$ as $\ell$ -- which is defined at iteration $t + 1$ as 
\begin{equation*}
    \ell_{\infty}^{(t+1)} =  \ell^{(t)} + \dfrac{1}{1+ a^{(t)}} \big(\ell^{(t+1)} - \ell^{(t)} \big),
\end{equation*}
where $a^{(t)} = \dfrac{\ell^{(t+1)} - \ell^{(t)}}{\ell^{(t)} - \ell^{(t-1)}}$. The algorithm is considered to have converged if $\ell_{\infty}^{(t+1)} - \ell^{(t)} < \epsilon$ \citep{McNMMcDF:2010}, where $\epsilon$ is an arbitrary small positive constant ($\epsilon = 10^{-6}$ in our experiments).

\newpage
\section{Simulation study: additional results} \label{sec: simulation_sup}
\begin{figure}
    \centering
    \includegraphics[width=\linewidth]{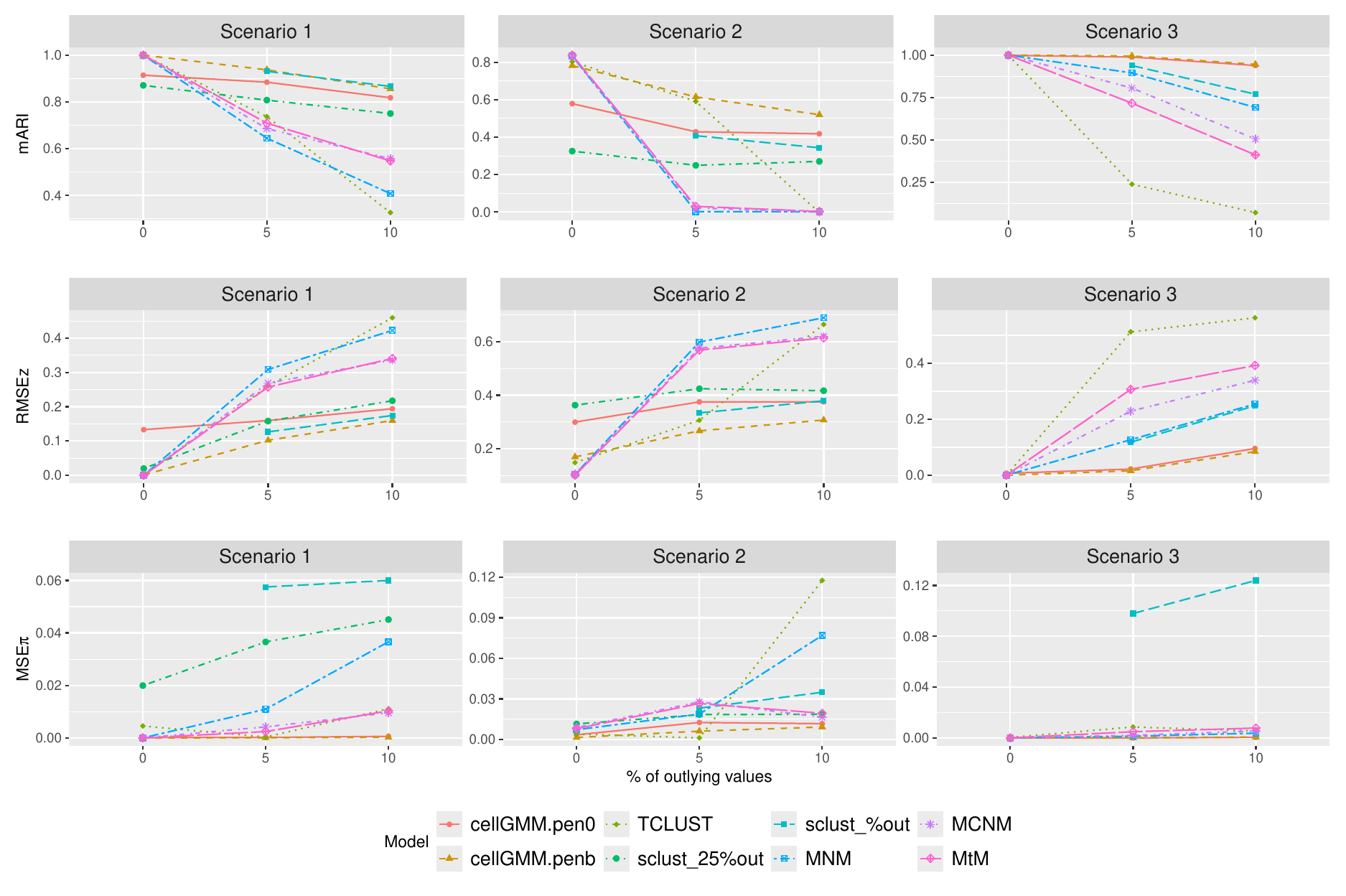}
    \caption{Results of the simulation study with less extreme outliers: mean of the Adjusted Rand Index (mARI), Root Mean Squared Error (RMSE) of the posterior probabilities and Mean Squared Error (MSE) of the prior probabilities per scenario, percentage of contamination, and model}
    \label{fig: simstudy_results123_sup}
\end{figure}

We present herein the results of several additional simulation studies beyond those reported in Section 3 of the main article. These include three scenarios with less extreme cellwise outliers and missing data (Section \ref{subsec: sim_less_sup}) and three scenarios with more extreme contamination (Section \ref{subsec: sim_more_sup}). For all nine scenarios considered, we provide the number of samples on which each model can be properly computed in Table \ref{tab: sim_paper_add_sup}. In addition to the indices already introduced in Section 3 of the main article for assessing the models' performance, we report the Adjusted Rand Index \citep[ARI, ][]{HA:1985}, the Root Mean Squared Error of the posterior probabilities and the Mean Squared Error of the prior probabilities. These indices for Scenarios 1-3 are reported in Figure \ref{fig: simstudy_results123_sup}. Computation times for the methodologies implemented in Section 3 of the main article are reported in Section \ref{subsec: sim_computingtime}, along with the computational complexity of cellGMM. Finally, in Section \ref{subsec: sim_20out_sup}, we present the results for Scenarios 1 and 2 with a higher percentage ($20\%$) of both less and more extreme outlying values, while a simulation study in which cellGMM is applied to data sets contaminated by structurally outlying cells is illustrated in Section \ref{subsec: sim_structural}.

\begin{table}
\centering
\caption{Number of samples on which each model can be properly run per scenario, percentage of contamination, and model. Dashes represent scenarios where models cannot be implemented or where implementation is not necessary}\label{tab: sim_paper_add_sup}
\resizebox{\linewidth}{!}{
\begin{tabular}{cl|cccccc|ccc}
\hline
& & \multicolumn{6}{c}{Less extreme outliers} & \multicolumn{3}{c}{More extreme outliers} \\
$\%$ out. & Model & Scenario 1 & Scenario 2 & Scenario 3 & Scenario 4 & Scenario 5 & Scenario 6 & Scenario 1 & Scenario 2 & Scenario 3\\ 
\hline
\multirow{7}{*}{0} & cellGMM.pen0 & 100 & 100 & 100 & 100 & 100 & 100 & - & - & - \\ 
& cellGMM.penb & 100 & 100 & 100 & 100 & 100 & 100 & - & - & - \\ 
& TCLUST & 100 & 100 & 100 & - & - & - & - & - & - \\ 
& sclust\_25 & 98 & 100 & 0 & - & - & - & - & - & - \\ 
& MNM & 100 & 100 & 100 & 100 & 100 & 100 & - & - & - \\ 
& MCNM & 100 & 100 & 100 & 80 & 78 & 13 & - & - & - \\ 
& M$t$M & 100 & 100 & 100 & 100 & 100 & 100 & - & - & - \\ 
\cmidrule{2-11}
\multirow{8}{*}{5} & cellGMM.pen0 & 100 & 100 & 100 & 100 & 100 & 100 & 100 & 100 & 100 \\ 
& cellGMM.penb & 100 & 100 & 100 & 100 & 100 & 100 & 100 & 100 & 100 \\ 
& TCLUST & 100 & 100 & 100 & - & - & - & 100 & 100 & 100 \\ 
& sclust\_25 & 97 & 94 & 0 & - & - & - & 31 & 0 & 0 \\ 
& sclust\_5 & 93 & 90 & 90 & - & - & - & 22 & 0 & 3 \\ 
& MNM & 100 & 100 & 100 & 100 & 100 & 100 & 98 & 91 & 99 \\ 
& MCNM & 100 & 100 & 100 & 16 & 24 & 0 & 96 & 61 & 76 \\ 
& M$t$M & 100 & 100 & 100 & 100 & 100 & 100 & 96 & 61 & 76 \\  
\cmidrule{2-11}
\multirow{8}{*}{10} & cellGMM.pen0 & 100 & 100 & 100 & 100 & 100 & 100 & 100 & 100 & 100 \\ 
& cellGMM.penb & 100 & 100 & 100 & 100 & 100 & 99 & 100 & 100 & 100 \\ 
& TCLUST & 100 & 100 & 100 & - & - & - & 100 & 100 & 100 \\ 
& sclust\_25 & 93 & 88 & 0 & - & - & - & 3 & 0 & 0 \\ 
& sclust\_10 & 88 & 63 & 53 & - & - & - & 3 & 0 & 0 \\ 
& MNM & 100 & 100 & 100 & 100 & 100 & 100 & 100 & 100 & 76 \\ 
& MCNM & 100 & 100 & 100 & 31 & 25 & 0 & 100 & 100 & 57\\ 
& M$t$M & 100 & 100 & 100 & 100 & 100 & 99 & 100 & 100 & 62\\ 
\hline
\end{tabular}%
}
\end{table}

\subsection{Less extreme outliers}\label{subsec: sim_less_sup}
The three additional scenarios with less extreme outliers are as follows: \textit{Scenario 4} and \textit{Scenario 5} with $n = 200, p = 5, G = 2$, non-spherical with $\sigu_{1} = [\sigma_{ij} = 0.9^{\lvert i-j \rvert}: i, j = 1, \ldots, p]$ and $\sigu_{2}$ obtained by an orthogonal rotation of $\sigu_{1}$, unbalanced ($\boldsymbol{\pi} = [0.3, 0.7]$) and well-separated and close components, respectively; \textit{Scenario 6} with $n = 400, p = 15, G = 4$, non-spherical with $\sigu_{1} = \sigu_{2} = [\sigma_{ij} = 0.9^{\lvert i-j \rvert}: i, j = 1, \ldots, p]$ and $\sigu_{3} = \sigu_{4}$ obtained by an orthogonal rotation of $\sigu_{1}$, unbalanced ($\boldsymbol{\pi} = [0.2, 0.2, 0.3, 0.3]$) and well-separated components. The component mean vectors are generated from uniform distributions. Specifically, $\mean_{1} = \vec{0}$, and the elements of $\mean_{g}, g = 2, \ldots, G,$ are drawn from a uniform distribution in $[0, 10]$ in the well-separated case, and from $[1, 3]$ in the close case. In the former, we assess whether the distance between the component mean vectors is less than 5, and re-generate them if this occurs. The components' configuration is controlled through the overlapping measure $\omega$ introduced by \citet{MM:2010}, where well-separated and close components correspond to $\omega_{\text{max}} < 0.01$ and $0.05 <\omega_{\text{max}} < 0.06$, respectively. For each scenario, we generate $100$ data matrices that we contaminate with $0\%, 5\%$, and $10\%$ of outlying cells randomly drawn from a uniform distribution in the interval $[-10, 10]$, ensuring that the contaminated observations do not lie within the $99$th percentile ellipsoid of any component. Moreover, we randomly remove $5\%$ of the cells which have not been contaminated to obtain samples with missing data. Due to the latter, the only models we can compare to cellGMM are those included in the R package \texttt{MixtureMissing}, i.e., MNM, MCNM, M$t$M, since they can handle missing values. 

\begin{figure}
    \centering
    \includegraphics[width=\linewidth]{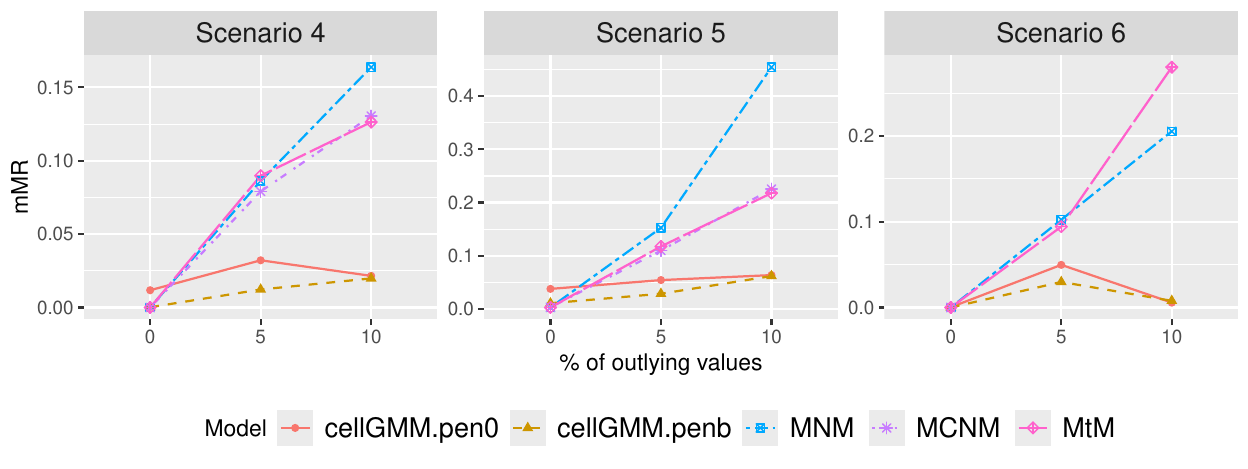}
    \caption{Results of the simulation study with less extreme outliers: mean of the Misclassification Rate (mMR) per scenario, percentage of contamination, and model}
    \label{fig: simstudy_results456_sup_MR}
\end{figure}

\begin{figure}[!htbp]
    \centering
    \includegraphics[width=\linewidth, height=0.6\textheight]{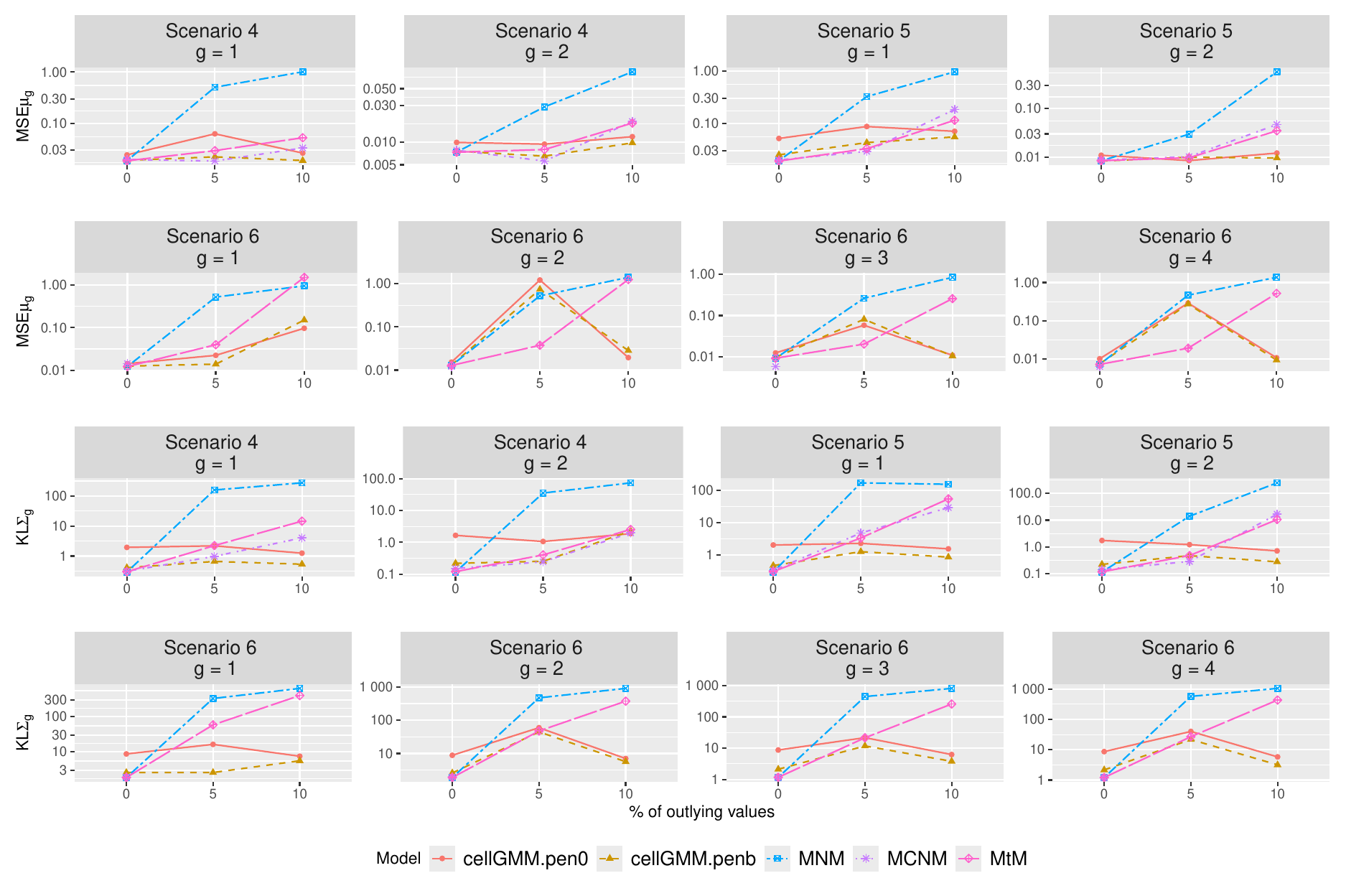}
    \caption{Results of the simulation study with less extreme outliers: Mean Squared Error (MSE) of the component mean vectors and Kullback-Leibler (KL) discrepancy for the component covariance matrices per scenario, percentage of contamination, and model. The values are represented via log-transformation, while the y-axis ticks are labeled using the original scale}
    \label{fig: simstudy_results456_sup_Parameters}
\end{figure}

\begin{figure}
    \centering
    \includegraphics[width=\linewidth]{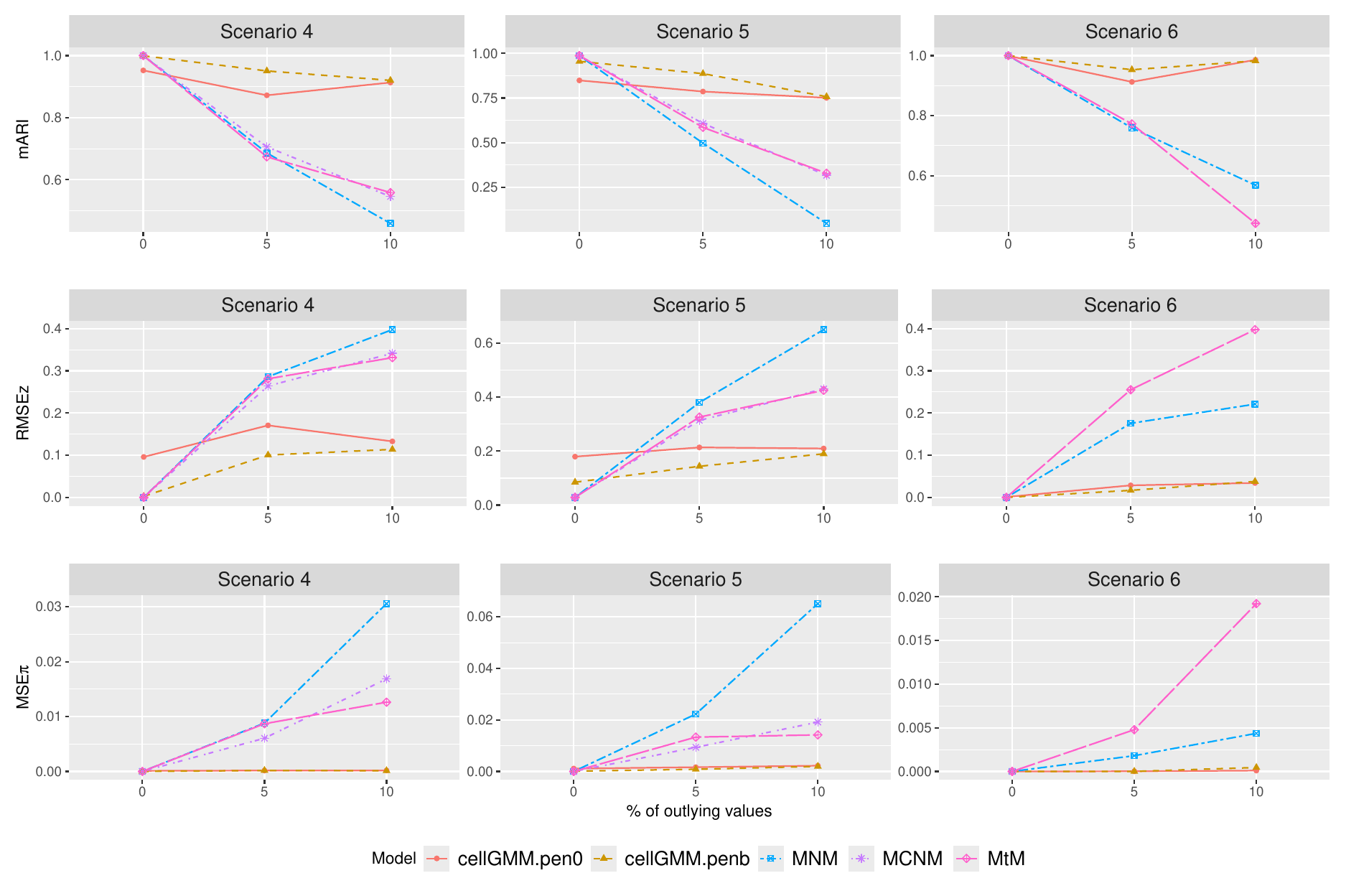}
    \caption{Results of the simulation study with less extreme outliers: mean of the Adjusted Rand Index (mARI), Root Mean Squared Error (RMSE) of the posterior probabilities and Mean Squared Error (MSE) of the prior probabilities per scenario, percentage of contamination, and model}
    \label{fig: simstudy_results456_sup2}
\end{figure}

\begin{table}[hbt!]
\centering
\caption{Outlier detection and imputation with less extreme outliers: percentage of True and False Positive ($\%$TP and $\%$FP), Mean Absolute and Root Mean Squared Error (MAE and RMSE) for comparing the original and imputed data matrices per scenario, percentage of contamination, and model}\label{tab: sim_W_Imp_Less_sup}
\resizebox{\textwidth}{!}{
\begin{tabular}{cr|cccc|cccc|cccc}
\hline
& & \multicolumn{4}{c|}{Scenario 4} & \multicolumn{4}{c|}{Scenario 5} & \multicolumn{4}{c}{Scenario 6} \\
\hline
$\%$ out. & Method & \%TP & \%FP & MAE & RMSE & \%TP & \%FP & MAE & RMSE & \%TP & \%FP & MAE & RMSE \\
\hline
\multirow{7}{*}{0} & cellGMM.pen0 & - & 19.63 & 0.24 & 0.65 & - & 19.67 & 0.24 & 0.60 & - & 20.00 & 0.19 & 0.45 \\ 
& cellGMM.penb & - & 2.16 & 0.04 & 0.20 & - & 2.41 & 0.05 & 0.23 & - & 2.63 & 0.04 & 0.19 \\ 
& MNM & - & - & 0.02 & 0.09 & - & - & 0.02 & 0.10 & - & - & 0.02 & 0.09 \\ 
& MCNM & - & 4.09 & 0.02 & 0.09 & - & 4.35 & 0.02 & 0.10 & - & 43.20 & 0.02 & 0.09 \\ 
& M$t$M & - & 22.76 & 0.02 & 0.09 & - & 22.67 & 0.02 & 0.10 & - & 19.88 & 0.02 & 0.09 \\ 
& cellMCD & - & 17.49 & 0.92 & 2.35 & - & 10.52 & 0.28 & 0.87 & - & 6.73 & 0.41 & 1.60 \\ 
& DI & - & 1.44 & 0.05 & 0.27 & - & 4.64 & 0.13 & 0.49 & - & 1.45 & 0.09 & 0.67 \\
\hline
\multirow{7}{*}{5} & cellGMM.pen0 & 92.24 & 19.60 & 0.30 & 0.99 & 94.76 & 19.59 & 0.26 & 0.72 & 95.35 & 19.98 & 0.23 & 0.68 \\ 
& cellGMM.penb & 92.98 & 6.89 & 0.09 & 0.47 & 95.00 & 7.35 & 0.09 & 0.37 & 93.95 & 6.89 & 0.07 & 0.35 \\ 
& MNM & - & - & 0.70 & 1.14 & - & - & 0.64 & 0.94 & - & - & 0.95 & 1.54 \\ 
& MCNM & 94.12 & 20.51 & 0.69 & 1.38 & 92.42 & 20.41 & 0.62 & 1.10 & - & - & - & - \\ 
& M$t$M & 99.54 & 37.85 & 0.67 & 1.33 & 99.34 & 37.11 & 0.63 & 1.09 & 93.93 & 48.76 & 0.79 & 1.38 \\ 
& cellMCD & 90.04 & 20.54 & 0.92 & 2.33 & 91.68 & 13.88 & 0.30 & 0.89 & 89.35 & 9.08 & 0.31 & 1.29 \\ 
& DI & 70.45 & 4.82 & 0.16 & 0.90 & 74.04 & 5.58 & 0.16 & 0.83 & 43.23 & 3.25 & 0.30 & 1.46 \\
\hline
\multirow{7}{*}{10} & cellGMM.pen0 & 94.50 & 19.55 & 0.25 & 0.80 & 94.68 & 19.61 & 0.22 & 0.66 & 94.03 & 20.00 & 0.19 & 0.54 \\ 
& cellGMM.penb & 93.50 & 12.09 & 0.17 & 0.71 & 94.50 & 13.09 & 0.14 & 0.51 & 92.72 & 12.95 & 0.12 & 0.46 \\ 
& MNM & - & - & 0.94 & 1.52 & - & - & 0.71 & 1.02 & - & - & 1.30 & 1.98 \\ 
& MCNM & 91.16 & 35.64 & 0.94 & 1.68 & 84.56 & 33.59 & 0.77 & 1.24 & - & - & - & - \\ 
& M$t$M & 97.84 & 46.95 & 0.88 & 1.66 & 91.82 & 44.90 & 0.74 & 1.19 & 79.41 & 54.32 & 1.06 & 1.69 \\ 
& cellMCD & 87.75 & 22.06 & 0.87 & 2.23 & 90.83 & 13.72 & 0.17 & 0.55 & 86.04 & 11.77 & 0.27 & 1.13 \\ 
& DI & 68.03 & 8.07 & 0.29 & 1.26 & 73.09 & 9.17 & 0.26 & 1.11 & 41.71 & 4.79 & 0.47 & 1.90 \\ 
\hline
\end{tabular}}
\end{table}

The results of the classification performance and parameter estimates are reported in Figures \ref{fig: simstudy_results456_sup_MR}, \ref{fig: simstudy_results456_sup_Parameters}, and \ref{fig: simstudy_results456_sup2}. As the level of contamination and the dimensionality of the data increase, the difference between cellGMM, specifically its penalized version (i.e. cellGMM.penb), and the competitors grows. Notably, in Scenario 6 with $10\%$ of cellwise outliers, cellGMM.penb improves the parameter estimates compared to cellGMM.pen0, although this improvement is not reflected in the averaged ARI over the $100$ samples. This is due to the distribution of the contaminated data being, by chance, well balanced among variables. In Table \ref{tab: sim_W_Imp_Less_sup}, we illustrate the results of the model on outlier detection and the corresponding value imputation for the aforementioned models, as well as cellMCD and DI. It should be noted that, unlike for Scenarios 1-3 analyzed in the main article, Table \ref{tab: sim_W_Imp_Less_sup} shows non-null values of indices for the imputation evaluation of the model-based clustering competitors, referring only to the missing values. Dashes indicate cases where the models cannot be run on any sample because errors occur in the code. Overall, cellGMM.penb outperforms the competitors by balancing good performance on TP\% and FP\% and data imputation.

\subsection{More extreme outliers}\label{subsec: sim_more_sup}
\begin{figure}
    \centering
    \includegraphics[width=\linewidth]{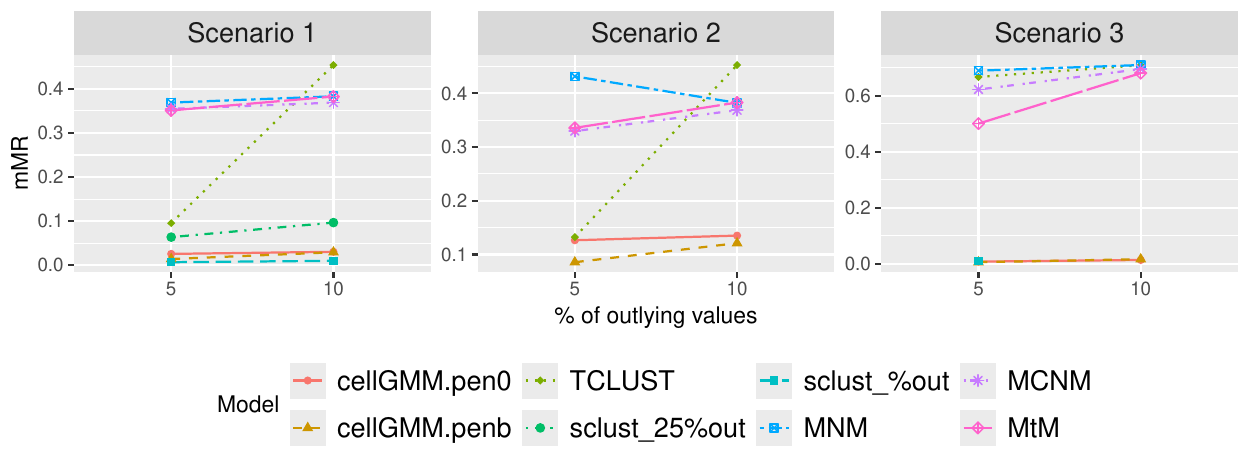}
    \caption{Results of the simulation study with more extreme outliers: mean of the Misclassification Rate (mMR) per scenario, percentage of contamination, and model}
    \label{fig: simstudy_results123_extreme_sup_MR}
\end{figure}

\begin{figure}
    \centering
    \includegraphics[width=\linewidth, height=0.6\textheight]{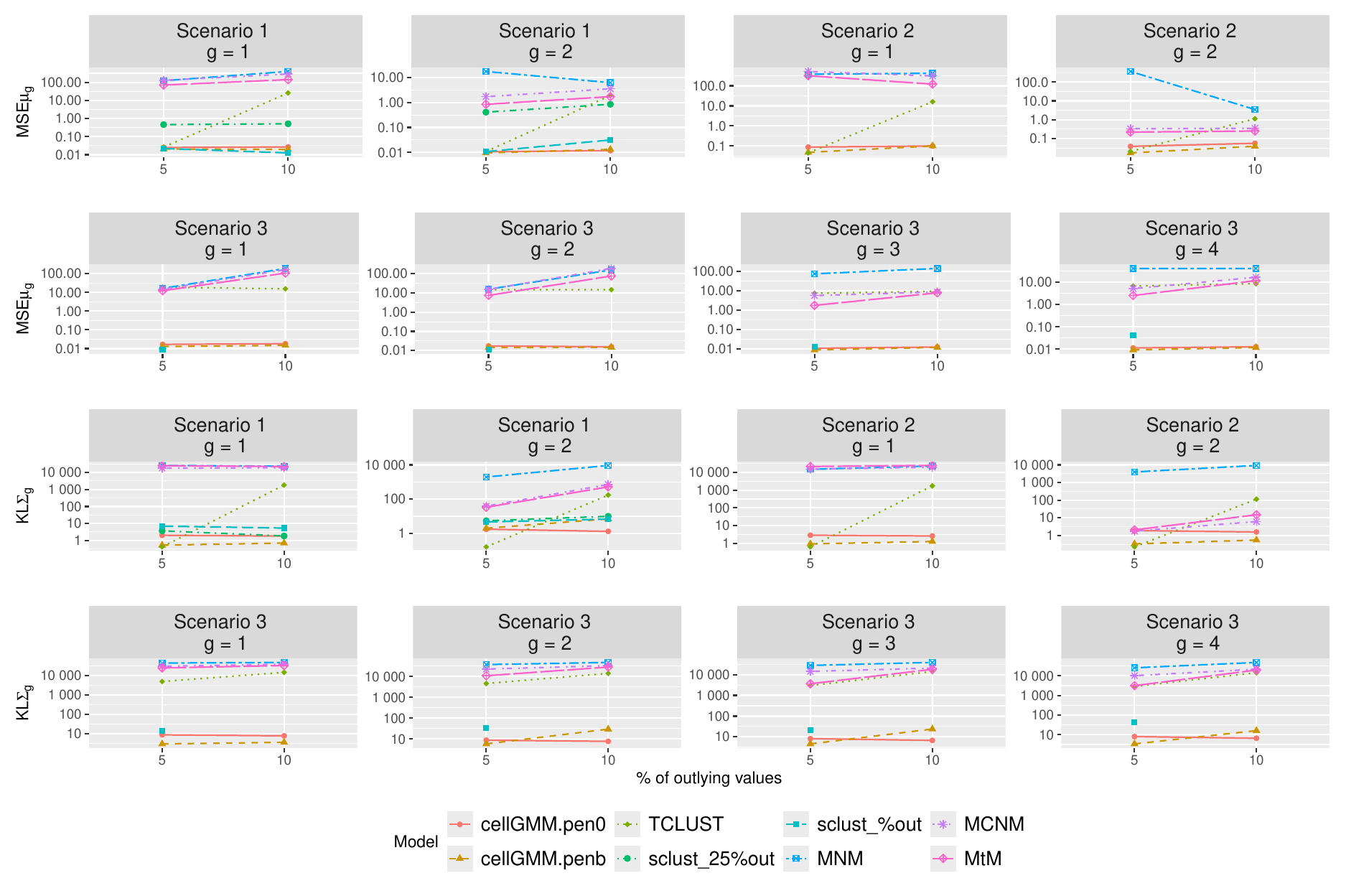}
    \caption{Results of the simulation study with more extreme outliers: Mean Squared Error (MSE) of the component mean vectors and Kullback-Leibler (KL) discrepancy for the component covariance matrices per scenario, percentage of contamination, and model. The values are represented via log-transformation, while the y-axis ticks are labeled using the original scale}
    \label{fig: simstudy_results123_extreme_sup_Parameters}
\end{figure}

\begin{figure}
    \centering
    \includegraphics[width=\linewidth]{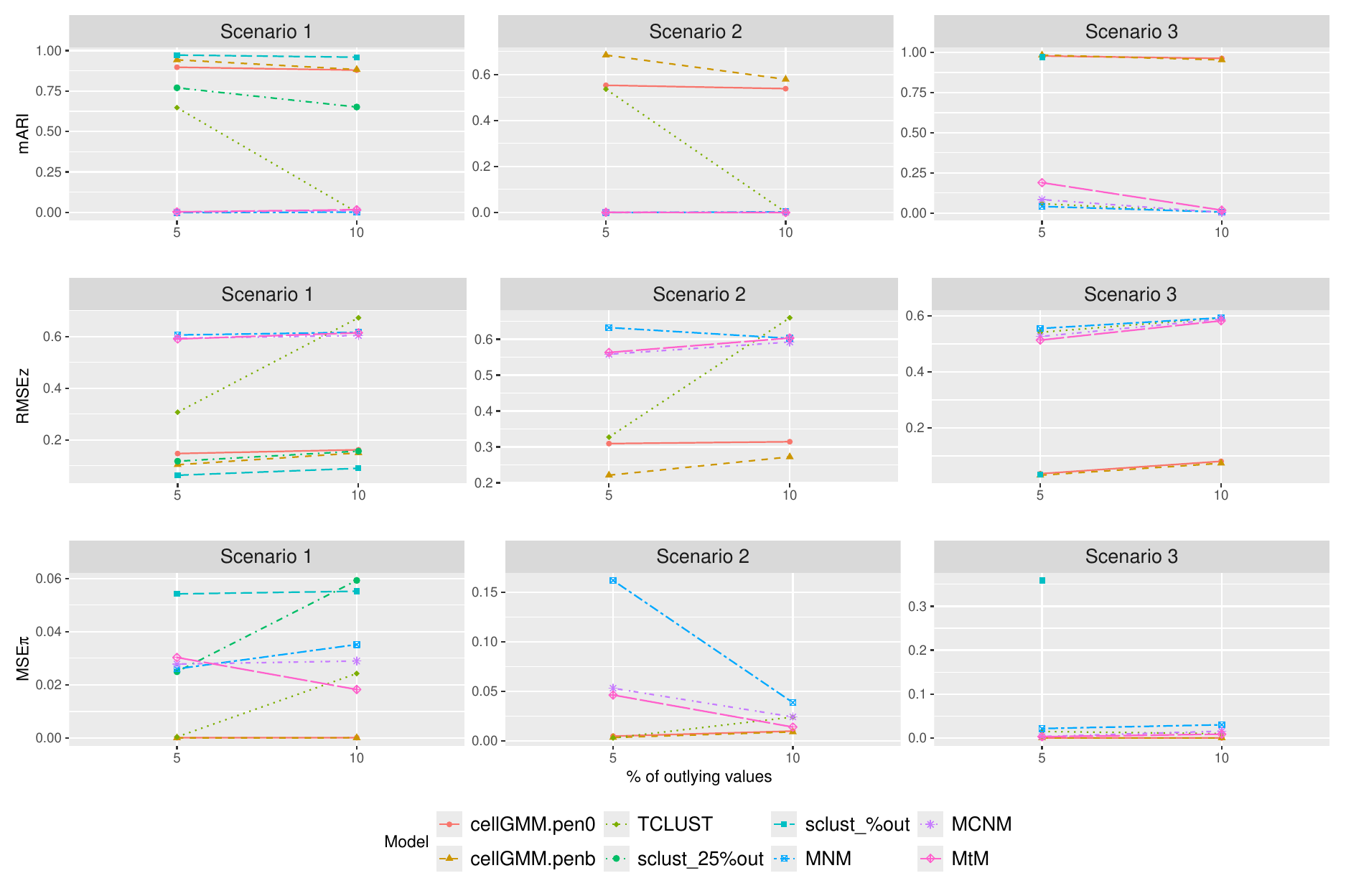}
    \caption{Results of the simulation study with more extreme outliers: mean of the Adjusted Rand Index (mARI), Root Mean Squared Error (RMSE) of the posterior probabilities and Mean Squared Error (MSE) of the prior probabilities per scenario, percentage of contamination, and model}
    \label{fig: simstudy_results123_extreme_sup2}
\end{figure}

The last three scenarios replicate Scenarios 1-3 of the main article, with $5\%$ and $10\%$ of cellwise outliers randomly drawn from a uniform distribution in the wider interval $[-100, 100]$ without any additional constraints. Looking at Figures \ref{fig: simstudy_results123_extreme_sup_MR}, \ref{fig: simstudy_results123_extreme_sup_Parameters}, and \ref{fig: simstudy_results123_extreme_sup2} and Table \ref{tab: sim_W_Imp_More_sup}, it is evident that the model-based clustering methodologies with heavy-tailed distributions and non-robust GMM break down. Generally, we can deduce similar conclusions to the three scenarios reported in the main article, although those illustrated here demonstrate the advantage of cellGMM in handling more extreme outliers.

\begin{table}[hbt!]
\centering
\caption{Outlier detection and imputation with more extreme outliers: percentage of True and False Positive ($\%$TP and $\%$FP), Mean Absolute and Root Mean Squared Error (MAE and RMSE) for comparing the original and imputed data matrices per scenario, percentage of contamination, and model}\label{tab: sim_W_Imp_More_sup}
\resizebox{\linewidth}{!}{
\begin{tabular}{cr|rrrr|rrrr|rrrr}
\hline
 & & \multicolumn{4}{c|}{Scenario 1} & \multicolumn{4}{c|}{Scenario 2} & \multicolumn{4}{c}{Scenario 3} \\
\hline
$\%$ out. & Method & \%TP & \%FP & MAE & RMSE & \%TP & \%FP & MAE & RMSE & \%TP & \%FP & MAE & RMSE \\
\hline 
\multirow{9}{*}{5} 
& cellGMM.pen0 & 98.92 & 21.11 & 0.22 & 0.58 & 98.86 & 21.11 & 0.20 & 0.48 & 99.37 & 21.09 & 0.23 & 0.61 \\ 
& cellGMM.penb & 98.94 & 3.92 & 0.10 & 0.53 & 98.76 & 5.05 & 0.10 & 0.43 & 99.16 & 4.24 & 0.08 & 0.41 \\ 
& TCLUST & 98.94 & 21.11 & - & - & 98.78 & 21.12 & - & - & 57.06 & 23.31 & - & - \\ 
& sclust\_25 & 96.65 & 21.23 & - & - & - & - & - & - & - & - & - & - \\ 
& sclust\_5 & 91.64 & 0.44 & - & - & - & - & - & - & 92.11 & 0.42 & - &  - \\ 
& MCNM & 57.19 & 10.19 & - & - & 75.31 & 13.81 & - & - & 40.01 & 15.03 & - & - \\ 
& M$t$M & 63.83 & 37.28 & - & - & 72.72 & 34.10 & - & - & 79.59 & 40.78 & - & - \\ 
& cellMCD & 98.68 & 13.78 & 0.52 & 1.46 & 98.68 & 2.18 & 0.04 & 0.21 & 98.74 & 3.94 & 0.23 & 1.14 \\ 
& DI & 98.36 & 3.83 & 0.12 & 0.43 & 98.62 & 1.91 & 0.04 & 0.21 & 98.72 & 8.12 & 0.37 & 1.28 \\ 
\hline
\multirow{9}{*}{10} 
& cellGMM.pen0 & 98.58 & 16.82 & 0.23 & 0.66 & 98.92 & 16.79 & 0.20 & 0.49 & 99.21 & 16.75 & 0.23 & 0.68 \\ 
& cellGMM.penb & 98.55 & 6.10 & 0.19 & 0.80 & 98.83 & 6.77 & 0.16 & 0.57 & 98.96 & 5.90 & 0.15 & 0.66 \\ 
& TCLUST & 65.86 & 20.46 & - & - & 65.84 & 20.46 & - & - & 47.54 & 22.50 & - & - \\ 
& sclust\_25 & 95.00 & 17.22 & - & - & - & - & - & - & - & - & - & - \\ 
& sclust\_10 & 93.00 & 0.78 & - & - & - & - & - & - & - & - & - & - \\ 
& MCNM & 70.77 & 22.93 & - & - & 70.96 & 23.04 & - & - & 47.14 & 25.09 & - & - \\ 
& M$t$M & 71.25 & 35.34 & - & - & 71.59 & 33.87 & - & - & 71.00 & 47.06 & - & - \\ 
& cellMCD & 98.72 & 13.41 & 0.53 & 1.47 & 98.82 & 1.96 & 0.06 & 0.24 & 98.61 & 2.84 & 0.18 & 0.93 \\ 
& DI & 98.06 & 1.60 & 0.07 & 0.27 & 98.84 & 1.92 & 0.06 & 0.24 & 98.72 & 2.57 & 0.12 & 0.57 \\ 
\hline
\end{tabular}}
\end{table}

\subsection{Higher level of contamination}\label{subsec: sim_20out_sup}
We present herein the results of the simulation study for Scenarios 1 and 2 with $20\%$ of both less and more extreme outlying values. Tables \ref{tab: sim_20out} and \ref{tab: sim_20out_statW} show that the performance of cellGMM deteriorates compared to scenarios with lower contamination levels, as expected. Indeed, $20\%$ of outlying cells represents a very high level of contamination in the cellwise framework, undoing the potential of the proposal when more observations, variables, and clusters are considered, as in Scenario 3.

This analysis empirically assesses the robustness of cellGMM, in addition to the scenarios reported in Section \ref{subsec: sim_more_sup}. The breakdown properties of cellGMM may not be directly derived from those of cellMCD in the single-population framework. This is also the case for TCLUST, which extends MCD to heterogeneous populations, where the notion of breakdown points depend on non-trivial \say{cluster separation} assumptions \citep[see][for further details]{RGGM:2013}. An in-depth theoretical discussion of these properties for cellGMM will be provided in future work. 

\begin{table}
\centering
\caption{Classification and parameter estimation results for the simulation study with $20\%$ of contamination per scenario, outlying value generation, and model}\label{tab: sim_20out}
\resizebox{0.75\linewidth}{!}{
\begin{tabular}{cl|cccccccrr}
\hline
Outliers & Model & \# samp. & mARI & mMR & RMSE$_{z}$ & MSE$_{\pi}$ & MSE$_{\mean_{1}}$ & MSE$_{\mean_{2}}$ & KL$_{\sigu_{1}}$ & KL$_{\sigu_{2}}$ \\
\hline
& & \multicolumn{9}{c}{Scenario 1}\\
\hline
\multirow{7}{*}{Less extreme} & cellGMM.pen0 & 100 & 0.85 & 0.04 & 0.19 & 0.00 & 0.09 & 0.02 & 0.96 & 0.72 \\ 
& cellGMM.penb & 96 & 0.71 & 0.08 & 0.25 & 0.01 & 0.57 & 0.05 & 1.91 & 4.66 \\ 
& TCLUST & 100 & 0.08 & 0.38 & 0.60 & 0.06 & 3.90 & 1.49 & 195.39 & 152.50 \\ 
& sclust\_25 & 63 & 0.73 & 0.07 & 0.25 & 0.05 & 0.66 & 0.61 & 39.15 & 26.31 \\ 
& sclust\_20 & 66 & 0.76 & 0.06 & 0.24 & 0.06 & 0.44 & 0.37 & 46.65 & 27.19 \\ 
& MNM & 100 & 0.22 & 0.27 & 0.51 & 0.08 & 1.32 & 0.28 & 410.90 & 181.13 \\ 
& MCNM & 100 & 0.19 & 0.28 & 0.51 & 0.03 & 0.97 & 0.05 & 133.85 & 17.65 \\ 
& MtM & 100 & 0.18 & 0.29 & 0.52 & 0.04 & 1.03 & 0.08 & 184.67 & 45.86 \\ 
\cmidrule{2-11}
\multirow{8}{*}{More extreme} & cellGMM.pen0 & 98 & 0.90 & 0.03 & 0.14 & 0.00 & 5.89 & 0.02 & 1.45 & 0.87 \\ 
& cellGMM.penb & 24 & 0.88 & 0.03 & 0.15 & 0.00 & 0.03 & 0.01 & 0.57 & 3.78 \\ 
& TCLUST & 100 & 0.00 & 0.50 & 0.70 & 0.08 & 15.63 & 23.81 & 5975.82 & 9165.80 \\ 
& sclust\_25 & 1 & 0.92 & 0.02 & 0.14 & 0.00 & 0.04 & 0.09 & 4.63 & 11.00 \\ 
& sclust\_20 & 0 & - & - & - & - & - & - & - & - \\ 
& MNM & 100 & 0.00 & 0.41 & 0.63 & 0.05 & 572.85 & 11.86 & 20993.78 & 28767.40 \\ 
& MCNM & 100 & 0.00 & 0.43 & 0.65 & 0.02 & 37.96 & 3.57 & 26589.25 & 2464.01 \\ 
& MtM & 100 & 0.00 & 0.46 & 0.67 & 0.04 & 16.41 & 7.29 & 15076.11 & 10341.98 \\ 
\hline
& & \multicolumn{9}{c}{Scenario 2}\\
\hline
\multirow{7}{*}{Less extreme} & cellGMM.pen0 & 100 & 0.44 & 0.17 & 0.35 & 0.01 & 0.09 & 0.07 & 1.99 & 0.81 \\ 
& cellGMM.penb & 100 & 0.43 & 0.18 & 0.34 & 0.01 & 0.18 & 0.05 & 1.76 & 0.62 \\ 
& TCLUST & 100 & 0.00 & 0.49 & 0.67 & 0.07 & 1.07 & 0.64 & 96.62 & 199.22 \\ 
& sclust\_25 & 70 & 0.28 & 0.24 & 0.41 & 0.03 & 0.18 & 0.12 & 7.92 & 5.41 \\ 
& sclust\_20 & 51 & 0.31 & 0.22 & 0.40 & 0.04 & 0.20 & 0.09 & 9.84 & 6.87 \\ 
& MNM & 100 & 0.00 & 0.50 & 0.68 & 0.04 & 1.81 & 1.03 & 103.24 & 436.66 \\ 
& MCNM & 100 & 0.01 & 0.45 & 0.64 & 0.03 & 1.10 & 0.50 & 238.58 & 112.21 \\ 
& MtM & 100 & 0.01 & 0.44 & 0.63 & 0.02 & 0.89 & 0.33 & 262.47 & 105.37 \\ 
\cmidrule{2-11}
\multirow{8}{*}{More extreme} & cellGMM.pen0 & 99 & 0.54 & 0.13 & 0.30 & 0.01 & 1.42 & 0.06 & 2.90 & 0.83 \\ 
& cellGMM.penb & 23 & 0.65 & 0.09 & 0.23 & 0.00 & 0.03 & 0.01 & 0.97 & 0.34 \\ 
& TCLUST & 100 & 0.00 & 0.49 & 0.69 & 0.07 & 10.36 & 20.71 & 5438.39 & 8969.34 \\ 
& sclust\_25 & 0 & - & - & - & - & - & - & - & - \\ 
& sclust\_20 & 0 & - & - & - & - & - & - & - & - \\ 
& MNM & 0 & - & - & - & - & - & - & - & - \\ 
& MCNM & 100 & 0.00 & 0.43 & 0.64 & 0.02 & 84.52 & 1.94 & 24365.74 & 1997.77 \\ 
& MtM & 100 & 0.00 & 0.46 & 0.66 & 0.04 & 10.91 & 5.04 & 15968.84 & 8942.05 \\ 
\hline
\end{tabular}%
}
\end{table}

\begin{table}
\centering
\caption{Outlier detection and imputation with $20\%$ of contamination per scenario, outlying value generation, and model}\label{tab: sim_20out_statW}
\resizebox{0.8\linewidth}{!}{
\begin{tabular}{cr|cccc|cccc}
\hline
& & \multicolumn{4}{c|}{Scenario 1} & \multicolumn{4}{c}{Scenario 2} \\
\hline
Outliers & Method & \%TP & \%FP & MAE & RMSE & \%TP & \%FP & MAE & RMSE \\
\hline
\multirow{9}{*}{Less extreme} & cellGMM.pen0 & 91.36 & 8.41 & 0.26 & 0.89 & 92.47 & 8.13 & 0.19 & 0.56 \\ 
& cellGMM.penb & 88.01 & 5.46 & 0.31 & 1.09 & 91.33 & 4.64 & 0.18 & 0.56 \\ 
& TCLUST & 46.24 & 19.69 & - & - & 47.31 & 19.42 & - & - \\ 
& sclust\_25 & 65.47 & 14.88 & - & - & 83.88 & 10.28 & - & - \\ 
& sclust\_20 & 61.53 & 9.62 & - & - & 79.33 & 5.17 & - & - \\ 
& MCNM & 74.44 & 39.76 & - & - & 48.28 & 23.83 & - & - \\ 
& MtM & 77.70 & 43.84 & - & - & 63.50 & 36.41 & - & - \\ 
& cellMCD & 87.38 & 9.57 & 0.51 & 1.43 & 90.53 & 1.59 & 0.12 & 0.38 \\ 
& DI & 83.94 & 2.22 & 0.19 & 0.67 & 90.77 & 1.96 & 0.12 & 0.39 \\ 
\cmidrule{2-10}
\multirow{9}{*}{More extreme} & cellGMM.pen0 & 98.52 & 6.62 & 0.18 & 0.68 & 98.67 & 6.58 & 0.15 & 0.42 \\ 
& cellGMM.penb & 98.27 & 3.52 & 0.18 & 0.73 & 98.50 & 3.38 & 0.14 & 0.48 \\ 
& TCLUST & 48.85 & 19.04 & - & - & 49.02 & 18.99 & - & - \\ 
& sclust\_25 & 95.50 & 7.38 & - & - & - & - & - & - \\ 
& sclust\_20 & - & - & - & - & - & - & - & - \\ 
& MCNM & 69.36 & 34.69 & - & - & 67.94 & 33.67 & - & - \\ 
& MtM & 72.03 & 41.24 & - & - & 71.86 & 39.79 & - & - \\ 
& cellMCD & 98.39 & 8.70 & 0.43 & 1.26 & 98.75 & 1.60 & 0.10 & 0.29 \\ 
& DI & 98.06 & 1.39 & 0.12 & 0.39 & 98.70 & 1.96 & 0.11 & 0.33 \\  
\hline
\end{tabular}%
}
\end{table}

\subsection{Structural outlier generation}\label{subsec: sim_structural}
In the simulation studies presented so far, we have considered random contamination affecting a certain percentage of the $n \times p$ cells in a data matrix. In this section, we also explore the performance of cellGMM in the presence of structural outliers. We focus on Scenarios 1 and 2 previously illustrated, where the prior probabilities are set to $\boldsymbol{\pi} = [0.7, 0.3]$. We replace $5\%, 10\%$, and $20\%$ of the cells in the submatrix corresponding to the observations generated from the first component with contaminated values, extending the generation scheme of \citet{RR:2023} to the heterogeneous population framework. Therefore, for each column of the data matrix, we randomly draw $5\%, 10\%$ or $20\%$ of the indices to contaminate. The variables corrupted for each observation of the first component are stored into the set $K = \{j_{1}, \ldots, j_{k} \}$, where $k = \lvert K \rvert$. It is worth noting that the contamination level is the same across variables but can differ for each unit. The $k$-dimensional vector of contaminated cells per observation is obtained as $ - \gamma \sqrt{k} \frac{\boldsymbol{\nu}_{K}}{\text{MD}\big(\boldsymbol{\nu}_{K}, \mean_{1[K]}, \boldsymbol{\Sigma}_{1[K]} \big)}$, where $\gamma$ quantifies the distance of the outlying cells from the center of the first component. Here, $\mean_{1[K]}$ and $\boldsymbol{\Sigma}_{1[K]}$ denote the mean vector and the covariance matrix of the first component, respectively, restricted to the variables in $K$, and $\boldsymbol{\nu}_{K}$ is the normed eigenvector of $\boldsymbol{\Sigma}_{1[K]}$ corresponding to its largest eigenvalue. We vary $\gamma$ in $\{1, 5, 10\}$. This generation scheme ensures that the resulting contaminated cells are structurally outlying in the corresponding subspace.

\begin{table}
\centering
\caption{Results of the simulation study in the presence of structural outlying cells per percentage of contamination and $\gamma$ level, for Scenario 1}\label{tab: sim_class_RR_well}
\resizebox{0.9\textwidth}{!}{
\begin{tabular}{cll|ccccrrcc}
\hline
$\%$ out. & $\gamma$ & Model & \# samp. & mMR & MSE$_{\mean_{1}}$ & MSE$_{\mean_{2}}$ & KL$_{\sigu_{1}}$ & KL$_{\sigu_{2}}$ & $\%$TP & $\%$FP \\ 
\hline
\multirow{6}{*}{5} & \multirow{2}{*}{1} & cellGMM.pen0 & 100 & 0.03 & 0.02 & 0.03 & 1.83 & 2.14 & 66.90 & 22.79 \\ 
& & cellGMM.penb & 100 & 0.02 & 0.01 & 0.03 & 0.27 & 1.74 & 52.34 & 4.85 \\ 
& \multirow{2}{*}{5} & cellGMM.pen0 & 100 & 0.03 & 0.01 & 0.02 & 1.74 & 1.84  & 99.26 & 21.09 \\
& & cellGMM.penb & 100 & 0.01 & 0.01 & 0.02 & 0.29 & 2.32 & 100.00 & 4.21 \\ 
& \multirow{2}{*}{10} & cellGMM.pen0 & 100 & 0.03 & 0.01 & 0.02 & 1.76 & 1.87 & 99.84 & 21.06 \\
& & cellGMM.penb & 100 & 0.02 & 0.01 & 0.02 & 0.30 & 3.00 & 100.00 & 4.73 \\ 
\cmidrule{2-11}
\multirow{6}{*}{10} & \multirow{2}{*}{1} & cellGMM.pen0 & 100 & 0.04 & 0.02 & 0.09 & 1.49 & 2.09 & 59.98 & 21.11 \\
& & cellGMM.penb & 100 & 0.06 & 0.02 & 0.10 & 0.33 & 6.56 & 48.83 & 8.10 \\
& \multirow{2}{*}{5} & cellGMM.pen0 & 100 & 0.03 & 0.02 & 0.03 & 1.32 & 1.40 & 98.38 & 16.85 \\
& & cellGMM.penb & 100 & 0.02 & 0.01 & 0.02 & 0.37 & 5.50 & 99.98 & 5.85 \\
& \multirow{2}{*}{10} & cellGMM.pen0 & 100 & 0.03 & 0.01 & 0.02 & 1.35 & 1.44 & 99.61 & 16.71 \\
& & cellGMM.penb & 100 & 0.03 & 0.01 & 0.03 & 0.38 & 6.60 & 100.00 & 6.64 \\
\cmidrule{2-11}
\multirow{6}{*}{20} & \multirow{2}{*}{1} & cellGMM.pen0 & 100 & 0.06 & 0.07 & 0.87 & 1.90 & 2.92 & 41.51 & 20.87 \\ 
& & cellGMM.penb & 100 & 0.16 & 0.09 & 2.50 & 2.80 & 27.29 & 24.99 & 15.76 \\
& \multirow{2}{*}{5} & cellGMM.pen0 & 100 & 0.04 & 0.04 & 0.26 & 0.88 & 0.63 & 96.52 & 7.12 \\
& & cellGMM.penb & 100 & 0.03 & 0.08 & 0.46 & 2.49 & 5.12 & 99.11 & 3.54 \\ 
& \multirow{2}{*}{10} & cellGMM.pen0 & 100 & 0.02 & 0.04 & 0.22 & 1.25 & 1.23 & 99.45 & 6.39 \\
& & cellGMM.penb & 97 & 0.03 & 0.05 & 0.45 & 0.89 & 2.72 & 100.00 & 3.23 \\
\hline
\end{tabular}%
}
\end{table}

\begin{table}
\centering
\caption{Results of the simulation study in the presence of structural outlying cells per percentage of contamination and $\gamma$ level, for Scenario 2}\label{tab: sim_class_RR_close}
\resizebox{0.9\textwidth}{!}{
\begin{tabular}{cll|ccccrrcc}
\hline
$\%$ out. & $\gamma$ & Model & \# samp. & mMR & MSE$_{\mean_{1}}$ & MSE$_{\mean_{2}}$ & KL$_{\sigu_{1}}$ & KL$_{\sigu_{2}}$ & $\%$TP & $\%$FP \\ 
\hline
\multirow{6}{*}{5} & \multirow{2}{*}{1} & cellGMM.pen0 & 100 & 0.14 & 0.08 & 0.08 & 2.15 & 2.87 & 66.52 & 22.81 \\ 
& & cellGMM.penb & 100 & 0.11 & 0.06 & 0.09 & 0.45 & 1.05 & 53.42 & 6.07 \\
& \multirow{2}{*}{5} & cellGMM.pen0 & 100 & 0.13 & 0.05 & 0.07 & 2.01 & 2.41 & 99.96 & 21.05 \\
& & cellGMM.penb & 100 & 0.09 & 0.03 & 0.07 & 0.44 & 0.87 & 100.00 & 5.28 \\
& \multirow{2}{*}{10} & cellGMM.pen0 & 100 & 0.13 & 0.04 & 0.07 & 2.05 & 2.37 & 99.96 & 21.05 \\
& & cellGMM.penb & 100 & 0.09 & 0.02 & 0.06 & 0.43 & 0.85 & 100.00 & 5.55 \\  
\cmidrule{2-11}
\multirow{6}{*}{10} & \multirow{2}{*}{1} & cellGMM.pen0 & 100 & 0.19 & 0.16 & 0.21 & 1.94 & 2.84 & 57.22 & 21.42 \\ 
& & cellGMM.penb & 100 & 0.21 & 0.18 & 0.30 & 0.77 & 1.87 & 47.02 & 9.06 \\ 
& \multirow{2}{*}{5} & cellGMM.pen0 & 100 & 0.13 & 0.04 & 0.06 & 1.62 & 2.01 & 99.97 & 16.67 \\ 
& & cellGMM.penb & 100 & 0.12 & 0.03 & 0.08 & 0.56 & 1.04 & 100.00 & 6.68 \\ 
& \multirow{2}{*}{10} & cellGMM.pen0 & 100 & 0.13 & 0.05 & 0.77 & 1.61 & 2.00 & 99.90 & 16.68 \\ 
& & cellGMM.penb & 100 & 0.13 & 0.03 & 0.09 & 0.59 & 1.12 & 100.00 & 7.31 \\
\cmidrule{2-11}
\multirow{6}{*}{20} & \multirow{2}{*}{1} & cellGMM.pen0 & 100 & 0.29 & 0.63 & 0.47 & 2.34 & 2.50 & 37.74 & 21.81 \\
& & cellGMM.penb & 100 & 0.34 & 0.60 & 0.86 & 2.58 & 3.02 & 31.24 & 12.81 \\
& \multirow{2}{*}{5} & cellGMM.pen0 & 100 & 0.09 & 0.02 & 0.47 & 0.78 & 0.87 & 99.73 & 6.32 \\
& & cellGMM.penb & 100 & 0.10 & 0.03 & 0.09 & 0.54 & 0.60 & 100.00 & 3.50 \\ 
& \multirow{2}{*}{10} & cellGMM.pen0 & 100 & 0.15 & 0.11 & 19.58 & 1.01 & 1.73 & 99.80 & 6.30 \\ 
& & cellGMM.penb & 100 & 0.10 & 0.04 & 0.09 & 0.58 & 0.82 & 100.00 & 3.56 \\
\hline
\end{tabular}%
}
\end{table}

\begin{table}
\centering
\caption{Percentage of contaminated observations that lie within the $99$th percentile ellipsoid of one component averaged over $100$ samples (\# out. in), and level of contamination estimated by cellGMM.penb (Est. cont.) averaged over the number of samples on which the model can be properly computed per scenario, percentage of contamination, and $\gamma$ level}\label{tab: sim_class_RR_interpr}
\resizebox{0.9\textwidth}{!}{
\begin{tabular}{lc|ccc|ccc|ccc}
\hline
& $\%$ out. & \multicolumn{3}{c}{$5\%$} & \multicolumn{3}{c}{$10\%$} & \multicolumn{3}{c}{$20\%$} \\
& $\gamma $ & 1 & 5 & 10 & 1 & 5 & 10 & 1 & 5 & 10 \\
\hline
\multirow{2}{*}{Scenario 1} & \# out. in &  56.92 & 0.00 & 0.00 & 53.10 & 0.00 & 0.00 & 46.23 & 0.00 & 0.00 \\
& Est. cont. & 7.23 & 9.00 & 9.49 & 12.18 & 15.27 & 15.98 & 17.60 & 22.66 & 22.58 \\
\hline
\multirow{2}{*}{Scenario 2} & \# out. in & 58.58 & 0.00 & 0.00 & 55.97 & 0.00 & 0.00 & 49.22 & 0.00 & 0.00 \\ 
& Est. cont. & 8.44 & 10.02 & 10.27 & 12.86 & 16.02 & 16.58 & 16.50 & 22.80 & 22.85 \\ 
\hline
\end{tabular}%
}
\end{table}

The results of this simulation study are reported in Tables \ref{tab: sim_class_RR_well} and \ref{tab: sim_class_RR_close}. As $\gamma$ increases, the classification performance of cellGMM.penb likely improves. Indeed, when $\gamma = 1$, several contaminated observations lie within the $99$th percentile ellipsoid of one component, as reported in Table \ref{tab: sim_class_RR_interpr}, making them hard to detect (see $\%$TP in Tables \ref{tab: sim_class_RR_well} and \ref{tab: sim_class_RR_close}). As discussed in the previous section, a $20\%$ of contamination is very high and may limit the potential of the proposed methodology. This is particularly evident in Scenario 2 with close components, especially when $\gamma = 1$ where the level of contamination is underestimated by cellGMM.penb. In all other cases, the results presented here provide evidence of the potential of cellGMM in detecting structural outlying values. 

\subsection{Computational aspects}\label{subsec: sim_computingtime}
\begin{table}
\centering
\caption{Computation time in seconds averaged over $10$ samples per scenario with different percentages of contamination and models. Dashes represent scenarios where models cannot be implemented or where implementation is not necessary}\label{tab: sim_comp_time}
\resizebox{\textwidth}{!}{
\begin{tabular}{cl|ccc|ccc|ccc}
\hline
\multirow{2}{*}{Model} &  & \multicolumn{3}{c|}{0\% out.} & \multicolumn{3}{c|}{5\% out.} & \multicolumn{3}{c}{10\% out.} \\
& & Scenario 1 & Scenario 2 & Scenario 3 & Scenario 1 & Scenario 2 & Scenario 3 & Scenario 1 & Scenario 2 & Scenario 3 \\
\hline
cellGMM.pen0 & & 11.82 & 19.81 & 339.33 & 13.80 & 20.13 & 372.76 & 15.19 & 24.42 & 410.48 \\ 
cellGMM.penb & & 1.07 & 4.36 & 36.89 & 2.21 & 9.05 & 59.91 & 4.41 & 16.22 & 126.34 \\ 
TCLUST & & 0.03 & 0.03 & 0.56 & 0.02 & 0.03 & 0.44 & 0.02 & 0.03 & 0.48 \\ 
sclust\_25 & & 0.43 & 0.51 & - & 0.35 & 0.38 & - & 0.65 & 1.11 & - \\ 
sclust\_5 & & - & - & - & 0.44 & 1.00 & 2.14 & - & - & - \\ 
sclust\_10 & & - & - & - & - & - & - & 0.32 & 1.46 & 2.35 \\ 
MNM & & 0.01 & 0.29 & 0.20 & 0.10 & 0.06 & 1.03 & 0.11 & 0.09 & 1.81 \\ 
MCNM & & 2.06 & 2.55 & 16.88 & 0.38 & 1.15 & 5.21 & 0.48 & 0.99 & 7.44 \\ 
M$t$M & & 2.05 & 2.06 & 10.63 & 0.35 & 1.37 & 9.31 & 0.42 & 1.94 & 14.41 \\ 
\hline
\end{tabular}%
} 
\end{table}

In this section, we provide indicative information on the computation time of the cellGMM algorithm compared to the other methodologies implemented in the simulation study for the three scenarios illustrated in the main article. The computation times reported in Table \ref{tab: sim_comp_time} are obtained using a MacBook Pro with a 6-Core Intel Core i7 (2.6 GHz) processor and 16GB DDR4 RAM (2667 MHz). The runtime is averaged over $10$ samples per scenario and percentage of contamination, for which all models have been properly implemented. 

Table \ref{tab: sim_comp_time} shows that our proposal is more time-consuming compared to the other methodologies. This difference increases significantly with the dimensionality of the data. Nonetheless, a higher computational cost often corresponds to better results in terms of classification, parameter estimation, and outlier detection, as illustrated in Section 3 of the main article. The greatest runtime is required for the update of $\vec{W}$. Algorithmic optimization, which may include the use of a compiled programming language, is part of future developments. It is worth noting that if the user has prior information to set the penalty term in advance, cellGMM.penb can be run directly, thus speeding up the computation time.

The computational complexity of cellGMM is $\mathcal{O}(n\log(n)p + nGp^2)$. Specifically, the C-step costs $\mathcal{O}(n\log(n))$ for each variable to sort $\{\Delta_{ij}\}_{i = 1}^{n}$ \citep{K:1998}. The computations in the E- and M-steps, which are also involved in the C-step, include the SWEEP operator, whose complexity $\mathcal{O}(nGp^2)$ \citep{G:1979} dominates both the covariance matrix inversion and the eigenvalue-ratio constraint. The latter have complexity $\mathcal{O}(Gp^3)$ \citep{GLS:1993} and $\mathcal{O}(G^2p^2)$, respectively, and the dominance holds since $n > G$ and $n > p$.

\section{Real data examples: additional results} \label{sec: app_sup}
In this section, we provide additional results on the application of cellGMM to the real data sets illustrated in the main article.

\subsection{Homogenized Meat Data Set}\label{subsec: data1_sub}
\begin{figure}[t]
\centering
\includegraphics[width=0.6\linewidth]{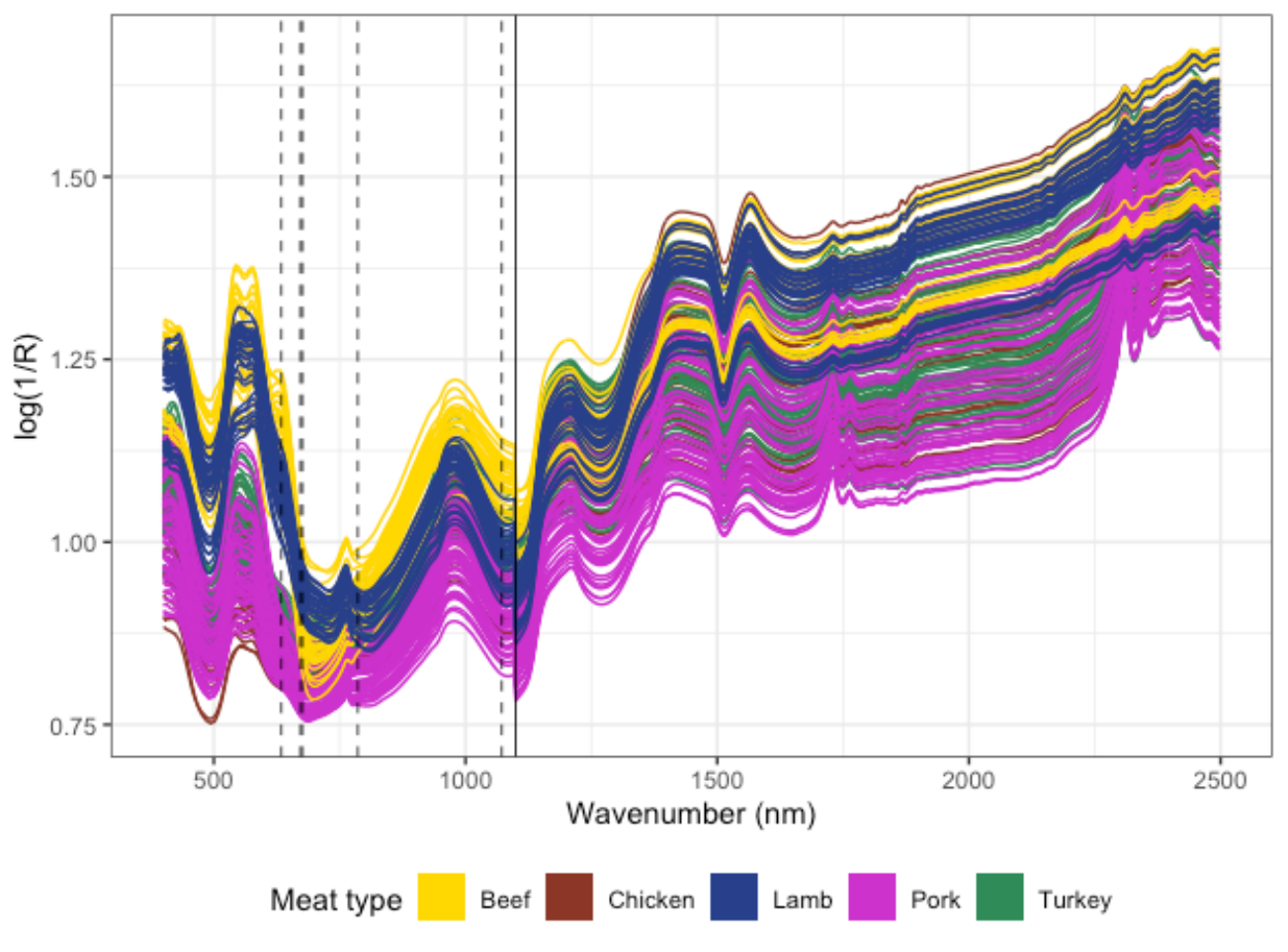}
\caption{Homogenized Meat data set: visible and near infrared spectra of five homogenized meat types (dashed lines: selected wavelengths; solid line: discontinuity).}
\label{fig: meat_spectra}
\vspace{0.5cm}
\end{figure}

Near-infrared reflectance spectroscopy is a method often used for collecting data in food authenticity studies \citep[][among others]{D:1996} since it is able to discern the properties and, therefore, the nature of food samples. For each of the $231$ homogenized meat samples  $1050$ reflectance measurements are collected with wavelengths between $400$ nm and $2500$ nm at $2$ nm intervals (Figure \ref{fig: meat_spectra}). Due to the number of measurements, one of the first tasks to handle is to reduce the dimensionality of the data. Methods for variable selection were applied on this data set by \citet{MDR:2010} and, in a robust framework, by \citet{CGM:2021}. We refer to the latter by considering five relevant wavelengths which span the protein spectral region. 

\begin{table}[t]
\centering
\caption{Adjusted Rand Index comparing the theoretical and the estimated classification of meat samples in two classes per model, and percentage of contamination and missing values, i.e. $(a\%, b\%)$} \label{tab: meat_ARI_G2_sub}
\resizebox{0.6\textwidth}{!}{
\begin{tabular}{lcccccc}
\hline
& cellGMM & TCLUST & sclust & MNM & MCNM & M$t$M \\
\hline
$(0\%, 0\%)$ & 0.95 & 1.00 & 1.00 & 1.00 & 1.00 & 1.00 \\
$(3\%, 0\%)$ & 0.95 & 0.86 & 0.85 & 0.81 & 0.80 & 0.75 \\
$(3\%, 2\%)$ & 0.95 & - & - & 0.80 & 0.78 & 0.75 \\
$(10\%, 0\%)$ & 0.93 & 0.45 & 0.73 & 0.48 & 0.25 & 0.36 \\
\hline
\end{tabular}
}
\end{table}

\begin{table}[t]
\centering
\caption{Average cellGMM classification results of the meat species with $G = 3, 4, 5$ (3\% of contamination)}\label{tab: class_data1_sub}
\resizebox{\textwidth}{!}{
\begin{tabular}{cccc|ccccc|cccccc}
\hline
& 1 & 2 & 3 & & 1 & 2 & 3 & 4 & & 1 & 2 & 3 & 4 & 5 \\ 
\hline
Beef & 0.88 & 0.03 & 0.09 & Beef & 0.97 & 0.00 & 0.00 & 0.03 & Beef & 0.97 & 0.03 & 0.00 & 0.00 & 0.00 \\ 
Lamb & 0.06 & 0.91 & 0.03 & Lamb & 0.06 & 0.94 & 0.00 & 0.00 & Lamb & 0.12 & 0.88 & 0.00 & 0.00 & 0.00 \\ 
\multirow{3}{*}{White meats} & \multirow{3}{*}{0.01} & \multirow{3}{*}{0.00} & \multirow{3}{*}{0.99} & Pork & 0.00 & 0.00 & 0.82 & 0.18 & Pork & 0.13 & 0.00 & 0.78 & 0.04 & 0.05 \\ 
& & & & \multirow{2}{*}{Poultry} & \multirow{2}{*}{0.01} & \multirow{2}{*}{0.00} & \multirow{2}{*}{0.15} & \multirow{2}{*}{0.84} & Chicken & 0.21 & 0.00 & 0.02 & 0.75 & 0.02 \\ 
& & & & & & & & & Turkey & 0.05 & 0.00 & 0.00 & 0.85 & 0.10 \\
\hline
\end{tabular}
}
\end{table}

As for the simulation study, we report herein the ARI between the theoretical and the estimated clustering structure for cellGMM and the other competitors. From Table \ref{tab: meat_ARI_G2_sub}, it is evident that as the level of contamination increases, the performance of the rowwise and non-robust model-based clustering competitors dramatically decreases, whereas that of cellGMM diminishes less, remaining good. Additionally, Table \ref{tab: class_data1_sub} depicts the average classification results for cellGMM (i.e. cellGMM.penb) when $G = 3$, i.e., pork, chicken and turkey are merged together representing white meats, $G = 4$, i.e., chicken and turkey meats are considered in the same group representing poultry, and $G = 5$, all in the case of $3\%$ of contamination. Although $G$ is also set to $5$, most of the chicken and turkey meats are grouped together, while the estimated cluster $5$ contains few observations that are residual from the other clusters. The structure in four groups better distinguishes between the meat species, even if there is an \say{overlapping} between pork and poultry meats, while the classification in three clusters discriminates well among beef, lamb and white meats.

\subsection{Top Gear Data Set}\label{subsec: data3_sub}
\begin{figure}
\centering
\includegraphics[width=\linewidth]{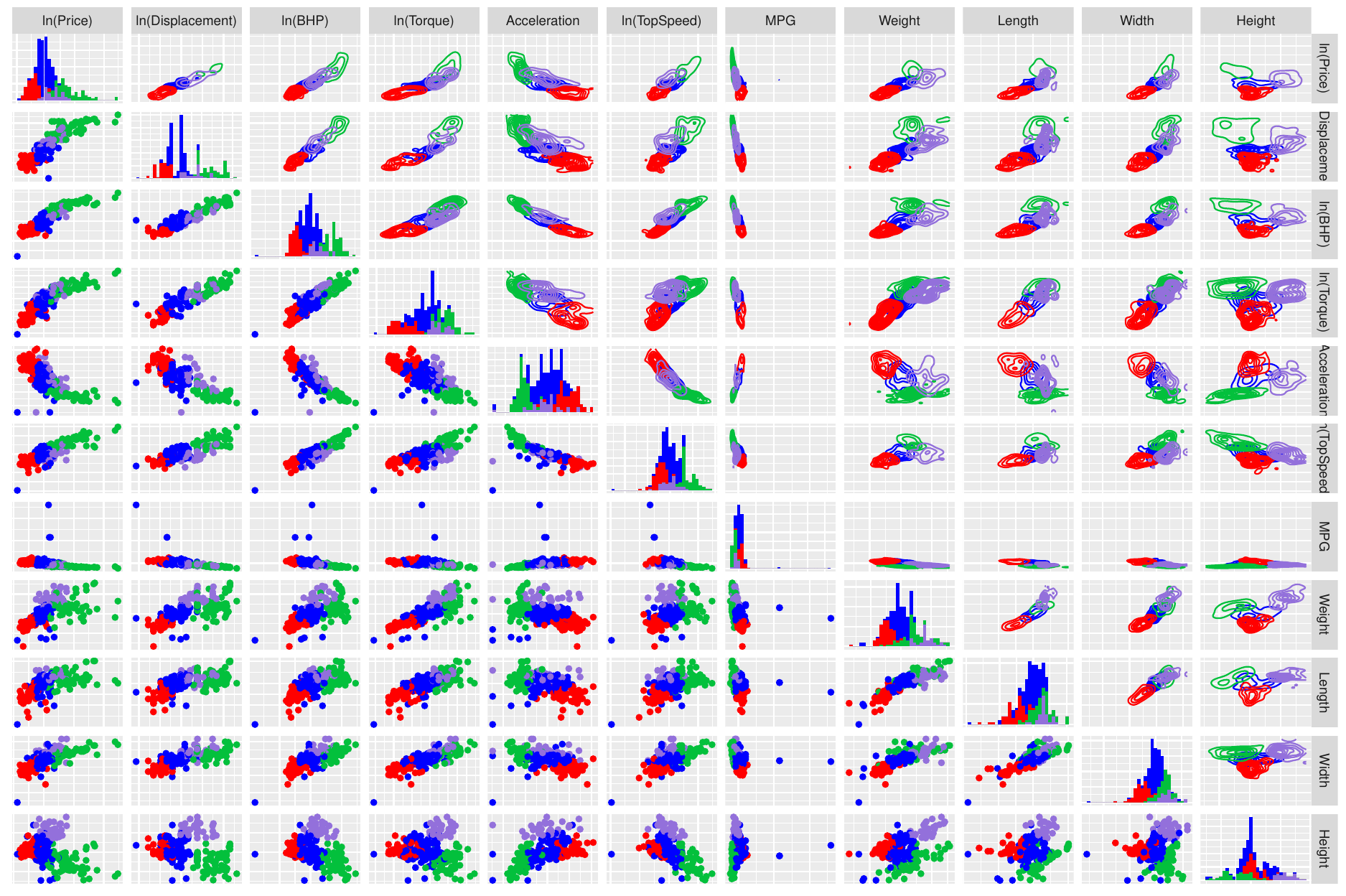}
\caption{Pair plots of the Top Gear data set with the estimated classification in four groups (Cluster 1: blue; Cluster 2: red; Cluster 3: green; Cluster 4: purple)}
\label{fig: Top Gear_sup}
\end{figure}

The scatterplot for pairs of variables referring to the original, potentially contaminated data with the estimated classification in four groups is reported in Figure \ref{fig: Top Gear_sup}, while the full list of four clusters' classification is detailed below.
\begin{enumerate}
    \item[Cluster 1.] Alfa Romeo Giulietta, Audi A3, Audi A4, Audi A4 Allroad, Audi A5, Audi A5 Sportback, Audi A6 Avant, Audi A6 Saloon, Audi Q3, Audi Q5, Audi TT Coupé, Audi TT Roadster, BMW 1 Series, BMW 1 Series Convertible, BMW 1 Series Coupé, BMW 3 Series, BMW 3 Series Convertible, BMW 4 Series Coupé, BMW i3, BMW X1, BMW Z4, Caterham CSR, Caterham Super 7, Chevrolet Cruze, Chevrolet Orlando, Chevrolet Volt, Chrysler Delta, Citroen C4, Citroen C4 Picasso, Citroen C5, Citroen DS3, Citroen DS4, Citroen DS5, Dacia Duster, Fiat Bravo, Fiat Doblò, Fiat Punto Evo, Ford C-Max, Ford Focus, Ford Focus Estate, Ford Focus ST, Ford Kuga, Ford S-MAX, Honda Accord, Honda Civic, Honda CR-V, Honda CR-Z, Hyundai i40, Hyundai ix35, Hyundai Veloster, Jaguar XF, Jaguar XF Sportbrake, Jeep Compass, Kia Cee'd, Kia Optima, Kia Soul, Kia Sportage, Land Rover Freelander 2, Land Rover Range Rover Evoque, Lexus CT 200h, Lexus GS, Lotus Elise, Mazda CX-5, Mazda Mazda3, Mazda MX-5, Mercedes-Benz A-Class, Mercedes-Benz B-Class, Mercedes-Benz C-Class, Mercedes-Benz CLS Shooting Brake, Mercedes-Benz E-Class, Mercedes-Benz E-Class Coupé, Mercedes-Benz SLK, Mini Countryman, Mini Cooper, Mini John Cooper Works, Mini Roadster, Mitsubishi ASX, Mitsubishi Outlander, Morgan 3 Wheeler, Nissan Juke, Nissan Leaf, Nissan Qashqai, Nissan X-Trail, Peugeot 207 CC, Peugeot 3008, Peugeot 308, Peugeot 308 CC, Peugeot 308 SW, Peugeot 5008, Peugeot RCZ, Renault Mégane, Renault Scénic/Grand Scénic, Renault Twizy, SEAT Altea, SEAT León, Skoda Octavia, Skoda Yeti, Subaru BRZ, Subaru Forester, Subaru Legacy Outback, Subaru XV, Suzuki Grand Vitara, Toyota Avensis, Toyota GT 86, Toyota Prius, Toyota RAV4, Toyota Verso, Vauxhall Ampera, Vauxhall Astra, Vauxhall Astra GTC, Vauxhall Astra VXR, Vauxhall Cascada, Vauxhall Corsa VXR, Vauxhall Insignia, Vauxhall Insignia Sports Tourer, Vauxhall Mokka, Vauxhall Zafira, Vauxhall Zafira Tourer, Volkswagen Beetle, Volkswagen CC, Volkswagen Eos, Volkswagen Golf, Volkswagen Golf Plus, Volkswagen Jetta, Volkswagen Passat, Volkswagen Scirocco, Volkswagen Tiguan, Volkswagen Touran, Volvo S60, Volvo S80, Volvo V40, Volvo V60, Volvo V70, Volvo XC70.
    
    \item[Cluster 2.] Alfa Romeo MiTo, Aston Martin Cygnet, Audi A1, Chevrolet Aveo, Chevrolet Spark, Chrysler Ypsilon, Citroen C1, Citroen C3, Citroen C3 Picasso, Dacia Sandero, Fiat 500, Fiat 500 Abarth, Fiat 500L, Fiat Panda, Ford B-Max, Ford Fiesta, Honda Insight, Honda Jazz, Hyundai i10, Hyundai i20, Hyundai i30, Hyundai ix20, Kia Picanto, Kia Rio, Kia Venga, Mini Clubman, Mini Convertible, Mitsubishi i-MiEV, Mitsubishi Mirage, Nissan Micra, Nissan Note, Perodua MYVI, Peugeot 107, Peugeot 207 SW, Peugeot 208, Proton GEN-2, Proton Satria-Neo, Proton Savvy, Renault Clio, Renault Twingo, Renault Zoe, SEAT Mii, SEAT Toledo, Skoda Fabia, Skoda Roomster, Smart fortwo, Suzuki Alto, Suzuki Jimny, Suzuki Splash, Suzuki Swift, Suzuki Swift Sport, Suzuki SX4, Toyota Auris, Toyota AYGO, Toyota iQ, Toyota Yaris, Vauxhall Adam, Vauxhall Agila, Vauxhall Corsa, Vauxhall Meriva, Volkswagen Polo, Volkswagen Up.
    
    \item[Cluster 3.] Aston Martin DB9, Aston Martin DB9 Volante, Aston Martin V12 Zagato, Aston Martin Vanquish, Aston Martin Vantage, Aston Martin Vantage Roadster, Audi A7 Sportback, Audi A8, Audi R8, Audi R8 V10, Audi RS4 Avant, Bentley Continental, Bentley Continental GTC, Bentley Flying Spur, Bentley Mulsanne, BMW 6 Series, BMW 6 Series Convertible, BMW 6 Series Gran Coupé, BMW 7 Series, BMW M3, BMW M6, Bugatti Veyron, Chevrolet Camaro, Chrysler 300C, Corvette C6, Ferrari 458, Ferrari California, Ferrari F12, Ferrari FF, Infiniti EX, Infiniti G37, Infiniti M, Jaguar F-Type, Jaguar XFR, Jaguar XJ Series, Jaguar XK, Lamborghini Aventador, Lamborghini Gallardo, Lexus IS, Lotus Evora, Lotus Exige S, Maserati GranTurismo, Maserati Quattroporte, McLaren MP4-12C, Mercedes-Benz C63 AMG, Mercedes-Benz CL-Class, Mercedes-Benz CLS-Class, Mercedes-Benz E63 AMG, Mercedes-Benz SL 63, Mercedes-Benz SL-Class, Mercedes-Benz SLS, Morgan Aero, Morgan Roadster, Nissan 370Z, Noble M600, Pagani Huayra, Porsche 911, Porsche Boxster, Porsche Panamera, Rolls-Royce Ghost, Rolls-Royce Phantom, Rolls-Royce Phantom Coupé, Vauxhall VXR8, Volkswagen Phaeton.

    \item[Cluster 4.] Audi Q7, BMW X3, BMW X5, BMW X6, Chevrolet Captiva, Chrysler Grand Voyager, Ford Galaxy, Hyundai i800, Hyundai Santa Fe, Infiniti FX, Jeep Grand Cherokee, Jeep Wrangler, Kia Sorento, Land Rover Defender, Land Rover Discovery 4, Land Rover Range Rover, Land Rover Range Rover Sport, Lexus RX, Mercedes-Benz G-Class, Mercedes-Benz GL-Class, Mercedes-Benz M-Class, Mercedes-Benz R-Class, Mercedes-Benz S-Class, Mitsubishi Shogun, Nissan Pathfinder, Porsche Cayenne, SEAT Alhambra, Ssangyong Rodius, Toyota Land Cruiser, Toyota Land Cruiser V8, Vauxhall Antara, Volkswagen Sharan, Volkswagen Touareg, Volvo XC60, Volvo XC90.
\end{enumerate}

As highlighted in the main article, we analyze this data set via a cluster-oriented approach. However, not all the cars have a perfect assignment to the corresponding cluster. Some of them have a maximum a posteriori probability lower than $0.80$, with a moderately high second probability. For instance, the Renault Twizy, Citroen DS3, Nissan Juke, and Volkswagen Golf Plus are assigned to Cluster 1 with probabilities of $0.57$, $0.66$, $0.71$, and $0.77$, respectively, and to Cluster 2 with probabilities of $0.34$, $0.34$, $0.29$, and $0.23$, respectively. Meanwhile, the Lotus Elise is assigned to Cluster 1 with a probability of $0.67$, despite having a $0.32$ probability of being associated to Cluster 4. On the other hand, the Suzuki SX4 and Toyota Auris, which belong to Cluster 2 with probabilities $0.66$ and $0.74$, respectively, would be grouped with compact and mid-size sedans and crossovers (Cluster 1) as a second choice. Finally, the SEAT Alhambra, which belongs to Cluster 4 with a probability of $0.71$, would be grouped secondarily with compact and mid-size sedans and crossovers. Cluster 3, on the other hand, has assignment probabilities higher than $0.95$.

\bibliographystyle{jabes}
\bibliography{technometrics-bib}